%% file: IGMF_Paper.tex
\documentclass[iop]{emulateapj}
\usepackage{enumerate}
\usepackage{threeparttable}

\usepackage{graphicx}
\usepackage{epsfig}
\usepackage{subfigure}
\usepackage{chngpage}
\usepackage{amsmath}
\usepackage{url}

\begin{document}
\title{Intergalactic Magnetic Fields and Gamma Ray Observations of
  Extreme TeV Blazars}

\author{Timothy C. Arlen\footnote{Email: arlen@astro.ucla.edu},
  Vladimir V. Vassilev} \affil{Department of Physics and Astronomy,
  University of California, Los Angeles, CA 90095, USA} \author{Thomas
  Weisgarber, Scott P. Wakely} \affil{Enrico Fermi Institute,
  University of Chicago, Chicago, IL 60637, USA} \author{S. Yusef
  Shafi} \affil{Department of Electrical Engineering and Computer
  Sciences, University of California, Berkeley, CA 94720, USA}

\begin{abstract}
  The intergalactic magnetic field (IGMF) in cosmic voids can be
  indirectly probed through its effect on electromagnetic cascades
  initiated by a source of TeV gamma-rays, such as active galactic
  nuclei (AGN). AGN that are sufficiently luminous at TeV energies,
  ``extreme TeV blazars'' can produce detectable levels of secondary
  radiation from inverse Compton (IC) scattering of the electrons in
  the cascade, provided that the IGMF is not too large. We review
  recent work in the literature which utilizes this idea to derive
  constraints on the IGMF for three TeV-detected blazars-1ES 0229+200,
  1ES 1218+304, and RGB J0710+591, and we also investigate four other
  hard-spectrum TeV blazars in the same framework. Through a recently
  developed detailed 3D particle-tracking Monte Carlo code,
  incorporating all major effects of QED and cosmological expansion,
  we research effects of major uncertainties such as the spectral
  properties of the source, uncertainty in the UV - far IR
  extragalactic background light (EBL), undersampled Very High Energy
  (VHE; energy $\geq 100$ GeV) coverage, past history of gamma-ray
  emission, source vs. observer geometry, and jet AGN Doppler
  factor. The implications of these effects on the recently reported
  lower limits of the IGMF are thoroughly examined to conclude that
  presently available data are compatible with a zero IGMF
  hypothesis.
\end{abstract}

\maketitle


\section{Introduction}\label{section_intro}

Numerous observations have established the presence of magnetic fields
in our own galaxy and in other galaxies and galaxy clusters on the
order of a micro Gauss (see e.g.~\cite{GrassoRubinstein01,
  Widrow02,Carilli2002}). Furthermore, there is increasing theoretical
evidence based on numerical simulations of structure formation that
nano Gauss order fields permeate filaments of the large scale
structure (see e.g.~\cite{Ryu_2008}). However, an unambiguous
detection of the intergalactic magnetic field (IGMF), presumed to
exist in cosmic voids, which represent a significant fraction of the
volume of the universe, remains elusive. Such a field could be
produced, for example, through astrophysical mechanisms such as bulk
outflows of magnetized material from radio galaxies
(\cite{Kronberg94,Kronberg2001}), although it is unclear whether such
processes could efficiently fill the entire volume of the voids
(\cite{Zweibel2006}). Alternatively, processes such as the Biermann
battery mechanism (\cite{Biermann1950}) operating during phase
transitions in the early universe could produce the IGMF, provided
that its correlation length is sufficiently large to overcome magnetic
diffusion and survive to the present day (\cite{GrassoRubinstein01}).
``Primordial origin'' hypotheses, such as this one, are particularly
attractive because the IGMF could then play the role of the seed field
necessary in magnetohydrodynamic models commonly invoked to explain
the fields observed in galaxies and clusters
(\cite{Widrow02,Kulsrud_Zweibel_2008}). Consequently, the detection of
the IGMF could provide important insights for solving outstanding
problems of its origin and role in both the cosmology and astrophysics
of structure formation.

The standard observational technique used to detect weak magnetic
fields in galaxies, measuring the Faraday rotation of light from
distant quasars, is inadequate for detecting the IGMF for two
reasons. The first is that Faraday rotation measures the integrated
magnetic field along the line of sight and therefore the determination
of the IGMF relies on the subtraction of the imperfectly measured
Galactic magnetic field (\cite{Kronberg1982, Blasi1999}). The second,
and perhaps more important one, is that a sufficiently weak IGMF
strength will produce a Faraday rotation measure that is below the
resolution limit of currently employed techniques. Existing Faraday
rotation measurements place an upper limit of B $< 10^{-9}$ Gauss on
the strength of an IGMF with a correlation length greater than 1 Mpc,
and this limit weakens as the correlation length decreases until the
limits due to Zeeman splitting measurements of absorption lines in
distant quasars, become more constraining (as summarized in
\cite{NeronovSemikoz_Sensitivity09}). As a result of these two
effects, until recently, only upper limits on the IGMF strength have
been established.

A new measurement technique has emerged during the past few years
which may become a more sensitive tool for the measurement of IGMF
characteristics. This technique relies on observations of blazars,
believed to be AGN whose jet is oriented along the line of sight to
Earth, in the gamma-ray energy range from 100 MeV to greater than 10
TeV and is described by several authors (\cite{NeronovSemikoz06,
  Eungwan09, Dolag09, ENS09}). Briefly, TeV-scale gamma-rays from the
blazar interact with the EBL, producing an electron-positron pair. The
electrons and positrons then undergo IC scattering on the Cosmic
Microwave Background (CMB) radiation, producing secondary gamma-rays
of a lower energy than the primary. As a result, an electromagnetic
cascade develops. Because the pairs' trajectories depend on the
interactions with the magnetic field, the temporal and angular
profiles of this cascade emission at the GeV scale carry information
on the strength of the magnetic field in which the cascading occurs.
If cascading develops in the voids,
then spectral, temporal, and angular
characteristics of the secondary radiation will depend on IGMF
properties. A number of studies have characterized the spectral
properties of the secondary photons as well as the temporal profile,
commonly referred to as an ``echo'' (\cite{Plaga95, Ichiki08,
  Murase08}), and the angular profile, or ``halo,''
(\cite{AharonianCoppi94, NeronovSemikoz_Sensitivity09, Ahlers2011}).
Comparison of the characteristics of secondary gamma-ray radiation with
existing data can therefore become the methodology for studying
properties of the IGMF.

Most published studies of the subject have focused on the spectral
properties of the secondary radiation. For instance, utilizing simple
geometric models for the cascade, \cite{Neronov2010} and
\cite{Tavecchio2010} demonstrated that a lower limit on the IGMF
strength can be placed by requiring that the secondary gamma-ray
GeV-band emission does not exceed current measurements.  In another
study, \cite{Tavecchio_extreme_2011} performed detailed modeling of
the spectral energy distribution of four blazars, thereby reducing the
dependence of the conclusions on assumptions about the properties of
individual sources. Furthermore, \cite{Dermer2010} relaxed the
assumption that the characteristic time to build up the secondary
gamma radiation is less than the duration of the Very High Energy
(VHE; energy $\geq$ 100 GeV) activity of the source, and derived a
less constraining lower limit on the IGMF. Expanding on these
simplified geometric models, \cite{HaoHuan2011} described a
semi-analytic model employing the energy-dependent distributions of
electron and positron energies in the cascade and accounted for the
effects of the source lifetime. Monte Carlo simulations were used by
\cite{Dolag2010} and \cite{Taylor2011} to confirm and improve the
results of simplified geometric semi-analytic models. In particular,
\cite{Taylor2011} employed a three-dimensional particle tracking
simulation to follow the cascade development and derived a lower bound
on the strength of the IGMF which appeared to be consistent for three
blazars studied.

This paper reviews IGMF constraints derived from previous studies
particularly focusing on the conclusions of \cite{Taylor2011},
\cite{Neronov2010}, \cite{Tavecchio_extreme_2011}, \cite{Dolag2010},
and \cite{Dermer2010}. Because of the scientific importance of these
results, we wish to systematically understand all the ways in which
these IGMF constraints may fail. Particular emphasis is given to the
effects of major uncertainties such as the spectral properties of the
source, history of VHE activity, geometrical characteristics of the
source, and uncertainty due to the poorly known EBL spectral energy
density. The conclusions are derived based on the newly developed
Monte Carlo code which is described in section \ref{section_sims}
together with models for the EBL, IGMF, and AGN source model. Section
\ref{section_data} describes the data utilized in this paper, while
section \ref{section_systematics} provides detailed analysis of
individual sources and a comparison of the results of prior
publications. Section \ref{section_conclusion}, the discussion,
concludes the paper with a brief review of alternative interpretations
of the data.

\section{Numerical Simulations}\label{section_sims}
\input{Sims}

\section{Gamma Ray Data, Instrumentation, and Strategy for Data Analysis}\label{section_data}
\input{Data}

\section{IGMF Constraints and Effects of Systematic
  Uncertainties}\label{section_systematics}
\input{Systematics}

\section{Discussion}\label{section_conclusion}
\input{Discussion}


\bibliographystyle{apj}                       
\bibliography{apj-jour,Bibliography}

\end{document}

%% file: Sims.tex

VHE photons escaping a source such as an AGN jet interact with
surrounding diffuse photon fields and generate electromagetic
cascades. Cascading occurs in morphologically complex environments of
photon and magnetic fields of the host galaxies, the large scale
structure filaments, voids, etc. The amplitudes of the magnetic fields
and the density of the photon fields in these structures sensitively
affect the temporal, angular, and spectral evolution of the cascading,
secondary photons.

For example, the highest energy of the escaping photon is determined
by the spectral energy density of background photons of the host
galaxy. A 30 TeV photon progagating through a Milky Way-like galaxy
will have an optical depth of $\sim 1$, based on rough estimates of
the energy density of the Galaxy in the far infrared ($\sim 100$ $\mu
m$). Photons with energies higher than this will either be absorbed by
interactions with the galactic light or the CMB, on spatial scales
less than the size of the galaxy ($\leq 1$ Mpc). These photons will
initiate cascades under the influence of galactic magnetic fields
($\sim 10^{-5} - 10^{-6} $G, see, e.g. \cite{Widrow02}) which are
strong enough to isotropize the secondary photons of the cascade
(\cite{AharonianCoppi94}).

Photons with energies low enough to escape the galaxy are expected to
predominantly interact with the EBL. If this interaction occurs in the
environments of either galaxy clusters with magnetic fields of order
$10^{-6} - 10^{-7}$ G (see, e.g. \cite{Widrow02}) or intervening large
scale structure filaments, with magnetic fields $\gtrsim$ $10^{-12}$
G, then the secondary electrons will be isotropized, thereby
dramatically attenuating the observable flux of the secondary photons
produced by them. Effects of the electromagentic cascading may become
observable with present day instrumentation when the interaction with
the EBL occurs in the voids with characteristic magnetic fields
$\lesssim 10^{-12}$ G.

In order to explore in detail the potentially observable spectral,
angular, and temporal effects of cascading in the voids of the large
scale structure, we have developed a fully 3-dimensional Monte Carlo
code. It propagates individual particles of the cascade in a
cosmologically expanding universe and accounts for all QED
interactions with the EBL and CMB without simplifications. In the
following, details regarding these numerical simulations are provided
and some of its capabilities are illustrated with several
characteristic examples.

\subsection{Cosmology and EBL Model}\label{subsection_cosmology_ebl}

\begin{figure}  
\centering 
  \includegraphics[width=3.4in]{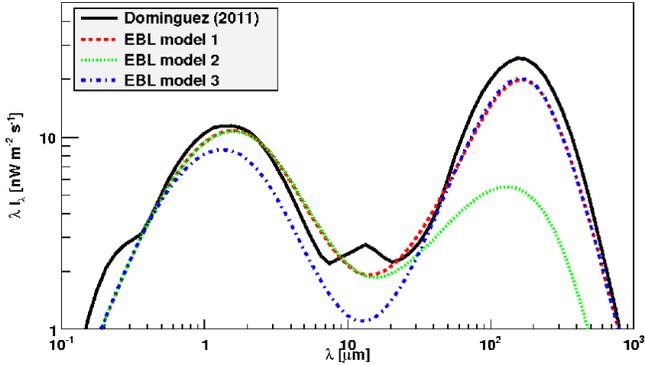}
  \caption{ \label{ebl_comparison_dominguez} Comparison between the EBL
      model of \cite{Dominguez_ebl_2010} and the 3 EBL models used in
      the present work-a typical EBL model (EBL model 1), one chosen
      with a low dust peak (EBL model 2), and one chosen with a low
      stellar peak (EBL model 3).}    
\end{figure}

Throughout this paper, a Friedmann Robertson Walker cosmology is used
with critical matter, cosmological constant, and radiation densities
given by $\Omega_M = 0.3$, $\Omega_{\Lambda}= 1.0 - \Omega_M$, and
$\Omega_R = 8.4\times10^{-5}$ respectively. The radiation density is
negligible for nearby sources of z $< 1$. The standard EBL model
chosen in the simulations at z = 0 is modeled with 6 energy density
points ($I_{\lambda}(240 \mu$m) = $15.9$ nW m$^{-2}$ sr$^{-1}$,
$I_{\lambda}(60) = 7.3$, $I_{\lambda}(12) = 2.0$, $I_{\lambda}(2.5) =
9.4$, $I_{\lambda}(0.38) = 3.2$, $I_{\lambda}(0.12) = 0.43$), and a
cubic spline is used to find the EBL energy density at intermediate
wavelengths as adopted in
\cite{Vassiliev2000}. Figure~\ref{ebl_comparison_dominguez} shows a
comparison of this EBL model (labeled EBL model 1) with one recently
proposed, which is based on various empirical observational data as
well as on EBL models existing in the literature
(\cite{Dominguez_ebl_2010}). Two additional EBL models are displayed,
one (labeled EBL model 2) with a substantially lower dust peak, and
another (labeled EBL model 3) with a reduced stellar peak. These two
models are explained in detail and used in
section~\ref{subsection_1es0229_analysis}.

In this study, we investigate sources with redshift $< 0.2$ and thus,
we neglect evolution of the EBL in the comoving reference frame; only
the effect of cosmological expansion on the EBL energy density and
energy of EBL photons is taken into account. By comparison with
\cite{Dominguez_ebl_2010}, we do note that evolutionary effects of the
EBL may need to be taken into account for sources with z $\gtrsim
0.3$.

\subsection{Magnetic Field
  Model}\label{subsection_magnetic_field_model}

The observational effects of magnetic fields in the gamma-ray signal
of extragalactic sources originate in the deflection of the charged
particles from the trajectory of the primary photon and subsequent
deflection of the direction of the secondary IC scattered
photons. Different regimes of influence of the IGMF can be determined
by exploring the interplay between the characteristic coherence
lengths of the IGMF, the e$^{\pm}$ IC cooling length, and the distance
from the interaction point to the observer. For the production of
secondary photons above 100 MeV, which is of relevance to this paper,
the pair-produced e$^{\pm}$ should have energies on the scale of
$\gtrsim$ 100 GeV. In the simulation, electrons are tracked down to
energies of 75 GeV. A 1 TeV electron loses its energy to IC scattering
on the CMB over a characteristic length of 0.4 Mpc, $\lambda_{IC}
\propto E_{e}^{-1}$. For the coherence length of magnetic fields,
$\lambda_{IGMF} \gg \lambda_{IC}$, the IC scattering effectively
happens in the environment of nearly constant B$_{IGMF}$. For
$\lambda_{IGMF} \ll \lambda_{IC}$, the deflection of charged particles
occurs in the non-coherent regime and leads to smaller deflection. In
this study, the coherence length of the IGMF is conservatively chosen
to be 1 Mpc, which corresponds to coherent scattering for all photons
of interest (with energies $>$ 100 MeV). It has been observed that the
reversal field length of the magnetic fields in clusters of galaxies
is on the scale of 10 - 100 kpc (\cite{GrassoRubinstein01}) and
reflects the spatial scales of the distribution of plasma. Thus, the
magnetic fields in the voids with significantly larger characteristic
plasma distribution scales, are likely to have coherence lengths much
larger than this.

The IGMF is modeled in the code as a system of cubic cells with a size
equal to the coherence length of the IGMF and magnetic field
amplitudes which are equal in value but randomly oriented in
direction. To preserve cosmic variance, each cell is assigned an
orientation when the first particle of the electromagnetic cascade
propagates through it. If the cascade develops over a large number of
these magnetic field cells, the observable effects of the IGMF are
randomized. However, if the distance to the observer is comparable to
the size of the magnetic field domain, the observational effects of
the randomly chosen direction become significant. For this study, we
analyze sources at distances greater than a few hundred Mpc (z $>$
0.1). The evolution of the IGMF is unknown, but is likely dominated by
cosmological expansion for z $<$ 1. Therefore, the size of the domains
and the magnetic field value are evolved with the standard
(1+z)$^{-1}$ and (1+z)$^{2}$ dependences. The values of the magnetic
field reported in this paper, refer to the values at z = 0.

\begin{figure*}[t] 
  \centering
  \mbox{\subfigure{\includegraphics[width=3.4in]{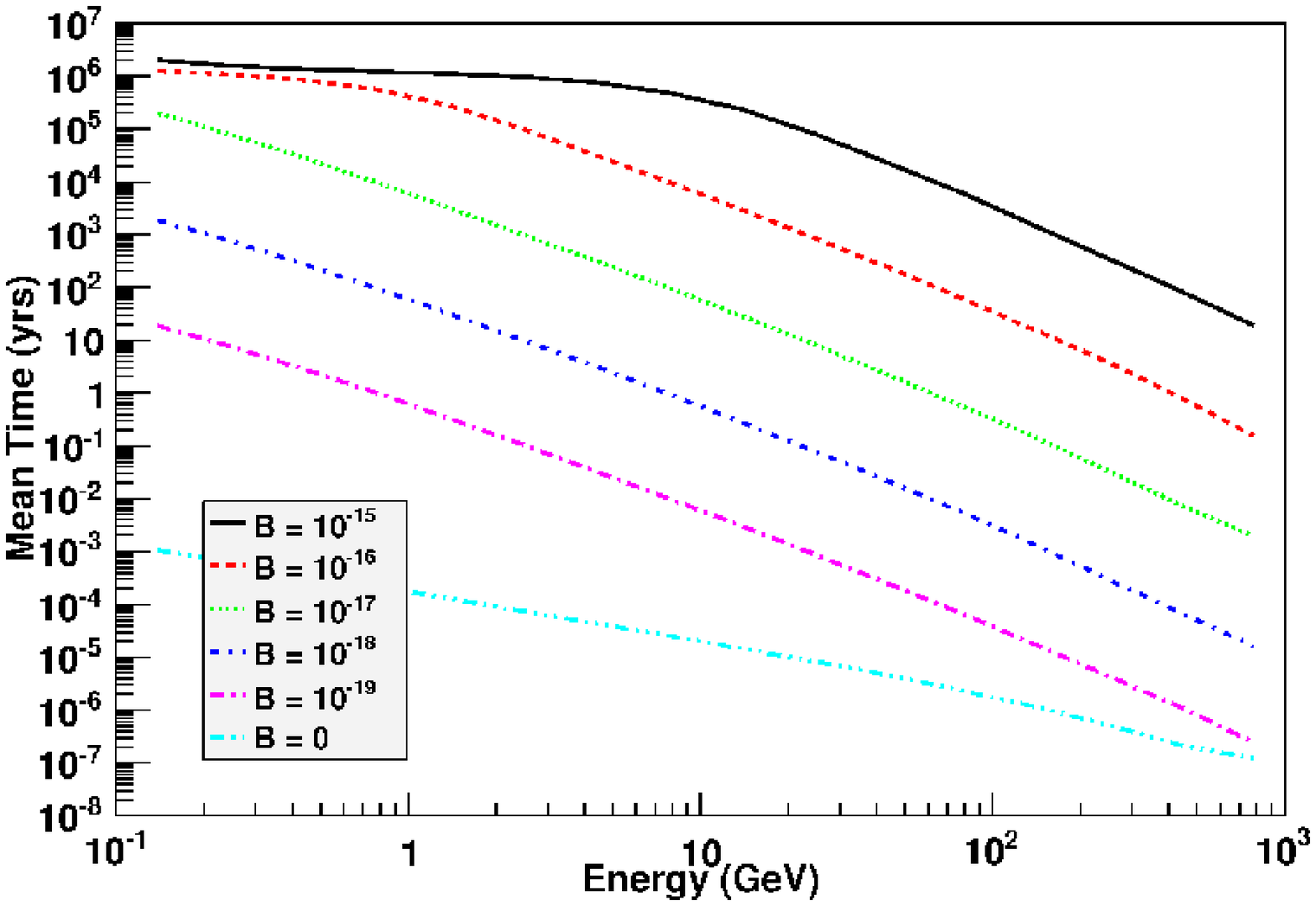}}\quad
    \subfigure{\includegraphics[width=3.4in]{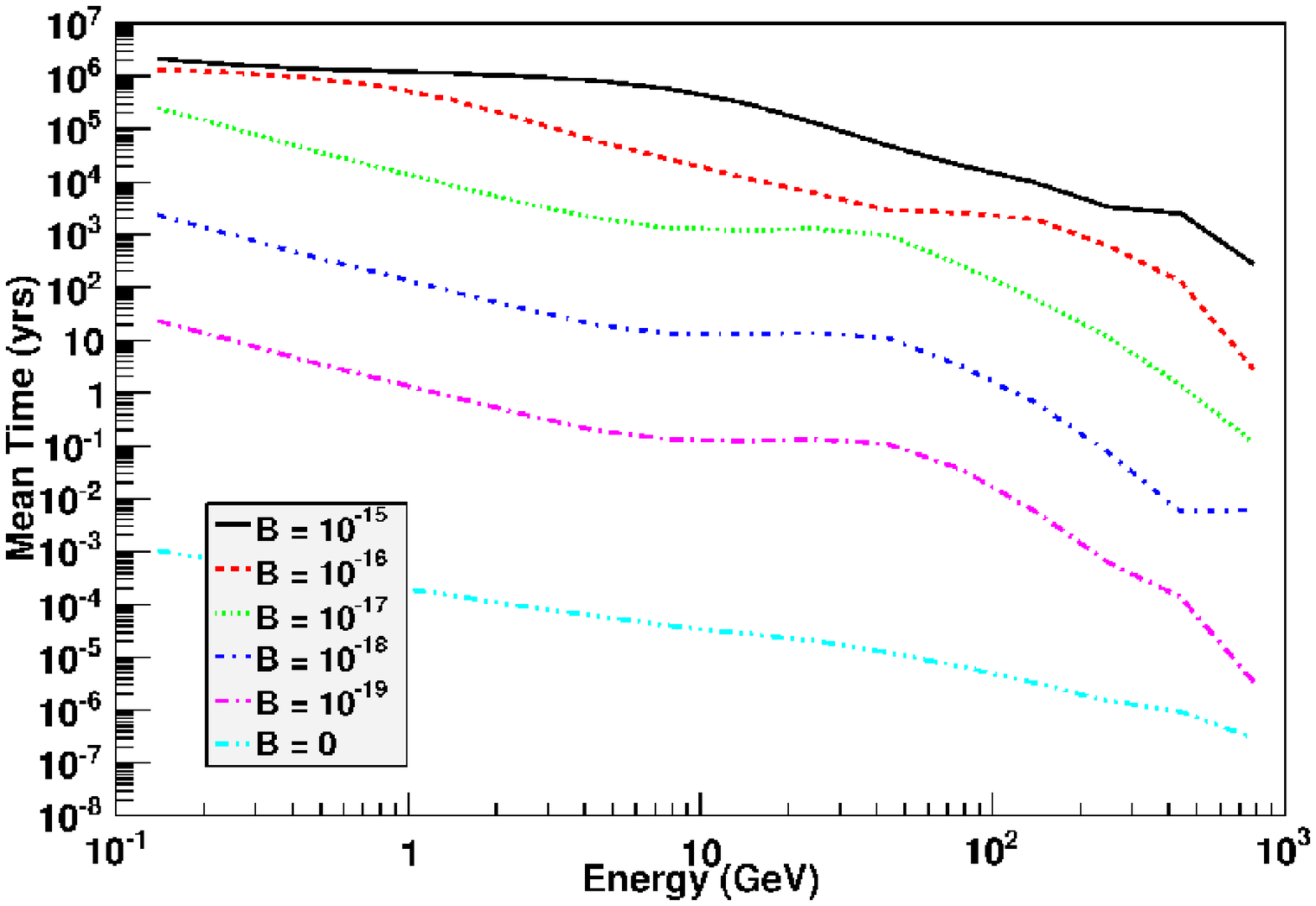} }}
\caption{ \label{mean_time_delay_compare} Mean time delay of secondary
  gamma-rays produced by a monoenergetic beam of 100 TeV photons at a
  redshift of $z = 0.13$, for a varying $B_{IGMF}$ with (a) (left) the EBL
  excluded as a target photon field in the IC scattering of electrons
  and (b) (right) with the EBL included as a target field.}
\end{figure*}

\begin{figure}
  \includegraphics[width=3.4in]{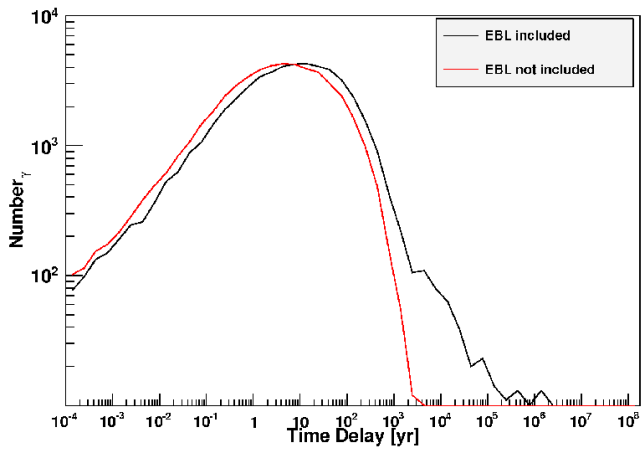}
  \caption { \label{mean_time_egy_bin_compare} Distribution of
    arrival times of secondary photons in a single energy bin of
    10.0 - 17.78 GeV at B$_{IGMF}$ = 10$^{-17}$ G with and without
    the EBL photons included as a target for the IC scattering
    computation.}
\end{figure}

Figure~\ref{mean_time_delay_compare} illustrates the mean time delay
of secondary photons produced by a monoenergetic beam of 100 TeV
primary photons at a redshift of z = 0.13. The photons arriving at the
observer are integrated over an aperture radius of 10.0$^\circ$. For
each of the energy bins (4 per decade), the mean delay time is
computed for 6 magnetic fields including the zero field case. The time
delay for the latter is due to QED scattering of the secondary
particles of the pair production and IC scattering processes. For a 10
GeV photon, the time delay amounts to about a half hour and for 0.1
GeV photons, the time delay is about 10 hours. The saturation effect
at low energies is due to the aperture
cutoff. Figure~\ref{mean_time_delay_compare}a shows the mean time
delay with no EBL photons as IC targets in the simulations. It is
consistent with \cite{Taylor2011} Figure 2, and follows the spectrum
of T$_{\mathrm{delay}}$ $\propto E^{-\frac{5}{2}}$, derived analytically by
making several simplifications, as explained
in~\cite{NeronovSemikoz_Sensitivity09}. Figure
~\ref{mean_time_delay_compare}b shows the result when the EBL field is
included. The time delay for a non-zero field of secondary photons
with energies above 10 GeV is significantly increased, because
electrons can move farther from the position of pair production and
still scatter higher energy EBL photons towards the observer
increasing the average time delay in a given energy bin. The effect
can be seen more clearly in Figure \ref{mean_time_egy_bin_compare},
which shows the distribution of arrival times of secondary photons in
a single energy bin of 10.0 - 17.78 GeV with and without the target
EBL photons for IC scattering.

\subsection{Gamma Ray Source Model}\label{subsection_source_model}
The gamma-ray source model employed in the simulations is based on
leading theoretical speculations about the nature of the TeV blazar
source (see, e.g., \cite{Urry_Padovani_1995}). In the reference frame
of the blazar jet, the VHE photons are distributed isotropically with
a power law spectrum of index $\alpha$. Once this distribution is
boosted into the reference frame of the host galaxy with a Doppler
factor of $\Gamma=(1-\beta^2)^{-1/2}$, it can be parameterized as

\[
\epsilon\frac{dF}{d\epsilon}(\epsilon,\theta_{\text{v}}) =
F_o\delta^{2}\left(\frac{\epsilon}{\epsilon_o\delta}\right)^{-\alpha+1}e^{-\epsilon/\epsilon_c},
\]

\noindent where $\epsilon$ is the photon energy in the reference frame
of the host galaxy, $\delta$ = $[\Gamma$(1 -
$\beta\cos\theta_{\text{v}}$)]$^{-1}$, $\theta_{\text{v}}$ is the
viewing angle from the blazar jet axis to the line of sight of the
observer, $F_o$ is a flux normalization factor, and $\epsilon_o$ is an
energy scale factor. To account for absorption of photons at the
highest energies inside the host galaxy, an exponential cutoff at
energy $\epsilon_c$ is introduced. This model is characterized by 5
physical parameters, and is sufficient to model the VHE part of the
spectrum ($\gtrsim$ 100 GeV).

Since the energy range of interest for this study covers more than 5
decades of energy (from 0.1 GeV to $>$ 10 TeV) the HE part of the
source spectrum ($\lesssim$ 100 GeV) is allowed to obey a power law
with a different spectral index $\gamma$

\begin{equation}\label{broken_power_law_model}
  \epsilon\frac{dF}{d\epsilon} = F_0\delta^{2}
\begin{cases}
  \left( \frac{\epsilon}{\epsilon_B\delta} \right)^{-\gamma+1}
  \text{exp}\left(\frac{-\epsilon}{\epsilon_{c}} \right) &
  \frac{\epsilon}{\epsilon_B\delta} < 1 \\
  \left( \frac{\epsilon}{\epsilon_B\delta} \right)^{-\alpha+1}
  \text{exp}\left(\frac{-\epsilon}{\epsilon_{c}} \right) &
  \frac{\epsilon}{\epsilon_B\delta} > 1
\end{cases}.
\end{equation}

\noindent The seventh parameter, $\epsilon_B$, introduced in
this equation, is the spectral break energy. This multi-parameter
gamma-ray source spectrum is given at the redshift of the host galaxy
and is necessary and sufficient to satisfy observational data of TeV
blazars in both the HE and VHE regimes.

\begin{figure*}[!t] 
  \centering
  \mbox{\subfigure{\includegraphics[width=3.5in]{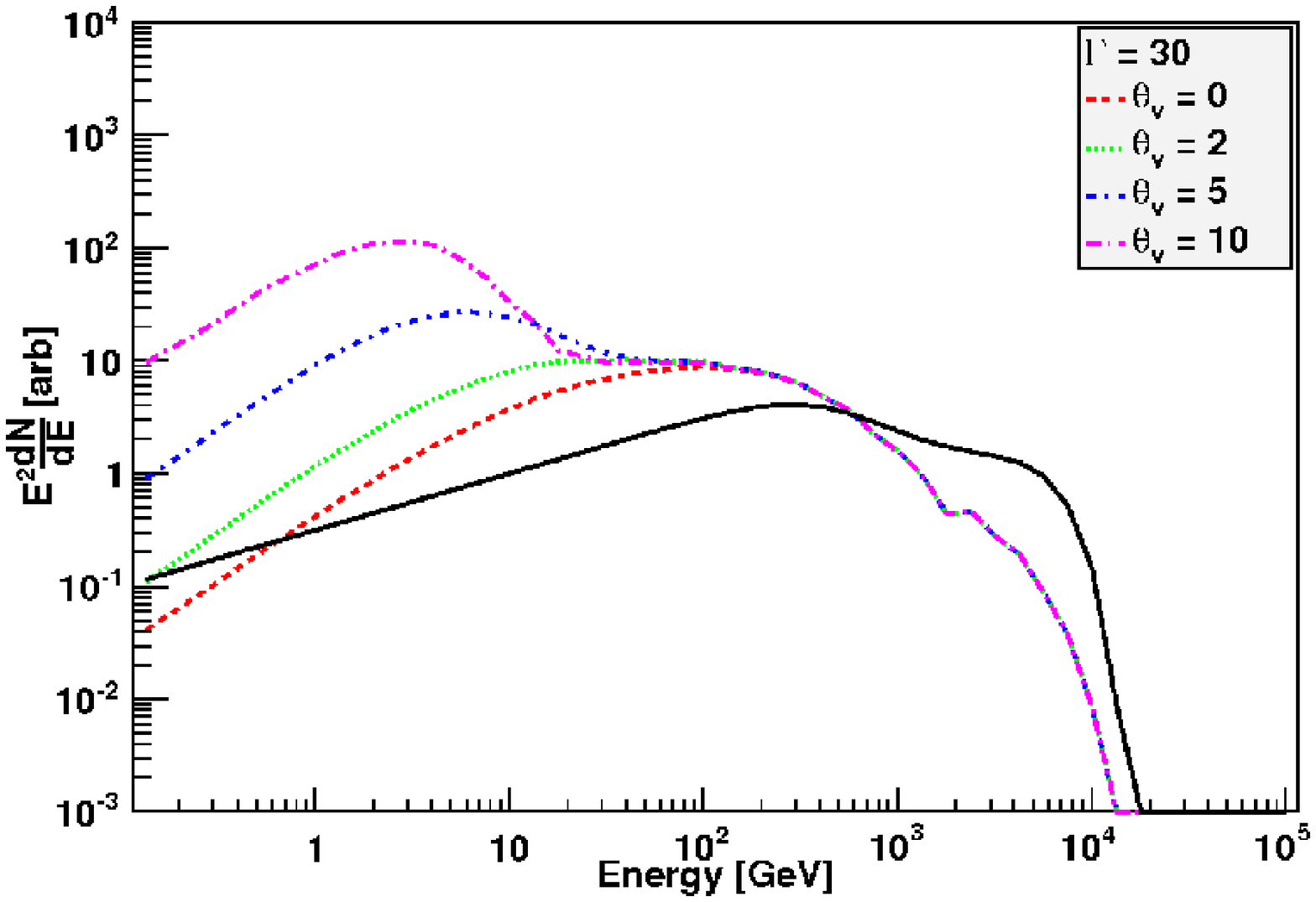}}\quad
    \subfigure{\includegraphics[width=3.5in]{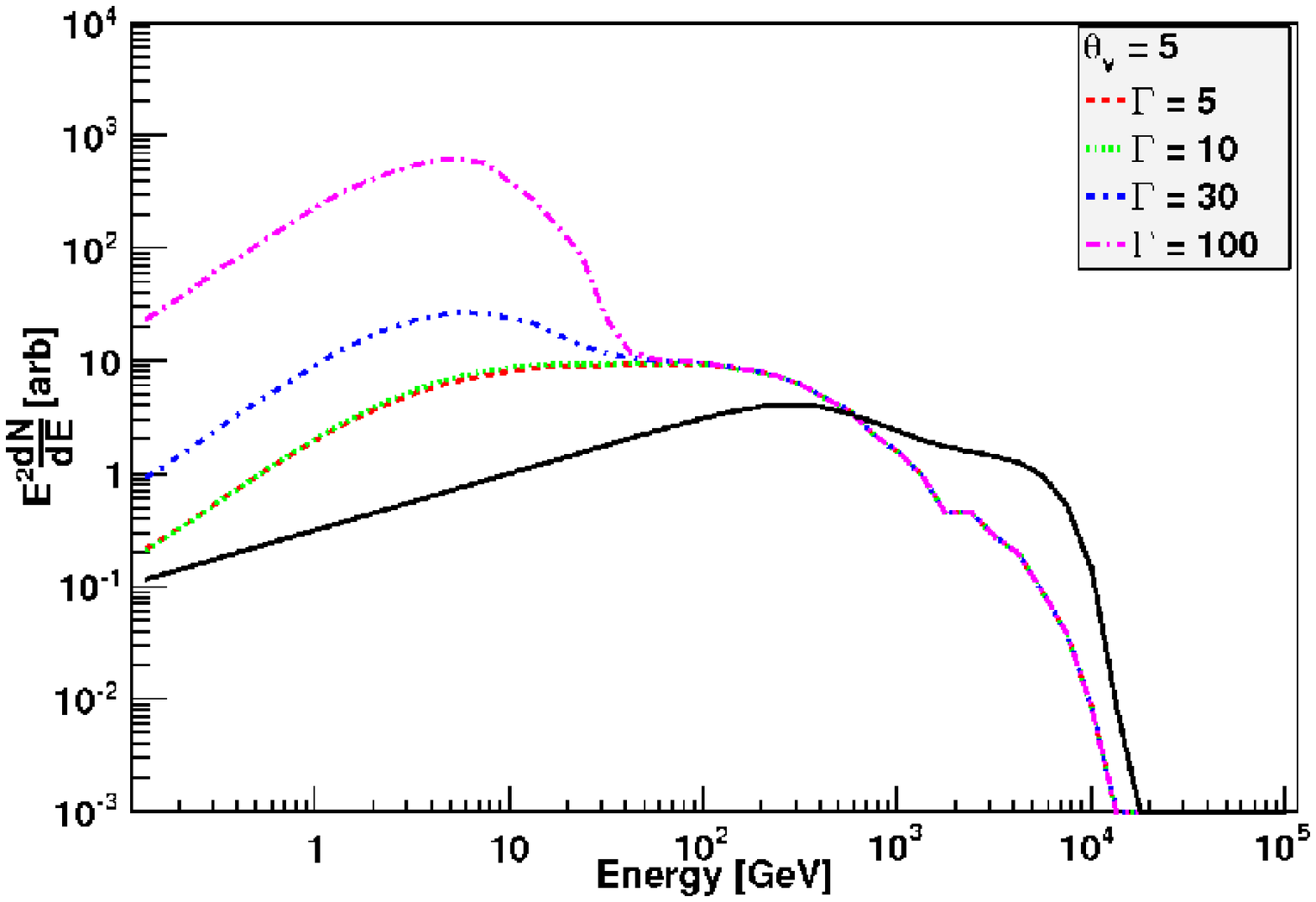}
    }}
\caption{ \label{spectra_geometry_scan} Simulations of a source at
  z=0.13, with $\alpha$ = $\gamma$ = 1.5, $\epsilon_c$ =
  30 TeV, and B$_{\mathrm{IGMF}}$=10$^{-15}$ for a) (left) $\Gamma$ =
  30 and four different viewing angles and b) (right)
  $\theta_{\text{v}}$ = 5$^\circ$ and four different jet boost
  factors.}
\end{figure*}

To illustrate the effects of different gamma-ray source model
parameters, simulations are shown for z=0.13, B=10$^{-15}$ G,
$\epsilon_c$ = 30 TeV, $\alpha$ = $\gamma$ = 1.5. Figure
\ref{spectra_geometry_scan} shows the observed spectra of a source
with the above parameters when it is viewed at different angles and
different jet Doppler factors. All photons arriving within an aperture
of 5$^\circ$ around the source are integrated. The black (solid) line
shows the spectrum of the prompt photons reaching the observer, which
is modified by absorption on the EBL. The prompt photon spectrum is
normalized to the same level for different viewing angles and
different Doppler factors. Figure \ref{spectra_geometry_scan}a shows
the spectrum of secondary photons when the source is viewed at
$\theta_{\text{v}}$ = 0$^\circ$, 2$^\circ$, 5$^\circ$, 10$^\circ$ and
Figure \ref{spectra_geometry_scan}b shows the spectrum of secondary
photons for $\Gamma$ = 5, 10, 30, 100. Viewing a source with the same
prompt spectral energy density (SED), but with increasing viewing
angle or Doppler factor implies the power in the jet must increase.

The higher energy end of the secondary photon spectrum ($\gtrsim$ 20
GeV) is unchanged for different viewing angles and jet Doppler factors
because photons at these energies are generated from primary photons
leaving the source with nearly zero deflection from the angle to the
observer and the flux of these photons is directly proportional to
that of the prompt photons. The lower energy end of the spectrum
($\lesssim$ 20 GeV) is generated by secondary electrons of lower
energies, the trajectories of which are significantly deflected from
that of the primary photon producing them. Additionally, the cooling
length due to IC losses increases inversely proportionally to energy,
allowing significantly larger deflection angles. The position of the
peak of the SED is correlated to the value of the magnetic field, and
its intensity is proportional to the overall power in the jet which
increases with larger viewing angle or Doppler factors. When the
characteristic angular size of the jet becomes larger than the viewing
angle ($1/\Gamma > \theta_{\text{v}}$), the SED is nearly
independent of the Doppler factor.

\begin{figure*}[!t]
  \begin{adjustwidth}{-0.0in}{}
      \centering
      \includegraphics[width=7in]{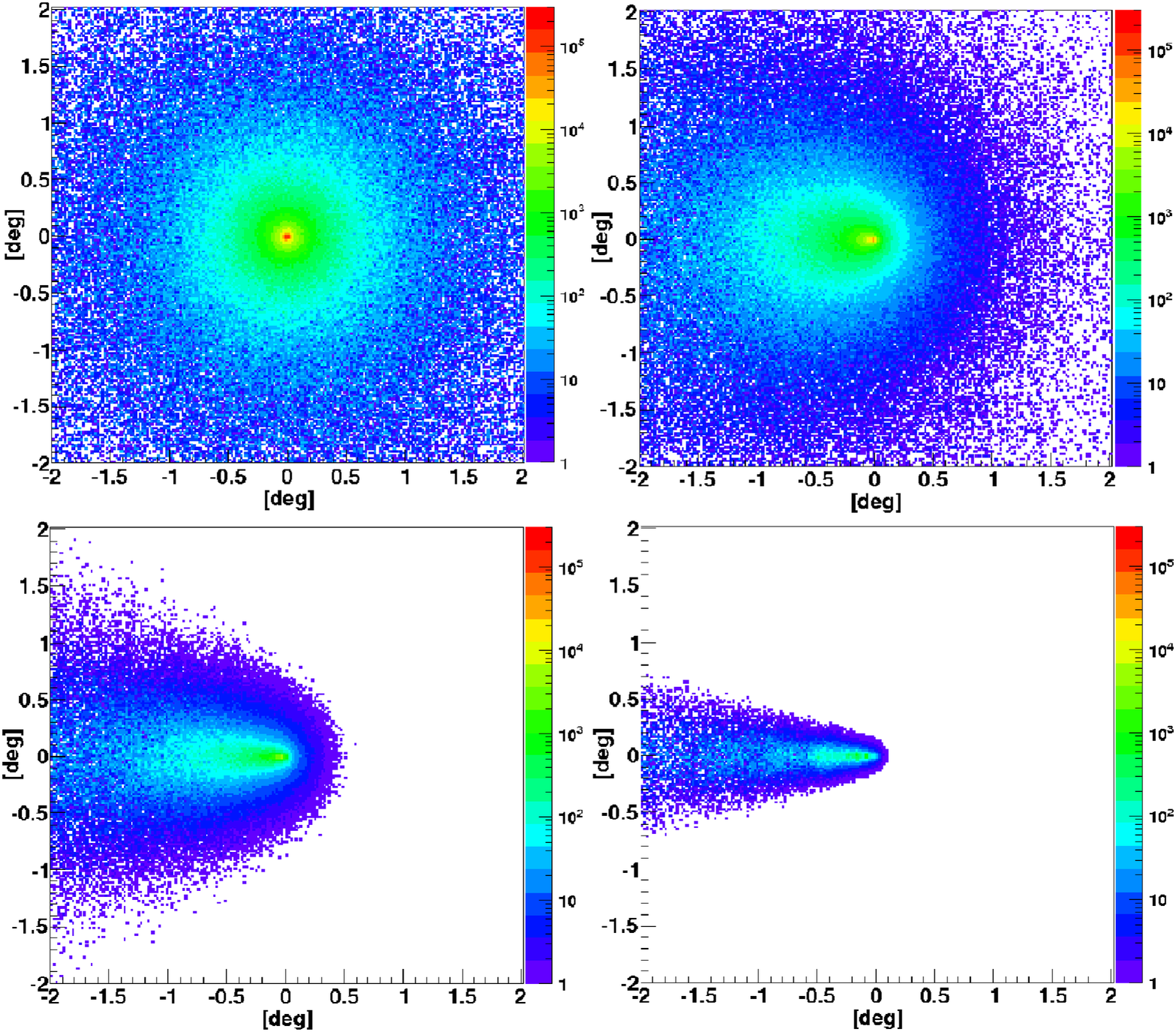}
      \caption { \label{he_skymap} Skymaps for source at z=0.13,
        $\alpha$ = 1.5, E$_{\mathrm{C}}$ = 30 TeV, and
        B$_{\mathrm{IGMF}}$=10$^{-15}$, $\theta_{\mathrm{obs}}$ =
        0$^{\circ}$ (upper left), 2$^\circ$ (upper right), 5$^{\circ}$
        (lower left), 10$^{\circ}$ (lower right) }
  \end{adjustwidth}
\end{figure*}

Figure \ref{he_skymap} displays a simulation of the photon arrival
distribution from a source at z = 0.13, with gamma-ray source model
parameters $\alpha$ = $\gamma$ = 1.5, $\epsilon_c$ = 30 TeV,
B$_{IGMF}$ = 10$^{-15}$ gauss, $\Gamma$ = 30, at four different
observing angles, $\theta_{\text{v}}$ = 0$^{\circ}$, 2$^{\circ}$
5$^{\circ}$, and 10$^{\circ}$. The main trend in these figures is that
the overall luminosity of prompt and secondary emission rapidly
declines as the observing angle increases, and the the photon
distribution around the source becomes increasingly axially
non-symmetric, when $\theta_{\text{v}} \sim 1/\Gamma$. The luminosity
of the secondary photons relative to the prompt emission rapidly
declines with increasing viewing angle, thus, detecting non-axially
symmetrical halos around AGN with existing instrumentation may prove
challenging. Figure \ref{he_skymap} is in qualitative agreement with
previously reported findings in the study of these effects
by~\cite{Neronov_DegreeScaleJets_2011}.

\subsection{Particle Propagation and
  Interaction}\label{subsection_propagation_and_interaction} The
development of cascades in intergalactic space intiated by VHE photons
is modeled by full 3-dimensional ray tracing and interaction of all
particles: electrons and photons. The processes of pair production and
IC scattering are treated with the use of full QED cross sections
(\cite{Jelley1966,Gould1967}) implemented without simplification. For
example, both the EBL and CMB are included in IC scattering as seed
photon fields and scattering includes the Klein Nishina regime. Both
pair production and IC scattering are treated in the code through the
sampling of marginal probability density functions. For example, the
mean free path $\lambda$ of electrons of energy $E$ due to IC
scattering is given by

\begin{equation}
\label{ics_mfp}
\frac{1}{\lambda} = \frac{1}{2}\left(\frac{m_e^2}{2E}\right)^2\frac{1}{\beta}\int_0^\infty d\epsilon\frac{n_\epsilon(\epsilon)}{\epsilon^2}\int_{\frac{2E\epsilon(1-\beta)}{m_e^2}}^{\frac{2E\epsilon(1+\beta)}{m_e^2}}dx\sigma_{\gamma e}(x)x,
\end{equation}
where $\beta$ is the speed of the electron, $\sigma_{\gamma e}(x)$ is
the differential IC cross section, $x=
(2E\epsilon/m_e^2)(1-\beta\cos\theta)$ where $\theta$ is the collision
angle and $n_\epsilon(\epsilon)$ is the spectral energy density of the
isotropic seed photon field. The code samples the propagation length
of the particle with the use of the mean free path, $\lambda$ to
determine the position of the interaction. It then generates the
marginal probability density for an interaction of the electron with a
given energy photon and samples the energy of the interacted
photon. The process is continued by sampling the interaction angle of
a given energy photon by generating a marginal probability density of
the angular distribution of seed photons. The azimuthal angle is
sampled randomly from a uniform distribution. At the completion of
this process, the kinematics of the interaction is fully determined,
allowing the computation of the energy-momentum vector of the outgoing
photon and electron. The pair production process is treated similarly.

Propagation of both photons and electrons is simulated in a
cosmologically expanding universe. Energy losses of photons are only
due to cosmological expansion, while energy losses of electrons also
include IC cooling. All photons with energies above 100 MeV are
tracked until they reach the z=0 surface, at which point the position
and momentum 4-vectors are saved. All electrons are tracked until
their energy decreases to less than 75 GeV, or until they reach the
surface of z=0. Particular care is given to the computation of time
delays. The time delay of individual particles is computed relative to
the arrival time of a putative photon propagating directly from the
source to the current position of the particle. This procedure is
adopted to maintain precision in numerical simulations down to minute
timescales. The computation accuracy of time delays was verified using
an arbritrary precision numerical integrator developed at Lawrence
Berkeley National
Laboratory\footnote{\url{http://crd-legacy.lbl.gov/$\sim$dhbailey/mpdist/}}. The
full solution for the equations of motion of electrons in a
cosmologically expanding universe and under the influence of constant
magnetic field are used to track the position and momentum 4-vectors
between IC interactions.

Previous models ranging from analytic to Monte Carlo codes have been
reported in the literature to generate constraints on the IGMF
(\cite{Neronov2010, Tavecchio2010, Dermer2010, Essey2010, Dolag2010,
Taylor2011, HaoHuan2011}). Each one has utilized various degrees of
simplification through the use of solutions of one or two dimensional
kinetic equations or simplified analytical approximations of QED
interactions or geometrical effects of cascade development. For
example, the VHE secondary photons produced by the highest energy
electrons are capable of generating second and higher orders of the EM
cascades, which are typically neglected in analytic and semi-analytic
codes, whereas the majority of Monte Carlo codes used in prior studies
of this subject, have neglected a full 3-dimensional simulation.

\begin{figure*}[t] 
  \centering
  \mbox{\subfigure{\includegraphics[width=3.4in]{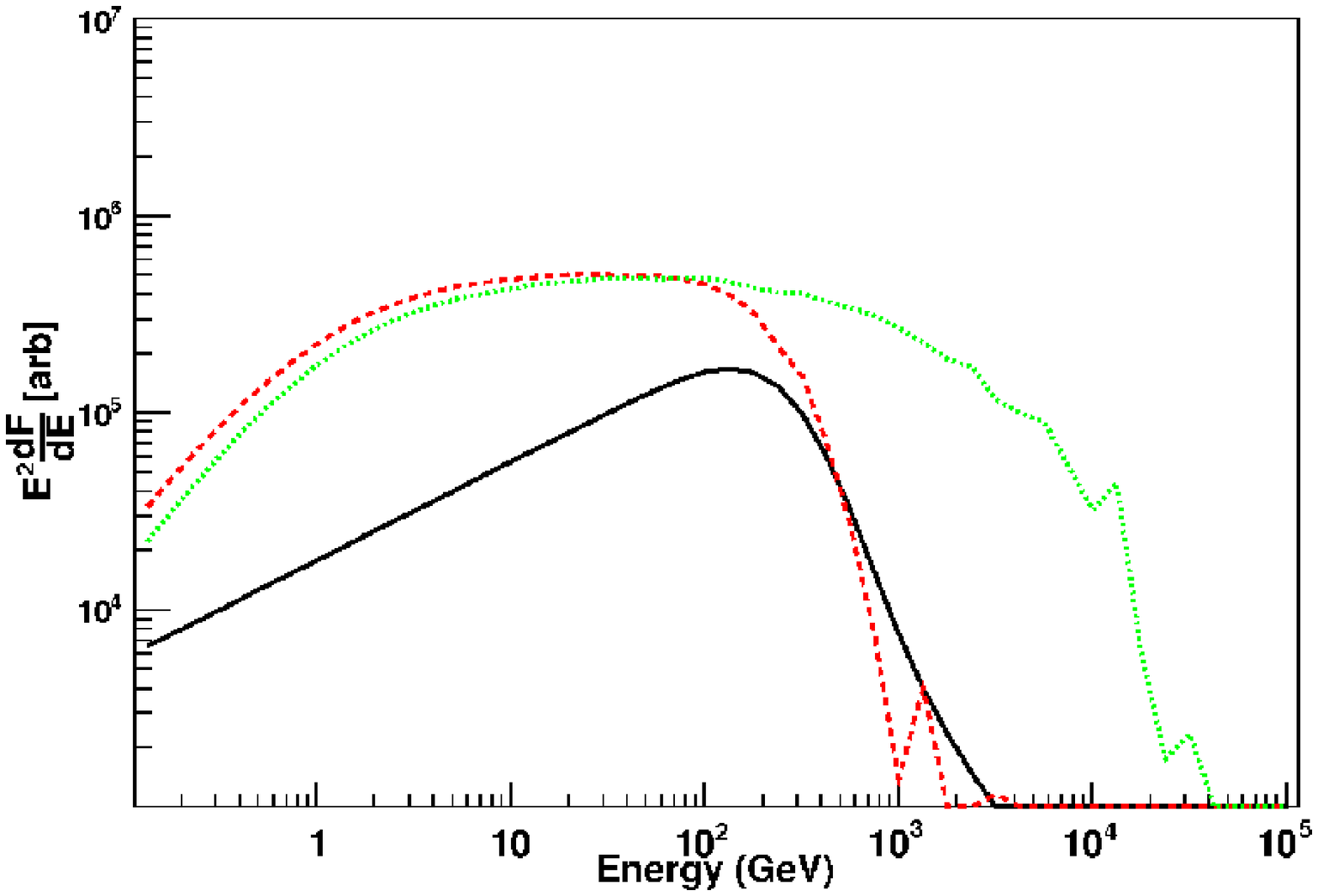}}\quad
    \subfigure{\includegraphics[width=3.4in]{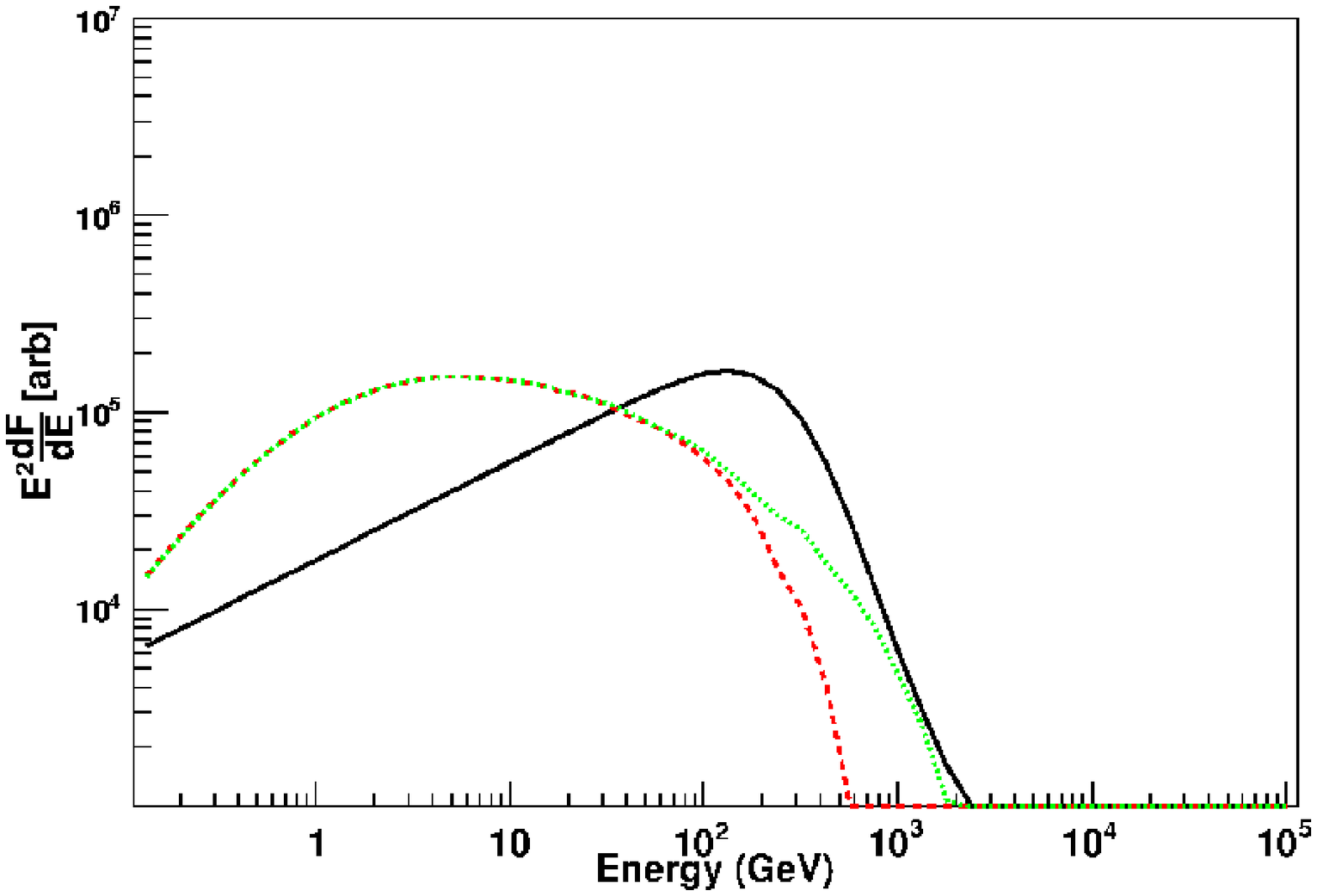}
    }}

  \caption{ \label{single_gen_compare} Comparison of spectra produced
    by both neglecting and including multiple generations in the
    cascade for a source at z=0.3, with B$_{IGMF}$ = $10^{-16}$ G,
    $\theta_{\text{v}}$ = 0, and $\Gamma$ = 10. (integrated over a 0.5
    deg aperture). The black (solid) line is the prompt, direct
    spectrum, the red (long dashed) line shows the secondary spectrum
    produced when multiple generations of the cascade are included,
    and the green (small dashed) line shows the secondary spectrum
    when only a single generation of cascading is allowed in the
    simulations. Two different source spectra were used a) (left)
    spectral index $\alpha$ = 1.5, cutoff energy, $\epsilon_c$ = 30
    TeV and b) (right) $\alpha$ = 1.5, $\epsilon_c$ = 5 TeV. }
\end{figure*}

Figure~\ref{single_gen_compare} illustrates the importance of higher
order cascading to correctly describe secondary photons with energies
$\gtrsim 200$ GeV. This figure shows the results of the simulations of
secondary emission produced for a source at z=$0.3$, with B$_{IGMF} =
10^{-16}$ G, $\theta_{\text{v}} = 0$, and $\Gamma = 10$. The two
panels display different intrinsic spectra for the source; panel a) is
obtained with $\alpha = 1.5$ and $\epsilon_c = 30$ TeV, and panel b)
corresponds to $\alpha = 1.5$ and $\epsilon_c = 5$ TeV. Both panels
show the direct differential flux energy density (dFED) of the source
together with the time-integrated secondary dFED, within 0.5$^\circ$
from the position of the source. The two lines of the secondary dFED
shown on the figure are obtained with second order cascading included
or not included in the simulation. It is evident that the secondary
dFED $\gtrsim 200$ GeV is strongly affected by this assumption and if
this simplification is made, the Monte Carlo simulations may
over-predict the total dFED in the VHE energy range.

%% file: Data.tex

The gamma-ray data used in this study are compiled from the data sets of
two types of instruments. In the HE regime, the data were obtained by
the \emph{Fermi}-LAT instrument and the data were processed utilizing
publically available tools, version v9r23p1, with the update from
November 6, 2011. 
For the VHE regime, data previously reported by the Imaging
Atmospheric Cherenkov Telescopes (IACT) instruments VERITAS and HESS
were used. In both the HE and VHE energy regimes, the software used
for analysis interprets the gamma-ray flux as originating from a point
source; no angular extension is assumed. Since the secondary photons
of intergalactic cascades inherently represent an extended source, the
methods of comparison of simulations and data are instrument-specific.

\subsection{VHE Data Considerations}

In most of the following discussion, the validity of the $B_{IGMF} =
0$ hypothesis (hereafter called $H0$) is examined. Under this
assumption, the source of the gamma-ray photons has an angular extent
due to the QED pair production and IC scattering processes. The
angular size of this extension is much smaller than the gamma-ray
point spread function (PSF) of both IACTs and the Fermi-LAT
instrument, and therefore, the gamma-ray sources are point-source
like. In the VHE regime, the point source assumption becomes invalid
for $B_{IGMF}$ $\gtrsim$ 10$^{-15}$ G because the PSF of IACTs and the
angular extension of the source become comparable. This requirement,
however, is dependent on the higher energy end of the spectral energy
density ($\gtrsim$ 10 TeV) of the source and its history of activity.

To process simulated data in the VHE energy regime, the instrumental
PSF of

\begin{equation}\label{hess_psf}
\frac{dP}{d\theta^2} = \frac{(1-\alpha)}{\theta_1^2}
\text{exp}\left(-\frac{\theta^2}{\theta_1^2} \right) + \frac{\alpha}{\theta_2^2}
\text{exp}\left(-\frac{\theta^2}{\theta_2^2} \right)
\end{equation}

\noindent is used as suggested in~\cite{Hess_crab_06}. The typical
values of parameters of the PSF are $\theta_1$ = 0.06505$^\circ$,
$\theta_2$ = 0.1697$^\circ$, $\alpha$ = 0.505. These parameters are
assumed to be energy independent in analyses of IACT data reported and
therefore, we make a similar assumption for processing simulated
data. For each simulated photon, Equation~\ref{hess_psf} is used to
find the probability for the given photon to be reconstructed within
the 68 \% containment radius from the position of the putative point
source. The effective point source flux from an AGN is estimated as
the total simulated flux within the 68 \% containment radius divided
by 0.68. This method of flux evaluation for each energy bin accurately
models the IACT data as reported in the literature.

The VHE data set used in this paper is summarized in Table
\ref{iact_data_table}. The data sets of the first three sources (RGB
J0710+591, 1ES 1218+304, and 1ES 0229+200) are identical to those used
in~\cite{Taylor2011}, except that an additional data set for 1ES
1218+304 obtained by the VERITAS Collaboration just before the launch
of the \emph{Fermi} satellite (on August 4, 2008 or MJD 54682) is
considered in this study. As reported in~\cite{VERITAS_1ES1218_2010},
the activity of the source is nearly identical during these
non-overlapping periods, except for an elevated flux of the source
peaking at the level of $\sim$20 \% Crab over a few nights of
observations. The data set for 1ES 0229+200 was also obtained prior to
the launch of the \emph{Fermi} satellite. Based on the report from
\cite{PerkinsPoster2010}, the activity of the source as measured by
VERITAS during the second year of the \emph{Fermi} mission, appeared
to resemble the reported SED by the HESS collaboration prior to the
launch of \emph{Fermi} satellite~\citep{HESS_1ES0229}. VERITAS has
continued monitoring this source since the Fermi launch and
tentatively detected flux variations on a yearly time scale (private
communication, J.S. Perkins and VERITAS collaboration).  Finally the
data sets for four other extreme TeV blazars (1ES 0347-121, 1ES
1101-232, H 2356-309, and RGB J0152+017) were taken from the discovery
publications by the HESS collaboration, and all of these sources were
observed prior to the start of the \emph{Fermi} mission.

\begin{deluxetable*}{ccccccc}
  \tablecolumns{7} \tablewidth{0pc}
  \tablecaption{ \label{iact_data_table} Data of RGB J0710+591 are
    taken by VERITAS~\cite{VERITAS_RGBJ0710}. Source 1ES 1218+304 was
    observed by VERITAS and the results were published in two separate
    submissions \cite{VERITAS_1ES1218_2010,VERITAS_1ES1218_2009}. This
    source was also detected during a six day period around MJD 53750
    by the MAGIC collaboration~\cite{MAGIC_1ES1218_2006}, at a level
    similar to that of the reported flux by VERITAS about one year
    later. Observations of 1ES 0229+200, 1ES 0347-121, 1ES 1101-232, H
    2356-309, and RGB J0152+017 were taken prior to the launch of the
    \emph{Fermi} satellite, and are reported by the HESS
    collaboration~\cite{HESS_1ES0229,HESS_1ES0347,HESS_1ES1101_232,HESS_H2356,HESS_RGBJ0152}. }
  \tablehead{
    \multicolumn{7}{c}{IACT data summary} \\
    \cline{1-7} \\
    \colhead{Source} & \colhead{z} & \colhead{IACT} & \colhead{Flux
      [10$^{-12}$cm$^{-2}$ s$^{-1}$]} & \colhead{index} & \colhead{MJD
        (approx)} & \colhead{Hrs} }
    \startdata 
    RGB J0710+591 & 0.125 & VERITAS & ($>$ 300 GeV) 3.9 $\pm$ 0.8 & 2.7 $\pm$ 0.3 & 54882-54892 & 22.1 \\ 
    1ES 1218+304 & 0.182 & VERITAS & ($>$ 200 GeV) 18.4 $\pm$ 0.9 & 3.1 $\pm$ 0.3 & 54829-54944 & 27.2 \\
    1ES 1218+304 & 0.182 & VERITAS & ($>$ 200 GeV) 12.2 $\pm$ 2.6 & 3.1 $\pm$ 0.1 & 54070-54220 & 17.4 \\ 
    1ES 0229+200 & 0.14 & HESS & ($>$ 580 GeV) 0.94 $\pm$ 0.24 & 2.5 $\pm$ 0.2 & 53614-53649, 53967-54088 & 41.8 \\ 
    1ES 0347-121 & 0.188 & HESS & ($>$ 250 GeV) 3.91 $\pm$ 1.1 & 3.1 $\pm$ 0.3 & 53948-54100 & 25 \\ 
    1ES 1101-232 & 0.186 & HESS & ($>$ 200 GeV) 4.5 $\pm$ 1.2 & 2.9 $\pm$ 0.2 & 53111-53445 & 43 \\ 
    H 2356-309 & 0.165 & HESS & ($>$ 200 GeV) 4.1 $\pm$ 0.5 & 3.1 $\pm$ 0.3 & 53157-53370 & 40 \\ 
    RBG J0152+017 & 0.08 & HESS & ($>$ 300 GeV) 2.7 $\pm$ 1.0 & 2.9 $\pm$ 0.5 & 54403-54418 & 15 \\
\enddata
\end{deluxetable*}

\subsection{HE Data Considerations}

For an AGN with significant power emitted at E $\gtrsim$ few TeV the
angular extent of the source due to cascade emission may become
comparable to the PSF of the \emph{Fermi}-LAT at fields as low as
$B_{IGMF}$ $\gtrsim$ 10$^{-17}$ G. For $B_{IGMF}$ magnitudes less than
this, an AGN is effectively a point source for the LAT. To evaluate
the effective point source flux of an AGN for the \emph{Fermi}-LAT, a
procedure similar to that of the IACT case was adopted. For every
simulated photon of a given energy, the energy-dependent PSF of the
\emph{Fermi}-LAT was used to determine the probability of
reconstruction of this photon within the 68 \% containment radius from
the position of the putative point source. The flux evaluated within
the 68 \% containment was again rescaled by 0.68$^{-1}$ to estimate
the effective point source flux in each energy bin. The PSF used for
this conversion was determined from a 2 year time-averaged sample of
AGN\footnote{\url{http://fermi.gsfc.nasa.gov/ssc/data/analysis/documentation/}
  \\ \url{Pass7\_usage.html}}.

To compare simulated differential fluxes from an effective point
source to the \emph{Fermi}-LAT data, the standard analysis tools were
applied but with a notable important distinction from previous
studies, which are cited in section
\ref{subsection_propagation_and_interaction}. Since in the HE regime,
the flux of gamma-ray photons can be dominated by either prompt or
secondary emission, we first derive from simulations, the spectral
index in each energy bin. This index is then used as a fixed parameter
for the maximal likelihood
evaluation\footnote{\url{http://fermi.gsfc.nasa.gov/ssc/data/analysis/}}
of the flux in each energy bin in the \emph{Fermi} data, within the
10$^\circ$ region of interest (ROI) which also includes all nearby
sources from the 2 year point source catalog and diffuse
backgrounds. The HE point source fluxes or upper limits are derived
using this procedure.

\subsection{Data and Model Comparison}\label{subsection_data_model_compare}
To compare the predictions of the Monte Carlo simulations to the data,
we combine both the HE and VHE regimes to compute a $\chi^2$-like
parameter, as further explained in this section. The simulated
effective point-source flux is used as the model expectation value,
and we assume that the statistical error of the vast amount of
simulations is negligible compared to the observational error. The
point-source fluxes derived with the use of the \emph{Fermi} tools as
explained above or obtained from IACT publications are used as
data. The observational error obtained or reported is taken to be the
primary source of discrepancy between the data and the model. A
$\chi^2$-like parameter is used to estimate the goodness-of-fit of a
given model and $\chi^2$ statistics is used to convert it to a
confidence level (or the probability of the model exclusion). For this
conversion, an explicit assumption is being made, that the errors in
both the HE and VHE regimes are dominated by statistics with a
gaussian distribution. Throughout the paper, our sole goal is to test
the null hypothesis, $H0$, that $B_{IGMF} = 0$ is incompatible with
the data and the simulations. The confidence level is the probability
that the measurement cannot be obtained with the assumption of the
given model. In what follows, we take the model to be incompatible if
the confidence level exceeds 95 \%, corresponding to a 2$\sigma$
deviation for a normal distribution. We do not claim any meaningful
interpretation of a higher confidence level, due to the unknown
behavior in the tails of the distribution of errors of each
instrument.

The model spectral energy density is derived based on the full Monte
Carlo simulations computed for monoenergetic primary photons with 8
bins per decade, of equal width in logarithmic space. The simulated
spectral energy density data are equally binned with 8 bins per decade
and each bin is centered on the energy of the primary photon
monoenergetic line. The VHE data are reported in different
publications at different energies and with different binning. We use
simulated data to interpolate the flux value to the reported positions
of the bins and their widths. In the 3 decades of the HE regime, 4-6
energy bins are generated depending on the given source luminosity and
statistics. The simulated data are then used to interpolate the flux
value to these energy bins. Moreover, we use the simulated data to
find the spectral index for the power law distribution of photons at
each energy bin. This spectral index is taken to be fixed when we find
the flux value and its error utilizing the \textit{Fermi} tools.

To compare the HE and VHE data of each source with the simulations,
the effective point source fluxes are generated for a set of gamma-ray
source models with fixed values of four parameters-$\alpha$,
$\epsilon_c$, $\Gamma$, and $\theta_{\text{v}}$. For each model,
$\alpha$ is chosen in the range (1.0, 2.5) with a step of 0.1, and
$\epsilon_c$ is chosen in the range (600 GeV, 60 TeV) with 8 bins per
decade, equally spaced in logarithmic energy. Each model is
characterized with default values of $\theta_{\text{v}}$ = 0, and
$\Gamma$ = 10, unless otherwise stated. Therefore, for each source, we
test 240 individual source models.

The remaining three parameters of the seven parameter source model
were determined as follows. Two of these parameters, the break energy
$\epsilon_B$ and the spectral index $\gamma$ are relevant only to the
HE part of the spectrum. They were chosen by minimizing the
$\chi_{HE}^2$ value, by allowing $\epsilon_B$ to vary between
$\gtrsim$ 10 GeV to the lower edge of the lowest energy VHE data bin
and $\gamma$ between -5 and 5. This interval for the break energy is
motivated by the fact that IACT instruments become insensitive in this
energy regime and \textit{Fermi} runs out of photon statistics,
therefore allowing a possible knee feature in the spectrum to be
undetectable. The spectral index $\gamma$ in the HE regime may or may
not be constrained by minimization of the $\chi_{HE}^2$ value. It is
evident that when secondary, cascade emission dominates in this energy
regime, it is sufficient for $\gamma$ to be larger than some value to
keep prompt emission negligible. The flux normalization factor, F$_0$,
is the only optimization parameter for a given model which is relevant
to the $\chi_{VHE}^2$ and which may or may not be relevant to the
$\chi_{HE}^2$, depending on the relationship between the prompt and
secondary emission of the source. We chose to determine $F_0$ by
minimization of the VHE part of $\chi^2$ by solving
$\partial \chi_{VHE}^2 / \partial F_0$ = 0 independently from the
behavior of $\chi_{HE}^2$. Effectively, this means that we have made a
stringent requirement of compatibility of the source model with the
VHE data and have excluded some models which would be highly
incompatible with the VHE measurements, but would allow statistical
compatibility with an overall $\chi^2$ = $\chi^2_{HE}$ +
$\chi^2_{VHE}$, just because of an increased number of data points and
therefore number of degrees of freedom. We view this weighting
procedure of HE and VHE parts of $\chi^2$ in determination of $F_0$ as
better physically motivated, since the highest energy data points in
the VHE regime are of extreme importance for the production of the
cascade emission, but from a statistical point of view, they are equal
to any other point of $N_{HE}$ or $N_{VHE}$ measurements.

The $\chi^2$ value for each model with four fixed
parameters ($\alpha$, $\epsilon_c$, $\Gamma$, and $\theta_{\text{v}}$)
was converted into a confidence level, using $\chi^2$ statistics with
$N_{HE} - 2 + N_{VHE} - 1$ degrees of freedom, assuming that three
parameters were optimized for each model ($F_0$, $\epsilon_B$,
$\gamma$). We make no attempt in our studies to evaluate the
confidence intervals of the latter three parameters of each model. Our
goal is exclusively to find a model or a set of source models which
are compatible with $H0$.

As an illustration of a typical result of the data and model
comparison, Figure~\ref{rgbj0710_B0_fit}a shows the $\chi^2$
confidence level of the dFED, assuming four fixed and three free
parameters for each model tested within the given range of $\alpha$
and $\epsilon_c$ parameters. The most favored model with value of
$\alpha$ = 1.8 and $\epsilon_c$ = 3.16 TeV is incompatible with the
data at the level of about 75 \%. If this choice of parameters
$\alpha$ and $\epsilon_c$ is considered as an optimization process, in
which case the number of free parameters in the model is 5, then this
model is incompatible with the data at the level of
88 \%. Figure~\ref{rgbj0710_B0_fit}b shows the data points for the
dFED and the best fit simulation result with these $\alpha$,
$\epsilon_c$ parameters. For this source's dFED, below 10 GeV, the
spectrum is dominated by cascade emission, while above 10 GeV, it is
dominated by prompt radiation.

%% file: Systematics.tex


The primary goal of this paper is to provide precise numerical
verification of the constraints on the IGMF as reported in
~\cite{Neronov2010,Tavecchio2010,Dolag2010,Dermer2010,Taylor2011},
since most of these results were obtained with various simplifications
in the analysis, and some were derived with semi-analytical approaches
to qualitatively verify more detailed computations
(\cite{Dermer2010}). Particular focus is given to establishing the
robustness of magnetic field limits when various systematics are taken
into consideration.

Three sources (RGB J0710+591, 1ES 1218+304, and 1ES 0229+200) are used
in this study which have been reported as having provided constraints
on the IGMF, in the comprehensive study of \cite{Taylor2011} (and
references within). Hereafter, we refer to this paper as TVN11. An
additional 4 hard-spectrum TeV blazars observed prior to the beginning
of the \emph{Fermi} mission are considered-1ES 0347-121, 1ES 1101-232,
H 2356-309, and RGB J0152+017. The \emph{Fermi}-LAT data for these
sources are re-analyzed and are collected from the mission start time
August 4, 2008 to February 14, 2012, and the updated P7SOURCE IRFs are
used along with the Pass 7 data. The models for extragalactic and
diffuse backgrounds were used together with the standard gamma-ray
selection constraint of zenith angle $<$ 100$^\circ$ which eliminates
earth limb gamma-rays.

Perhaps the main source of uncertainty in placing constraints on the
IGMF stems from the unknown duty cycle of TeV blazars and
particularly, the history of the highest energy TeV emission, as has
been pointed out in~\cite{Dermer2010}. The sampling of the VHE
activity of these sources reported by IACTs is limited to a few tens
of hours dispersed over a period of a few weeks to a few years. In the
regime of very low IGMF ($B_{IGMF}$ $<$ 10$^{-20}$ G), most of the
secondary radiation from intergalactic cascades with energy $>$ 100
MeV, which originates from the primary VHE flux sampled by IACTs,
would have reached the earth and would be detected by the
\emph{Fermi}-LAT (see Fig.~\ref{mean_time_delay_compare}) within a few
hours. This assumes that the HE flux from a given source sampled by
the \emph{Fermi}-LAT over the period of the mission (about 4 years)
could be viewed as ``contemporaneous'' to IACT measurements, for the
purposes of verification of $H0$. This explicitly assumes, however,
that the duty cycle of a given source in the VHE regime = 1 over this
same period.

\begin{figure*}[t] 
  \centering 
  \mbox{ \subfigure{\includegraphics[width=3.4in]{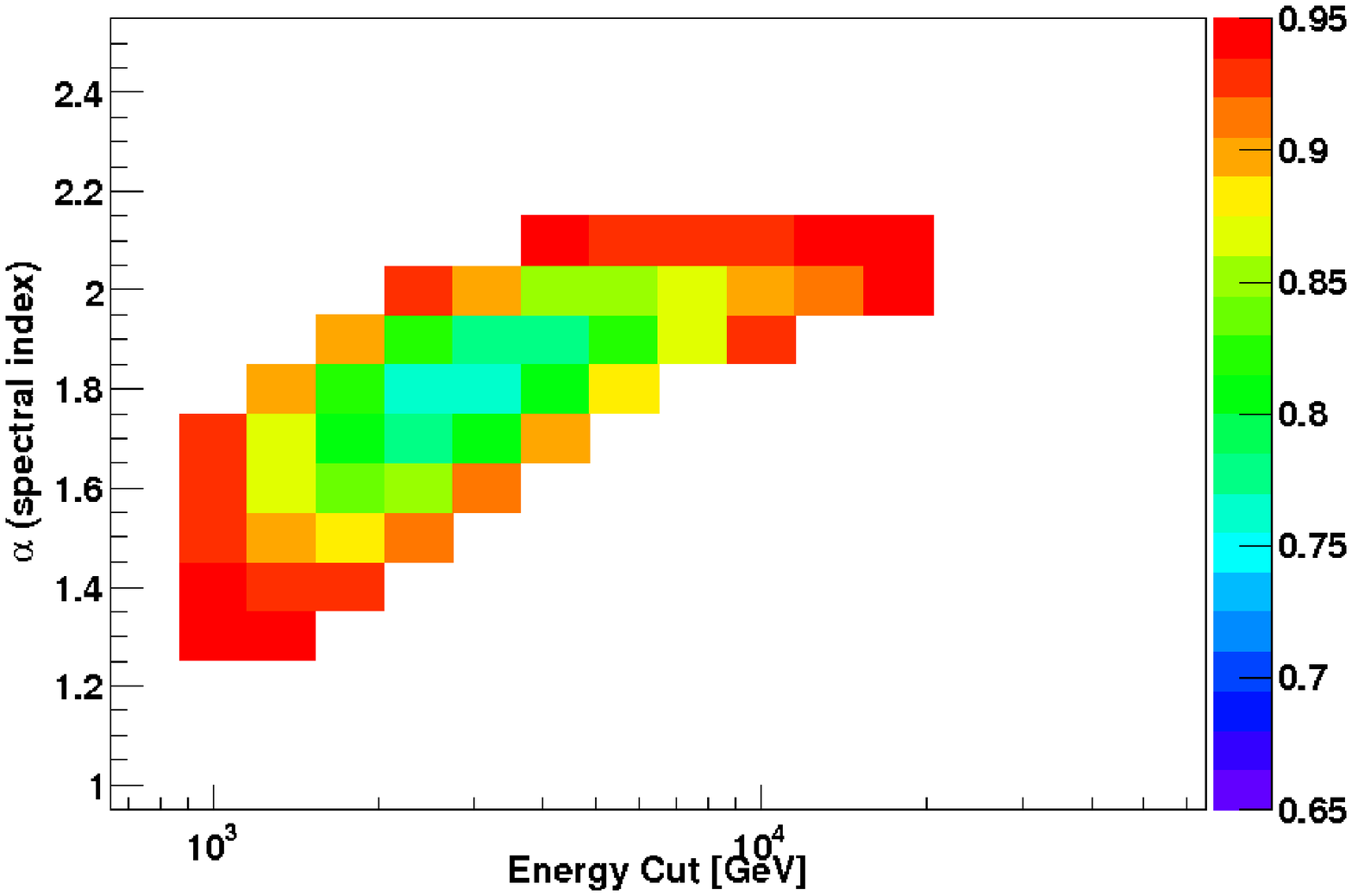}
      \subfigure{\includegraphics[width=3.4in]{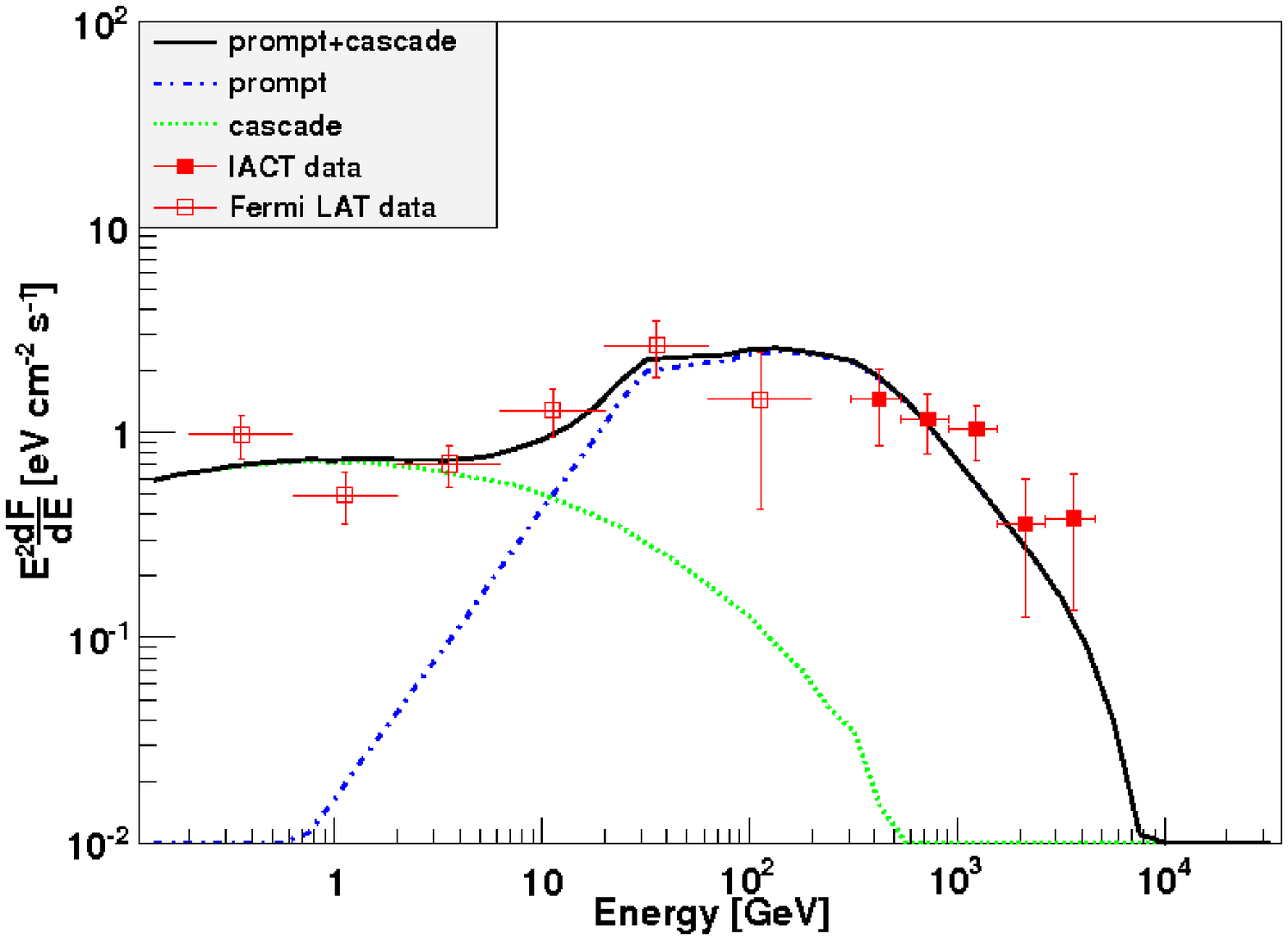}}\quad }}
\caption { \label{rgbj0710_B0_fit} Source RGB J0710+591 analyzed
    under the assumption of $H0$. (a) (left) Confidence level, or the
    probability of exclusion of the gamma-ray source model (with fixed
    $\alpha$ and $\epsilon_c$), where the remaining three parameters
    of the model ($\gamma$, $\epsilon_B$, and F$_0$, see
    Eq.~\ref{broken_power_law_model}) are chosen so as to minimize
    $\chi^2$. (b) (right) Simulated dFED of the best fit model, of
    $\alpha$ = 1.8, $\epsilon_c$ = 3.16 TeV, showing both the prompt
    and secondary cascade contributions to the total dFED, along with
    the HE and VHE data.}
\end{figure*}

\subsection{Analysis of RGB J0710+591 Data}
The VHE observations of RBG J0710+591 are summarized in
Table~\ref{iact_data_table}. They include 5 energy data points
reported by the VERITAS collaboration in \cite{VERITAS_RGBJ0710},
observed during the time period December 2008 - March 2009 for a total
of 22 hours. For the HE regime, 6 energy bins were used spanning from
200 MeV to 200 GeV. In each energy bin, if TS $>$ 9, the flux point
and 1$\sigma$ error bars are displayed. Otherwise, an upper limit was
computed. It is important to note that with this strategy and extended
data set as compared to previously used in TVN11, the flux in the
lowest energy bin now constitutes a flux point rather than an upper
limit. This data point had been critical for rejecting $H0$.

Figure~\ref{rgbj0710_B0_fit}a shows the confidence level for rejecting
$H0$, for a set of models characterized by the range of $\epsilon_c$
and $\alpha$ described in
section~\ref{subsection_data_model_compare}. It identifies the best
fit model with values of $\alpha$ = 1.8 and $\epsilon_c$ = 3.16 TeV,
which is incompatible with $H0$ at $<$ 75 \% confidence level assuming
three free parameters (and $<$ 88 \% assuming five free
parameters). The range of models in the vicinity of this point is not
incompatible with $H0$. The best fit of the simulated dFED and
observations is shown in Figure~\ref{rgbj0710_B0_fit}b. It appears
that the conclusion of TVN11 that $H0$ is ruled out at the 98.8 \%
level is invalidated due to two factors. First, the \emph{Fermi}-LAT
dataset underwent revision from the old pass 6 version (P6) to the
current pass 7 (P7) and this allowed a detection to be made in the
lowest energy bin, which is higher than the previously computed upper
limit. The second factor may be due to the fitting algorithm applied
in this present work which is different from that adopted in TVN11, in
which the flux or upper limit determination in a given energy bin was
fixed to the best fit index over the entire energy range.

Furthermore, the geometrical orientation of the jet with respect to
the observer and the jet boost factor represents another source of
uncertainty, and tuning these parameters can further improve the
goodness of the $\chi^2$ fit. For example, TVN11 assumes a viewing
angle of 2$^{\circ}$ and an effective jet opening angle of
6$^{\circ}$, corresponding to a boost factor of $\sim$ 10. As shown in
Figure~\ref{spectra_geometry_scan}, however, lower boost factors or
smaller viewing angles lead to lower total power of the jet at the
highest energies, and therefore lead to reduced secondary flux.

\subsection{Analysis of 1ES 1218+304 Data}
The VHE observations of 1ES 1218+304 are summarized in
Table~\ref{iact_data_table}, which includes 2 data sets. The first
set, obtained during December 2008 - May 2009, is based on 27 hours of
data and has 9 energy data points reported by the VERITAS
collaboration in~\cite{VERITAS_1ES1218_2010}. This data set was used
in TVN11, and for consistency it is also used in this study. The
\emph{Fermi}-LAT data for this source were produced in the same way as
for RGB J0710+591, with 6 energy bins spanning from 200 MeV to 200
GeV. This source has excellent statistics in each \emph{Fermi}-LAT
energy bin, with TS $>$ 25. It is important to note that the extended
exposure and updated pass 7 data set used in this work shows no
statistically significant difference compared to that of TVN11.

\begin{figure*}[t] 
  \centering
    \mbox{
      \subfigure{\includegraphics[width=3.4in]{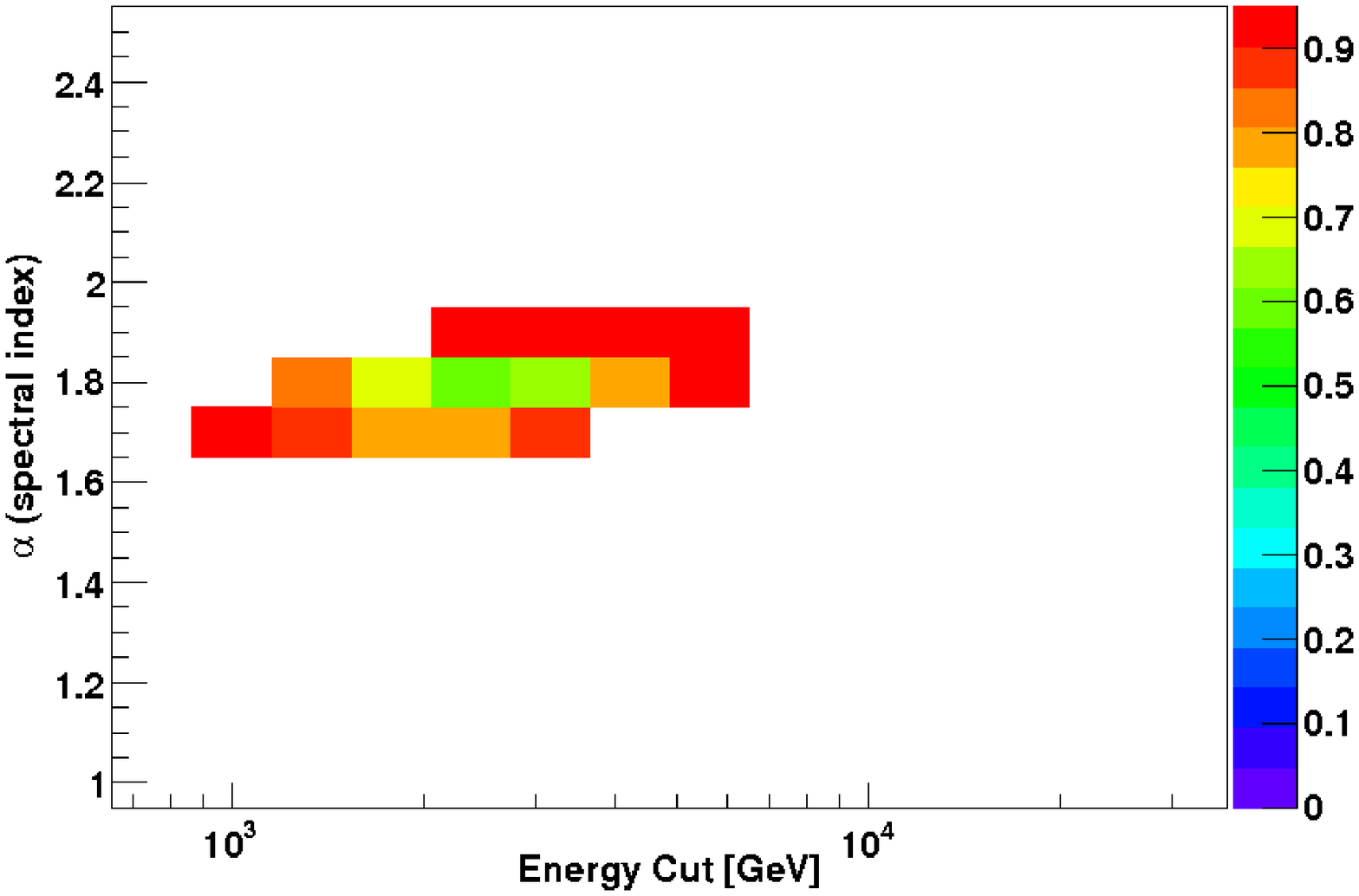}
        \subfigure{\includegraphics[width=3.4in]{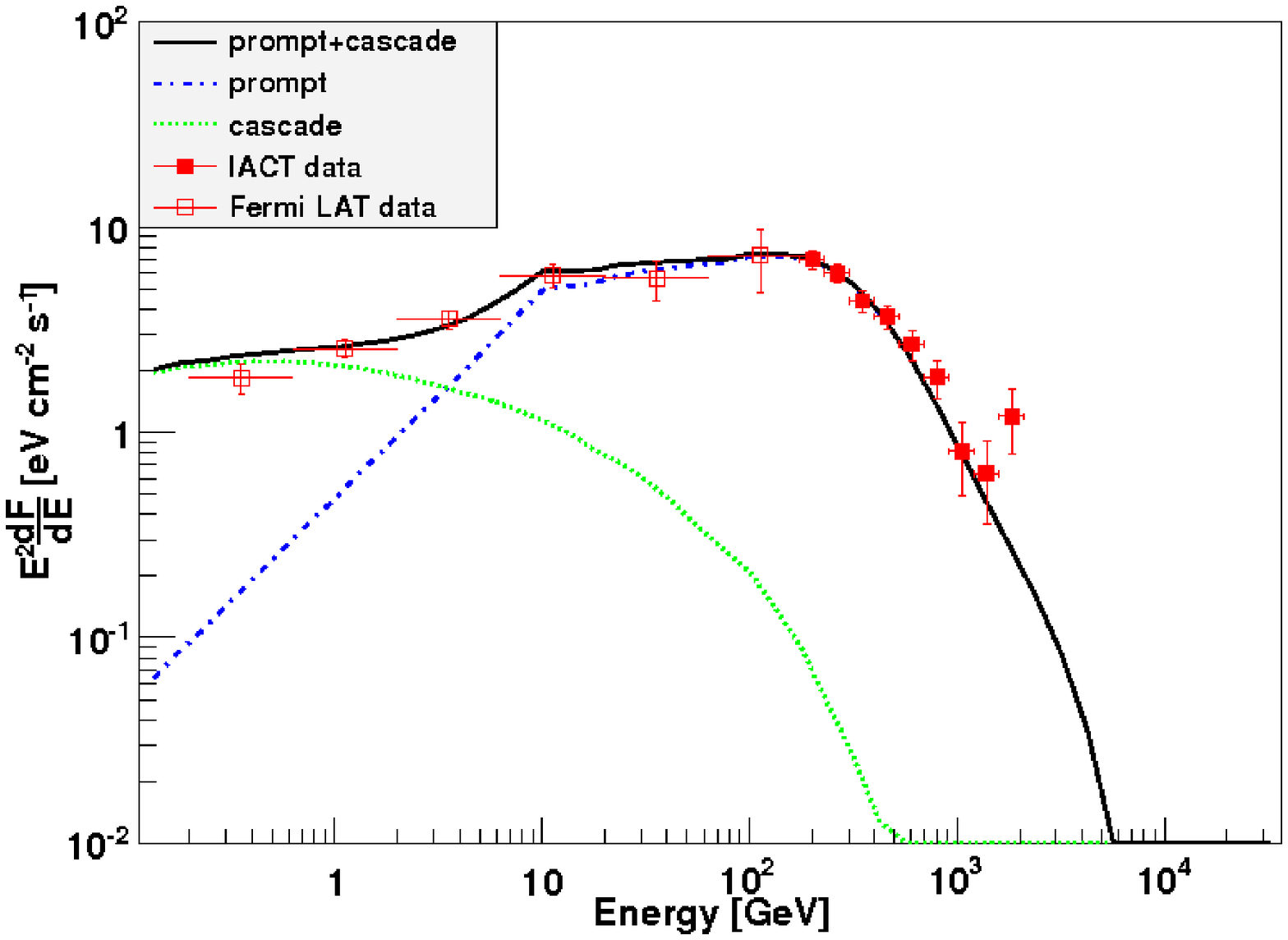}}\quad
    }}
  \caption { \label{1es1218_B0_fit} Source 1ES 1218+304 analyzed under
    the assumption of $H0$. (a) (left) Confidence level, or the
    probability of exclusion of the gamma-ray source model (with fixed
    $\alpha$ and $\epsilon_c$), where the remaining three parameters
    of the model ($\gamma$, $\epsilon_B$, and F$_0$, see
    Eq.~\ref{broken_power_law_model}) are chosen so as to minimize $\chi^2$.
    (b) (right) simulated dFED of the best fit model, of $\alpha$ =
    1.8, $\epsilon_c$ = 3.16 TeV, showing both the prompt and
    secondary cascade contributions to the total dFED, along with the
    HE and VHE data.}
\end{figure*}

Figure~\ref{1es1218_B0_fit}a shows the confidence level for rejecting
the $H0$ hypothesis on the $\alpha$-$\epsilon_c$ plane. It suggests a
best fit model with values of $\alpha$ = 1.8, $\epsilon_c$ = 3.16 TeV,
a spectral break energy of $\epsilon_B$ = 10 GeV, and an index below
the break energy of $\gamma$ = 1.0, which is incompatible with $H0$ at
the less than 65 \% confidence level, assuming three (and less than
80 \% assuming five) free model parameters. The fit of the simulated
dFED and observations is shown in Figure~\ref{1es1218_B0_fit}b. It is
evident that the conclusion of TVN11, that $H0$ is ruled out with more
than 99.99 \% probability, is purely due to the assumption that a
single source spectral index holds for over five orders of magnitude
in energy. Allowing a spectral break energy and an intrinsic spectral
index below the break energy to vary as detailed in
section~\ref{subsection_source_model} makes it possible to interpret
the 1ES 1218+304 data set as consistent with $H0$.

The amount of secondary radiation strongly depends on the power output
of the TeV blazar at the highest energies. For this source, more so
than the others, IACT observations demonstrate strong variability in
the VHE spectrum. The VERITAS collaboration reports that the 1ES
1218+304 data were sampled sparsely over a period of 115 days in late
2008 - 2009, and while the majority of the data are consistent with a
steady baseline flux, the data set also includes a statistically
significant flare which peaked at $\sim$ 20 \% Crab, and lasted a few
nights. The flux at the peak of the flare was 3-4 times higher than
the baseline flux and it significantly increases the average flux
value observed over the entire period. Furthermore, evidence for
variability of this source can be inferred from the VERITAS
publication covering its 2 year activity which occurred prior to the
\emph{Fermi} mission (\cite{VERITAS_1ES1218_2009}). The flux observed
at that time constitutes about 60-70 \% of that reported in the second
data set (see Table~\ref{iact_data_table}). Overall, the IACT data to
date suggest that the average observed VHE flux of this souce could be
lower than used in TVN11, yet the assumption of the higher average VHE
flux is still compatible with $H0$.

\subsection{Analysis of 1ES 0229+200
  Data}\label{subsection_1es0229_analysis}

The parameters of the data set of 1ES 0229+200 are given in
Table~\ref{iact_data_table} and include two sets of observations by
the HESS collaboration during the period from September 1, 2005 to
December 19, 2006, accumulating 41.8 hours exposure
(\cite{HESS_1ES0229}). This data set provides the time-averaged dFED
for 8 bins over the energy range spanning from 500 GeV to 16 TeV. This
same data set was used in the previous study of TVN11. The
\emph{Fermi}-LAT dFED for 1ES 0229+200 utilizes four evenly spaced
bins in log space in the range from 420 MeV to 300 GeV. Only the first
energy bin in this data set provides a strong detection (TS $> 25$),
for all simulated models in which the secondary flux dominates the
total flux. The TS for all other energy bins is typically found at $>
9$ for the majority of simulated models but in some cases, only the
upper limit can be established (TS $< 9$). This indicates that the
source detection in the HE regime is weak. Perhaps more so than for
any other source, the dFED of 1ES 0229+200 does not resemble a power
law in the HE energy regime, making the $\chi^2$ fits relatively poor.

\begin{figure*}
  \centering
  \mbox{\subfigure{\includegraphics[width=3.4in]{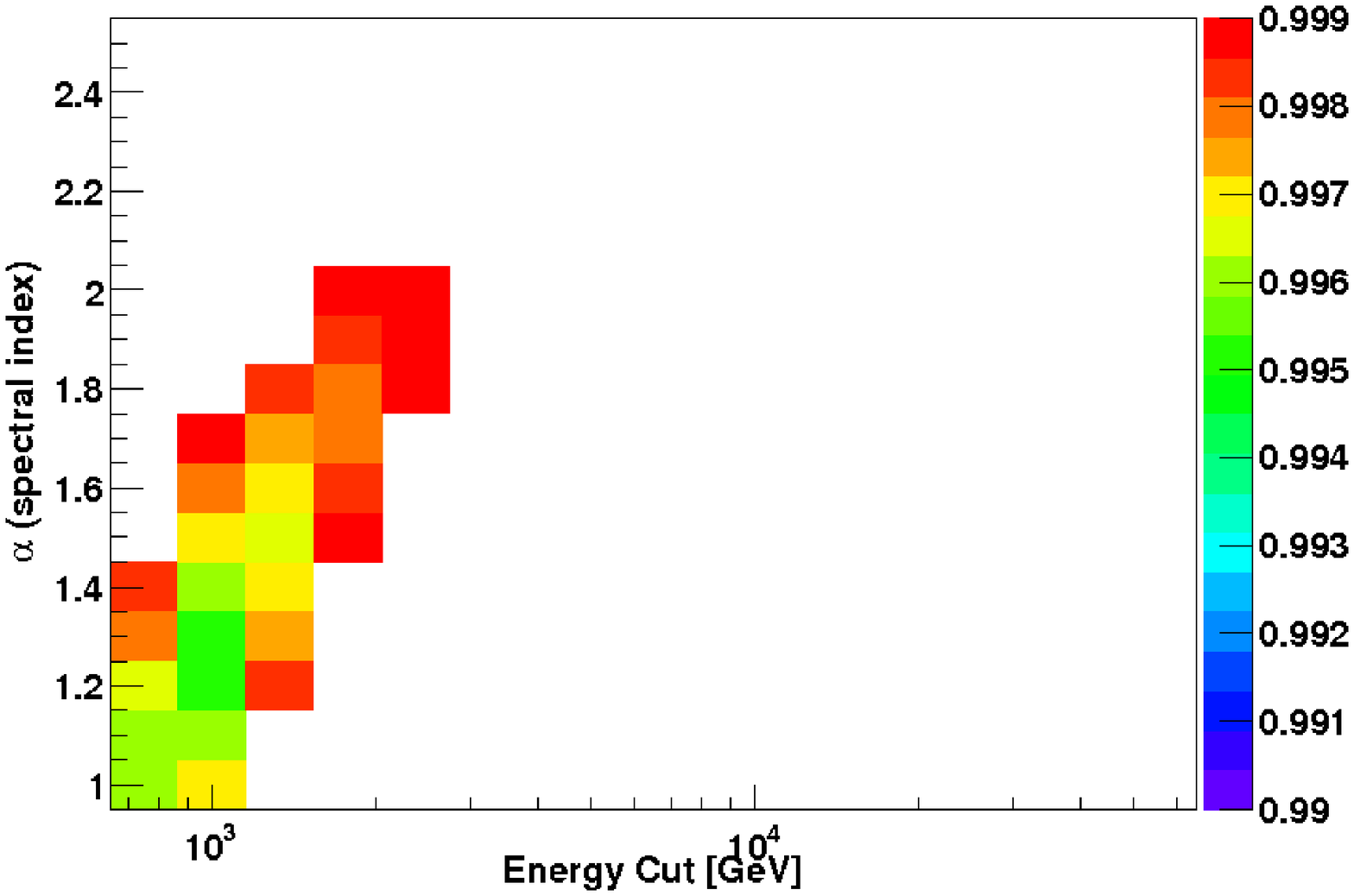}}\quad
    \subfigure{\includegraphics[width=3.4in]{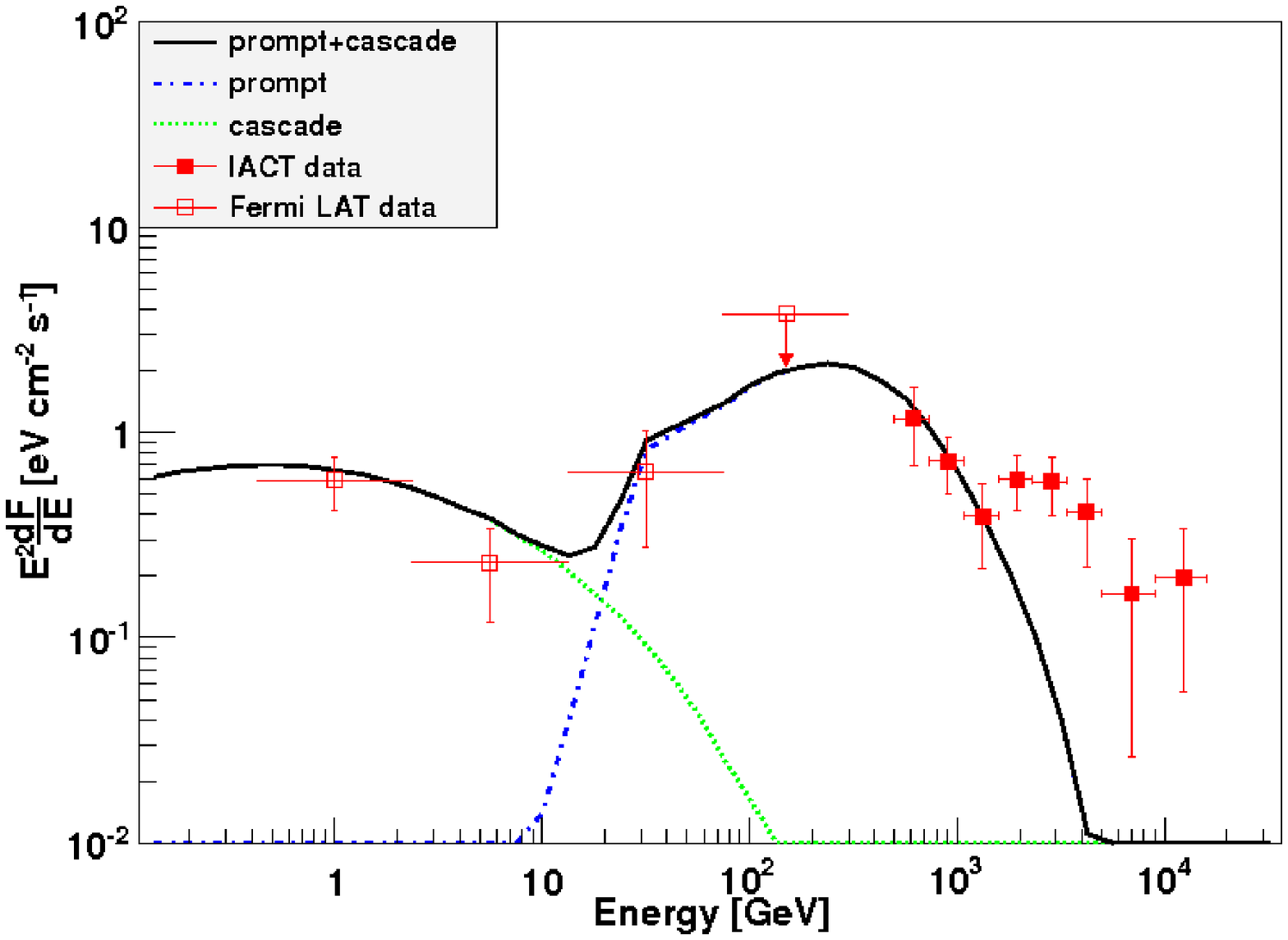}
    }}
  \caption { \label{1es0229_B0_fit} Source 1ES 0229+200 analyzed under
    the assumption of $H0$. (a) (left) Confidence level, or the
    probability of exclusion of the gamma-ray source model (with fixed
    $\alpha$ and $\epsilon_c$), where the remaining three parameters
    of the model ($\gamma$, $\epsilon_B$, and F$_0$, see
    Eq.~\ref{broken_power_law_model}) are chosen so as to minimize
    $\chi^2$. (b) (right) Simulated dFED of the best fit model (C.L. =
    0.995) of $\alpha$ = 1.3, $\epsilon_c$ = 1 TeV, showing both the
    prompt and secondary cascade contributions to the total dFED,
    along with the HE and VHE data.}
\end{figure*}

To evaluate the goodness of fit of the data to the Monte Carlo
simulations, we assume that the data accumulated by the HESS
collaboration over 2005-2006 is representative of the source activity
during the first 3.5 years of the \emph{Fermi}-LAT data used in this
work. The $\chi^2$ fit obtained under this assumption and for $H0$ is
shown in Figure~\ref{1es0229_B0_fit}a. We confirm the result of TVN11
that this source does not have a viable source model that explains the
combined HE-VHE data set and agree that $H0$ is ruled out at the 99.5
\% confidence level. The best fit ($\alpha = 1.3$, $\epsilon_c = 1$
TeV) model requires a dramatic spectral break just below 100 GeV and
the dFED of 1ES 0229+200 below this energy is completely dominated by
secondary flux as shown in Figure~\ref{1es0229_B0_fit}b.

The effects of two systematic uncertainties, viewing angle
$\theta_{\text{v}}$ and Doppler factor $\Gamma$, were considered. The
default assumptions in the analysis are $\theta_{\text{v}} =
0^{\circ}$ and $\Gamma = 10$. Increasing the viewing angle, e.g.
$\theta_{\text{v}} = 2^{\circ}$ as used in TVN11, would further
overproduce radiation in the HE regime (see
Figure \ref{spectra_geometry_scan}a). Increasing the Doppler factor,
$\Gamma$, combined with the $\theta_{\text{v}} = 0^{\circ}$ assumption
would imply that the overall luminosity of the jet in the VHE band
should be lower to fit the VHE observations since the jet is
collimated into a smaller angle. Rescaling of the prompt radiation to
fit the VHE data will however equally rescale secondary emission in
the HE regime if the angular distribution of the prompt photons is
significantly wider than the characteristic scattering angles acquired
in the QED processes. Therefore, for reasonable values of $\Gamma <
100$, the change in the secondary radiation above 100 MeV for the best
fit model was found to be negligible. The effect becomes of order 10
\% in the lower part of the HE spectral range only for $\Gamma$ $\sim$
$10^4$ - $10^5$, and it cannot be used to reconcile the observational
data of 1ES 0229+200 with $H0$.

\begin{figure*}
  \centering
  \mbox{\subfigure{\includegraphics[width=3.4in]{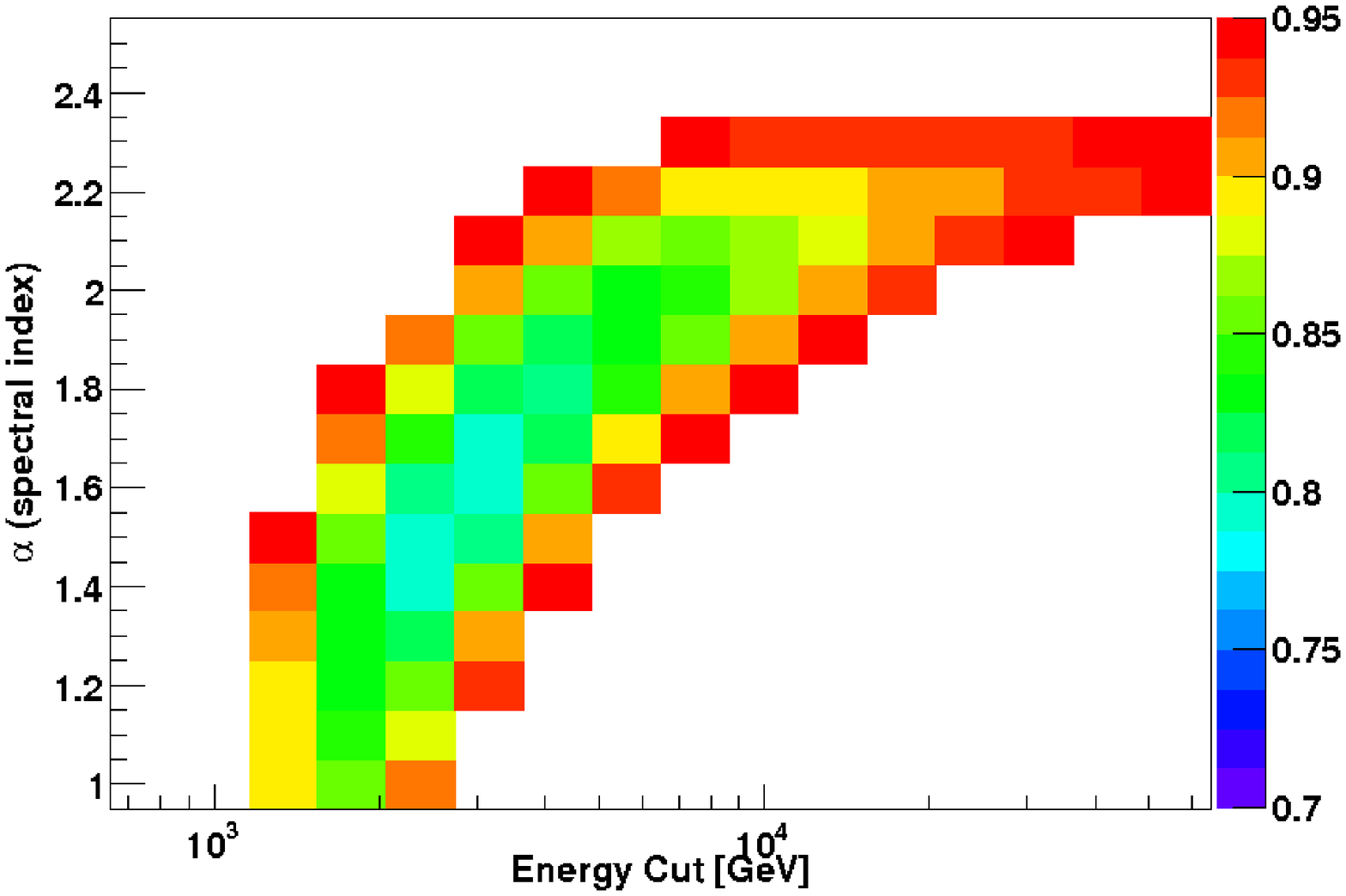}}\quad
    \subfigure{\includegraphics[width=3.4in]{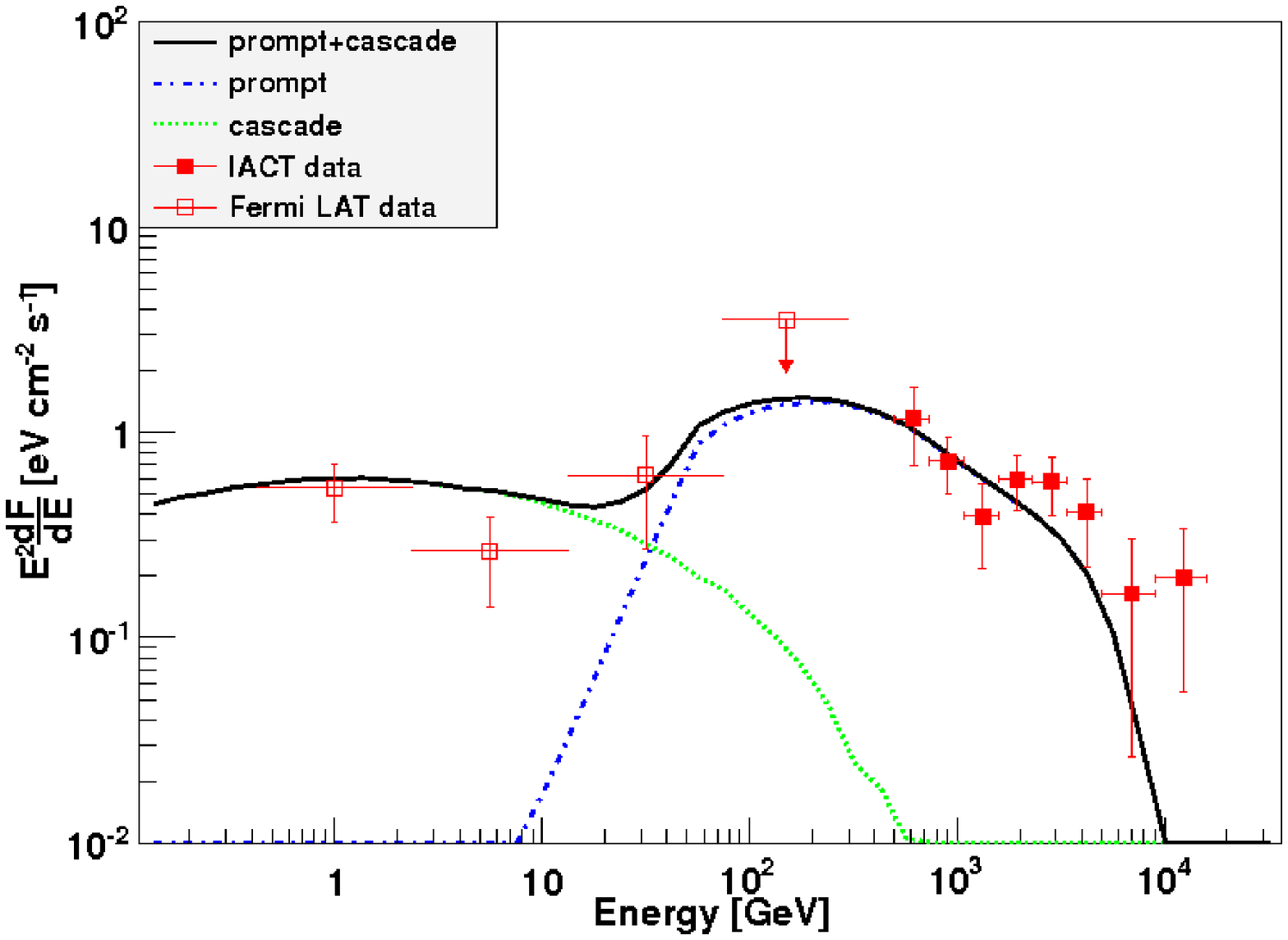} }}
  \caption { \label{1es0229_B0_fit_lowebl} Source 1ES 0229+200
    analyzed, assuming EBL Model 3 and $H0$. (a) (left) Confidence
    level, or the probability of exclusion of the gamma-ray source
    model (with fixed $\alpha$ and $\epsilon_c$), where the remaining
    three parameters of the model ($\gamma$, $\epsilon_B$, and F$_0$, see
    Eq.~\ref{broken_power_law_model}) 
    are chosen so as to minimize $\chi^2$. (b) (right)
    Simulated dFED of the best fit model (C.L. = 0.78) of $\alpha$ =
    1.6, $\epsilon_c$ = 3.16 TeV, showing both the prompt and
    secondary cascade contributions to the total dFED, along with the
    HE and VHE data.}
\end{figure*}

Since the energy density of the EBL directly affects the propagation
length of VHE photons, the uncertainty in the EBL model represents
perhaps the most important source of systematic error. To investigate
this, two additional EBL models were generated, EBL model 2 \& 3,
shown in Figure~\ref{ebl_comparison_dominguez}. EBL model 2 is
characterized by a considerably lower energy density in the far
infrared peak of the dust emission. This model is motivated by the
recently resolved lower limit on the EBL which is based on the galaxy
counts in the data obtained with the Spitzer
(\cite{Bethermin_Spitzer_2010,Dole_Spitzer_2006}), Herschel
(\cite{Berta_Herschel_2010}), and AKARI (\cite{Matsuura_Akari_2011})
satellites. This model corresponds to the lowest possible far infrared
EBL energy density allowed within $2 \sigma$. It was found that even
with such an extreme assumption about the far infrared EBL, the
decrease of the secondary radiation in the model of 1ES 0229+200 was
negligible. This conclusion is due to the fact that the source models
providing the best $\chi^2$ fits have high energy cutoffs,
$\epsilon_c$, below $\sim 5$ TeV, and are thus insensitive to EBL
photon wavelengths $\gtrsim 25 \mu$m (kinematic threshold of pair
production).

The EBL model 3 is based on the resolved EBL energy density of the
starlight peak, which is derived from galaxy counts utilizing data
from the HST (\cite{MadauPozzetti_2000}) in the visible (from 0.36 -
2.2 $\mu$m), Spitzer in the near-IR (3.6 - 8 $\mu$m)
(\cite{Fazio_IRAC_2004}), and ISO in the mid-IR (15-24 $\mu$m)
(\cite{Elbaz_ISOCAM_2002,Papovich_Spitzer_2004}). As compared to the
default model 1, this EBL model has an energy density in the visible
reduced by about 25\% which is compatible with the galaxy counts
results to within $1 \sigma$. The energy density in the near IR is
reduced by about 50 \%. At 3.6 $\mu$m, model 3 has an energy density
of 4.0 nW m$^{-2}$ sr$^{-1}$ which is within the $2 \sigma$ limit (3.5
nW m$^{-2}$ sr$^{-1}$) from the galaxy counts result derived
by \cite{Fazio_IRAC_2004}. At 4.5 $\mu$m, model 3 has an energy
density of 3.0 nW m$^{-2}$ sr$^{-1}$, which is within the $1 \sigma$
limit from the Fazio et. al. analysis. Given the uncertainty in the
5.8 and 8.0 $\mu$m galaxy counts by Fazio et. al. at the bright
fluxes, \cite{Franceschini2008} reanalyzed the Spitzer data to
conclude that at 8.0 $\mu$m the energy density of the EBL is 1.92 nW
m$^{-2}$ sr$^{-1}$ with a $2 \sigma$ lower bound of 1.23 nW m$^{-2}$
sr$^{-1}$. The energy density of EBL model 3 is 1.4 nW m$^{-2}$
sr$^{-1}$ which is within the $2 \sigma$ bound from
the \cite{Franceschini2008} result. As has been pointed out by several
authors (\cite{MazinRaue07,Kneiske2010,Dominguez_ebl_2010}) the 5.8
$\mu$m result of \cite{Fazio_IRAC_2004} is likely to have been also
contaminated by excessive contributions from bright local galaxies,
but it has not yet been re-analyzed, unlike the 8 $\mu$m
point. Nevertheless, all of these authors have recognized that that
the Fazio et. al. result at 5.8 $\mu$m should be corrected, and many
have used the EBL energy density at this point, significantly lower
than 3.6 nW m$^{-2}$ sr$^{-1}$ reported by \cite{Fazio_IRAC_2004}. The
1$\sigma$ lower bound used by \cite{MazinRaue07} is 2.4 nW m$^{-2}$
sr$^{-1}$ and it is 2.5 nW m$^{-2}$ sr$^{-1}$
in \cite{Dominguez_ebl_2010}. The 2.1 nW m$^{-2}$ sr$^{-1}$ assumed in
model 3 is within the $2 \sigma$ error bar from these later
results. Thus, EBL model 3 is compatible with the galaxy counts
results to within $2 \sigma$ but it effectively does not allow any
additional contribution to the EBL from unresolved or unknown sources.

The results of the simulations of intergalactic cascading for model 3
with $H0$ are shown in Figure~\ref{1es0229_B0_fit_lowebl}a. It was
found that a number of 1ES 0229+200 source models are compatible with
the combined VHE and HE data set, and one of the best fit examples
characterized by $\alpha = 1.6$, $\epsilon_c = 3.16$ TeV is shown in
Figure~\ref{1es0229_B0_fit_lowebl}b.

\begin{figure}
  \begin{adjustwidth}{-0.0in}{} 
  \centering 
  \includegraphics[width=3.4in]{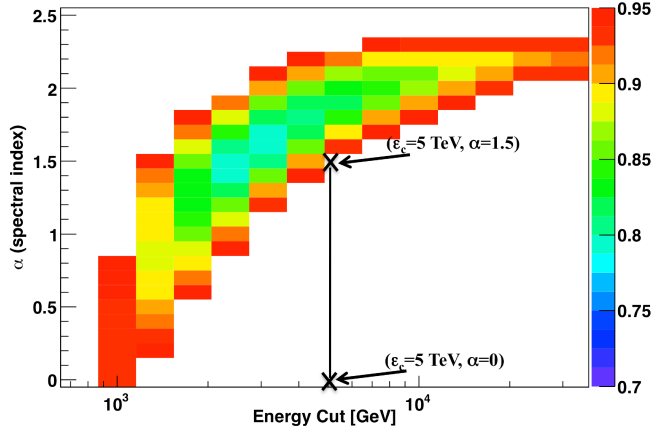}
  \caption{ \label{eblmodel6_compare_vovk} Confidence level, or the
      probability of exclusion of the gamma-ray source model (with
      fixed $\alpha$ and $\epsilon_c$), where the remaining three
      parameters of the model ($\gamma$, $\epsilon_B$, and F$_0$, see
      Eq.~\ref{broken_power_law_model}) are chosen so as to minimize
      $\chi^2$, for source 1ES 0229+200 obtained under the assumption
      of EBL Model 3. The black line terminated with crosses
      represents the range of models analyzed
      in \cite{Vovk_EBL_IGMF_2012}. } \end{adjustwidth}
\end{figure}

The strong sensitivity of the secondary photon flux to the EBL energy
density in the near-IR combined with the conspicuous lack of a
non-trivial EBL absorption feature in the VHE energy band ($\sim$ 200
GeV - 5 TeV) is due to the peculiar behavior of the EBL in this
wavelength range. For $\lambda I_{\lambda}$ $\propto$ $\lambda^{-1}$,
the optical depth is independent of the energy of a VHE photon (gray
opacity). A small deviation from this proportionality results in a
logarithmically slow dependence of the optical depth on the energy of
the VHE photon, producing a power law, rather than exponential-like
change in a blazar spectrum as discussed in \cite{Vassiliev2000}. The
behavior of the SED in the mid-IR exactly satisfies this condition and
explains the ``invisibility'' of EBL absorption effects in a blazar
dFED. The effect however is strong and reflected in the change of the
spectral index of this blazar. In fact, this feature was used to
derive upper limits on the EBL energy density in the mid-IR, which are
taken to be the values of model 1, by assuming that the spectral index
of the source 1ES 1101-232 cannot be harder than 1.5
(e.g. \cite{Aharonian_Nature_2006}). The 25 - 50\% lower mid-IR
density of model 2 significantly softens the intrinsic spectrum of 1ES
0229+200 reducing the total energy available for the development of
the intergalactic cascade, and therefore the flux of the secondary
photons. Thus, there are a range of EBL models with mid-IR energy
density bounded by the lower limits on the EBL to some SED slightly
below that of model 1 which are compatible with the EBL lower limits
and $H0$.

In a recent study of the dual constraints on the EBL and IGMF it was
found that an EBL model similar to model 3, is still incompatible with
$H0$ and would require a lower bound of $B_{IGMF}$ = $6\times
10^{-18}$ G (\cite{Vovk_EBL_IGMF_2012}). All source models analyzed in
that paper had a single cutoff energy $\epsilon_C$ = 5 TeV and varying
spectral index $\alpha$ in the range of 0 - 1.5 with a single power
law dFED over the entire VHE and HE energy range. A similar assumption
of a steady flux from 1ES 0229+200 over the lifetime of the
\emph{Fermi}-LAT was made. Figure \ref{eblmodel6_compare_vovk}
illustrates the confidence level for a wider range of source models
analyzed in this work together with the range of models considered by
\cite{Vovk_EBL_IGMF_2012}. The figure confirms the incompatibility of
$H0$ with the data given assumptions about the source model used in
that paper. However, in an extended parameter space of the 1ES
0229+200 models, $H0$ can be reconciled with observations of this
source.

\begin{figure}   
 \begin{adjustwidth}{-0.0in}{}
    \centering
    \includegraphics[width=3.4in]{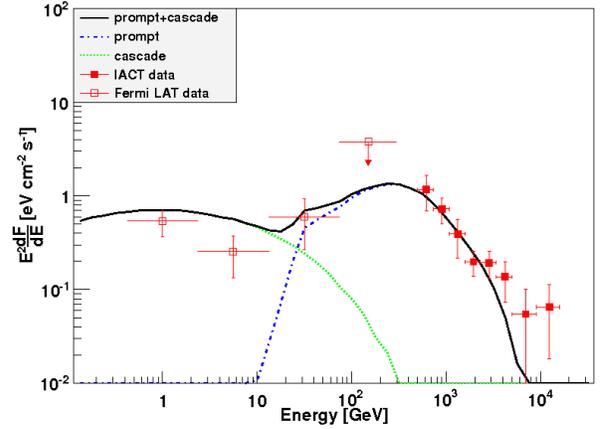}
    \caption { \label{1es0229_scale_half} Spectral model of 1ES
      0229+200 of $\alpha$ = 1.3, $\epsilon_{c}$ = 1.78 TeV
      (C.L. $\lesssim$ 0.8), utilized for the study of the effect of
      the duty cycle of the TeV data. The five highest energy data
      points are scaled down by 10$^{-1/2}$ of their flux values to
      that reported in \cite{HESS_1ES0229}.}  
    \end{adjustwidth}
\end{figure}

Another avenue to make the HE-VHE observational data of 1ES 0229+200
compatible with $H0$ is to question whether or not the VHE flux of the
source reported is representative of the average flux during the
\emph{Fermi} observations. The HESS measurements were accumulated in 2005
(6.8 hrs) - 2006 (35 hrs) and are not strictly contemporaneous with
the \emph{Fermi} HE spectrum. The significance of the detection in
2005 reported is 2.7$\sigma$, while in 2006 it is 6.1$\sigma$ with
average photon fluxes above 580 GeV of 6.8 $\times 10^{-13}$ cm$^{-2}$
s$^{-1}$, and 10 $\times 10^{-13}$ cm$^{-2}$ s$^{-1}$,
respectively. Due to the low flux of this source (1-2\% of the Crab
nebula flux), and the small data set in the original 1ES 0229+200
discovery paper, statistically significant flux variability as
observed in 2005 and 2006 was not detected. Although these
observations are compatible with a constant flux, the variability
hypothesis cannot be ruled out, based on the statistical and
systematic errors reported. Furthermore, observations of this source
in 2009, as reported by VERITAS (\cite{PerkinsPoster2010}), were
compatible with the average flux value of the HESS data set. However,
the average flux obtained was dominated by a period of significantly
higher ``flaring'' activity during a single dark run. In general, VHE
observations above a few TeV (relevant for secondary photon
production) require considerable integration time and so far, they are
too sparse to claim that the HESS value of the flux is representative
of the average flux during the \emph{Fermi} mission. Further
communication with the VERITAS Collaboration suggests that the flux
level of 1ES 0229+200 has been steadily declining from 2009 - 2012
(private communication). To investigate the effect of a reduced duty
cycle for 1ES 0229+200, the spectrum of this source was modified at
the highest 5 energy points to half an order of magnitude of their
reported values. It was found that the VHE - HE data set combined in
this way does not rule out $H0$ at more than 95\% confidence
level. One of these compatible models, with $\alpha$ = 1.3
$\epsilon_c$ = 1.78 is shown in
Figure~\ref{1es0229_scale_half}. Therefore, the conclusion that the
$H0$ is ruled out with high significance heavily rests on the
assumption that the HESS measurements are representative of the
average flux for E $\gtrsim 2$ TeV and a $\sim$ 10$^{-1/2}$ change of
the highest energy part of the spectrum invalidates this conclusion.

\subsection{Analysis of 1ES 0347-121, 1ES 1101-232, H 2356-309, and
  RGB J0152+017}
\begin{figure*}
  \mbox{\subfigure{\includegraphics[width=3.4in]{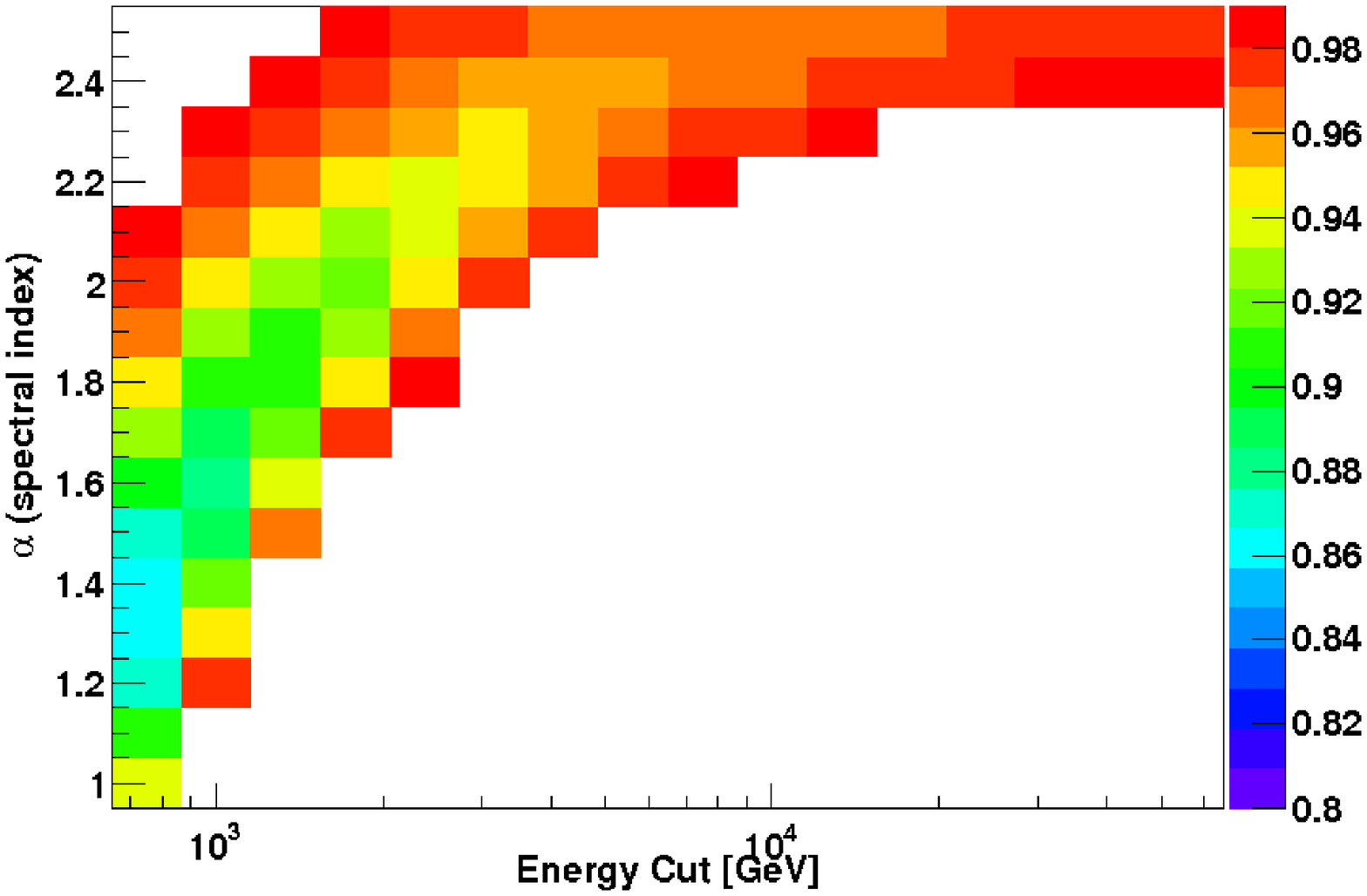}}\quad
    \subfigure{\includegraphics[width=3.4in]{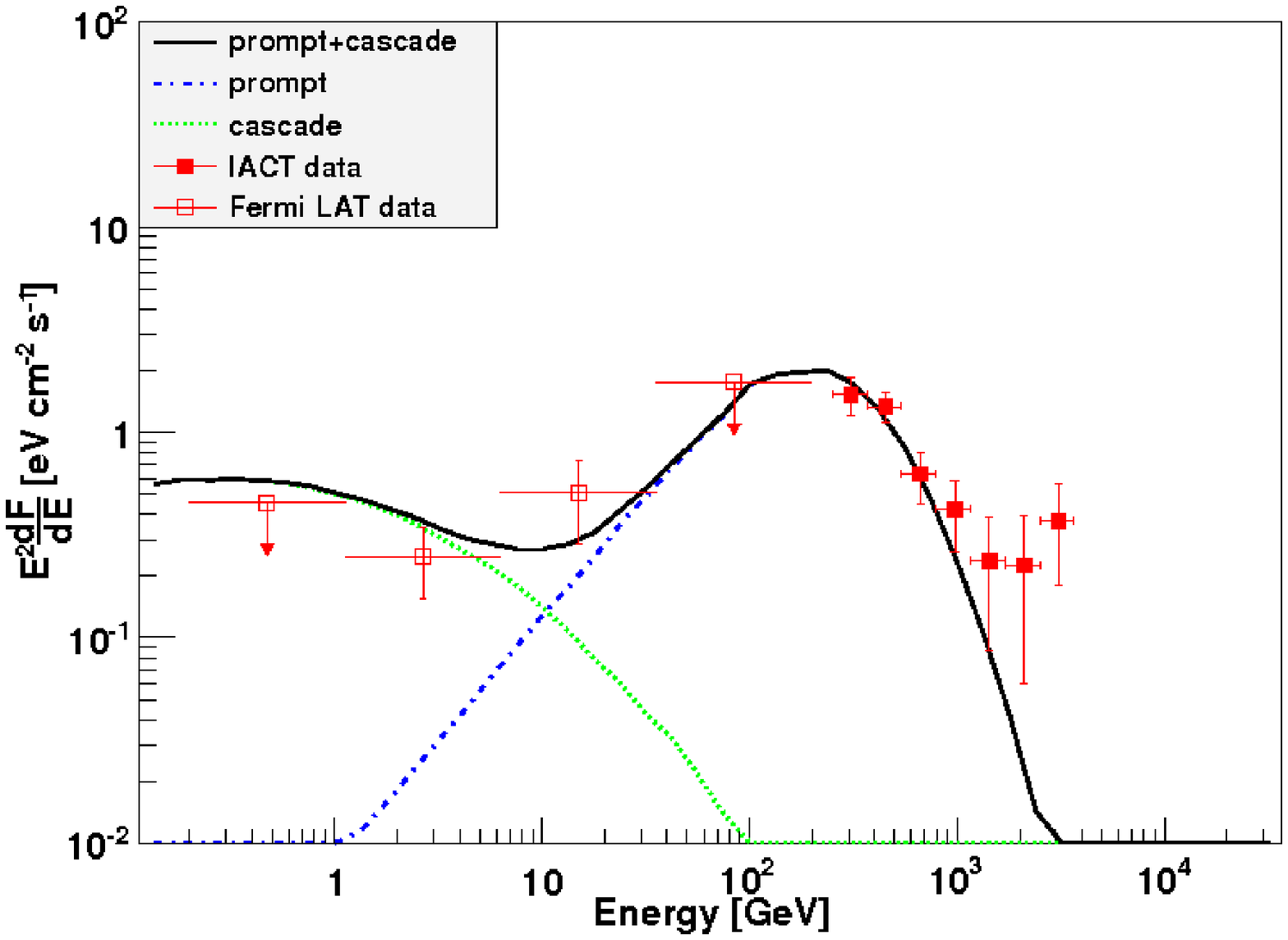} }}

\caption { \label{1es0347_B0} Source 1ES 0347-121 analyzed under the
    assumption of $H0$. (a) (left) Confidence level, or the
    probability of exclusion of the gamma-ray source model (with fixed
    $\alpha$ and $\epsilon_c$), where the remaining three parameters
    of the model ($\gamma$, $\epsilon_B$, and F$_0$, see
    Eq.~\ref{broken_power_law_model}) are chosen so as to minimize
    $\chi^2$. (b) (right) Simulated dFED of the best fit model of
    $\alpha$ = 1.3, $\epsilon_c$ = 0.75 TeV, showing both the prompt
    and secondary cascade contributions to the total dFED, along with
    the HE and VHE data.}
\end{figure*}

\begin{figure*}
 \mbox{\subfigure{\includegraphics[width=3.4in]{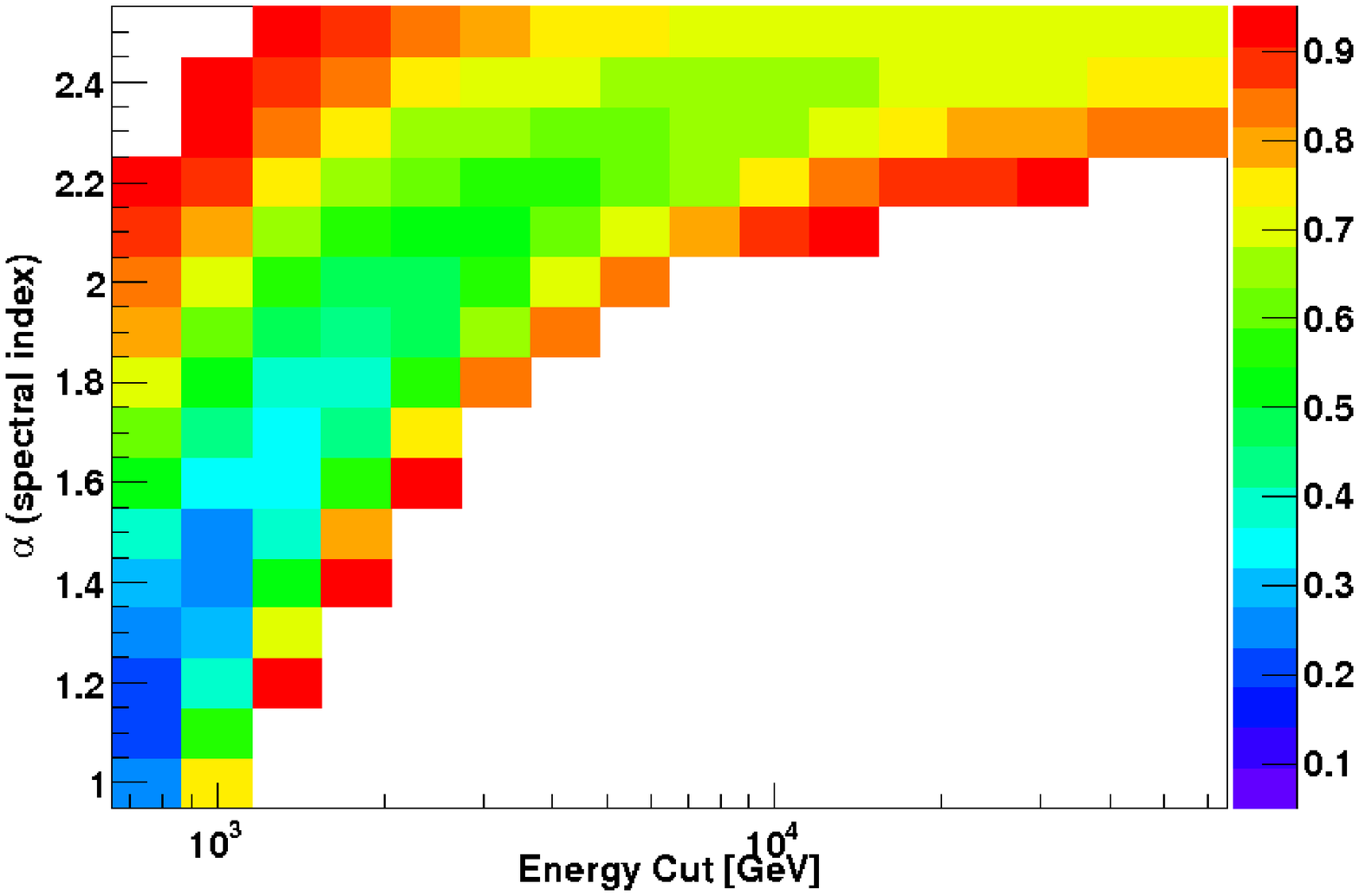}}\quad
    \subfigure{\includegraphics[width=3.4in]{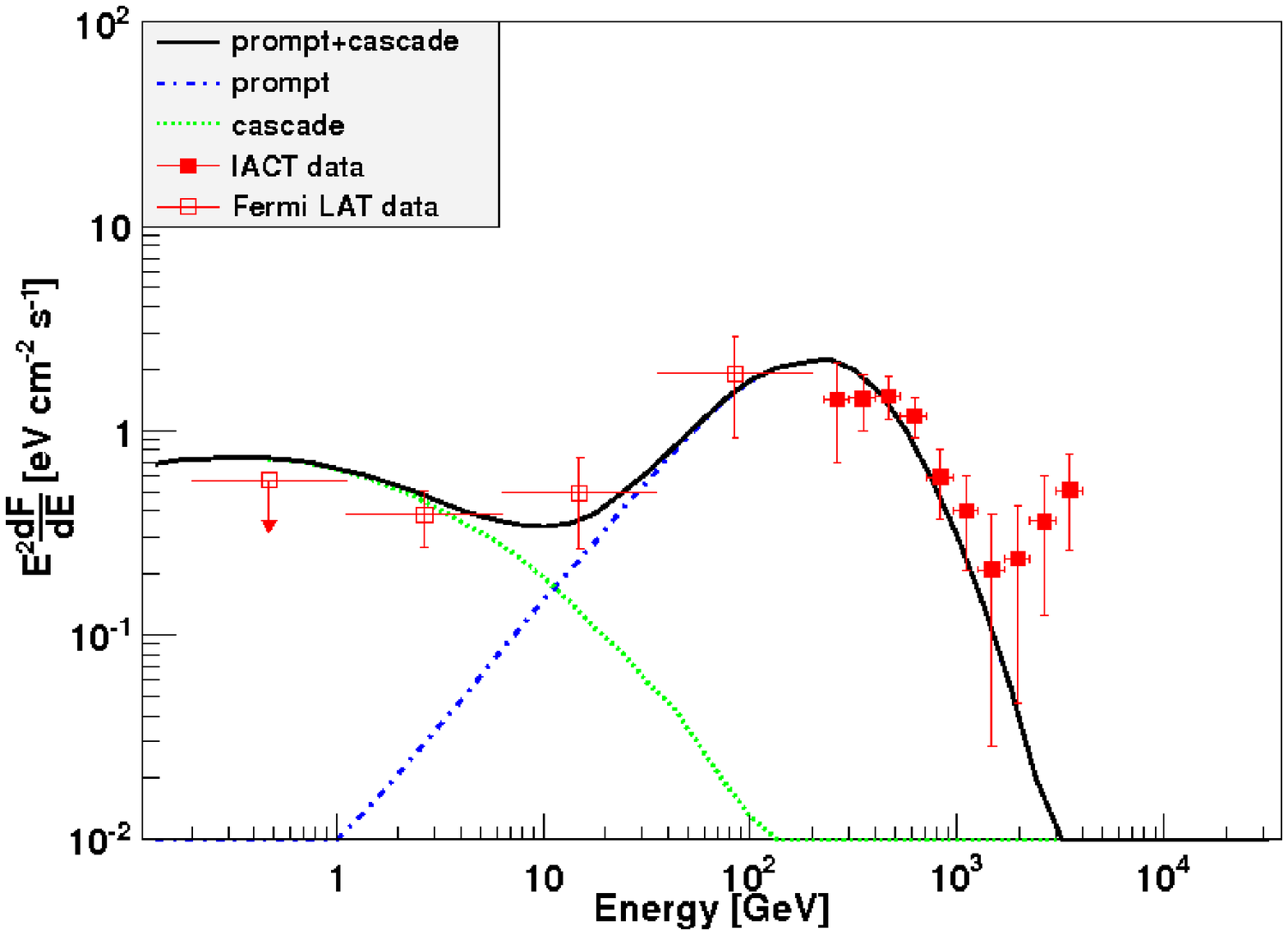} }}
\caption { \label{1es1101_B0} Source 1ES 1101-232 analyzed under the
  assumption of $H0$. (a) (left) Confidence level, or the probability
    of exclusion of the gamma-ray source model (with fixed $\alpha$
    and $\epsilon_c$), where the remaining three parameters of the
    model ($\gamma$, $\epsilon_B$, and F$_0$, see
    Eq.~\ref{broken_power_law_model}) are chosen so as to minimize $\chi^2$. (b)
    (right) Simulated dFED of the best fit model of $\alpha$ = 1.2,
    $\epsilon_c$ = 0.75 TeV, showing both the prompt and secondary
    cascade contributions to the total dFED, along with the HE and VHE
    data.}
\end{figure*}

Previously published IGMF studies (\cite{Neronov2010, Essey2010,
  Tavecchio_extreme_2011}) have used additional extreme TeV blazars to
derive constraints on the IGMF, four of which are considered in this
section, 1ES 0347-121, 1ES 1101-232, H 2356-309, and RGB
J0152+017. All of these sources were observed by the HESS
collaboration prior to the beginning of the \emph{Fermi}
mission. Therefore, to investigate $H0$, it is necessary to assume
that the VHE activity of these sources as characterized by the HESS
collaboration typically during the period of 2004 - 2007 is
representative of the VHE activity over the duration of the
\emph{Fermi} mission. The parameters of these data sets are summarized
in Table \ref{iact_data_table}. As before, the \emph{Fermi}-LAT
time-averaged dFED for all sources was derived from the beginning of
the mission until February 14, 2012, and depending on the strength of
the source, was computed for either 4 or 6 evenly spaced logarithmic
bins over the energy range of 200 MeV - 200 GeV.

The VHE data set of 1ES 0347-121 consists of a set of observations
over the period of August - December 2006, during which time 25.4
hours exposure was acculumlated. A time-averaged dFED for 7 bins over
the energy range from 250 GeV - 3.67 TeV was derived
(\cite{HESS_1ES0347}). The flux found in the first and fourth (last)
bins is weak (TS $<$ 9) for all spectral indices tested, allowing only
an upper limit to be established. The flux in the second and third
bins is typically found with $9 <$ TS $< 25$ for the simulated models
where the secondary flux dominates the total
flux. Figure~\ref{1es0347_B0}a shows the confidence level of simulated
models obtained for $H0$. The best fit models in the $\alpha$ -
$\epsilon_C$ plane are found at $\epsilon_C$ values near 1 TeV, and
the dFED for one of these models is illustrated in
Figure~\ref{1es0347_B0}b. The relatively high confidence level of the
1ES 0347-121 simulated models is partially due to the poor fit of the
highest energy bins of the VHE regime where the reported dFED
tentatively exhibits a feature of increasing energy density. This
trend in the dFED is not accounted for in the set of simulated models
investigated. A similar spectral feature appears to be even more
pronounced in the VHE data set of 1ES 1101-232, which perhaps may
signal unmodeled physics process(es) or a systematic error in the data
analyses.

The VHE data set of 1ES 1101-232 consists of 3 periods of observations
spanning from April 2004 - March 2005, for a total of 43 hours. The
time-averaged dFED is reported for 10 bins over the energy range 225
GeV - 4 TeV.  The \emph{Fermi}-LAT dFED for 1ES 1101-232 is especially
weak, and in fact, the first bin allows a determination of only an
upper limit (TS $<$ 9) for all simulated models tested. Figures
\ref{1es1101_B0}a and b illustrate the compatibility of the 1ES
1101-232 VHE and HE data sets with $H0$, despite the fact that the
previously described feature in the highest energy bins of the VHE
regime is not well fitted by the models.

\begin{figure*}
  \mbox{\subfigure{\includegraphics[width=3.4in]{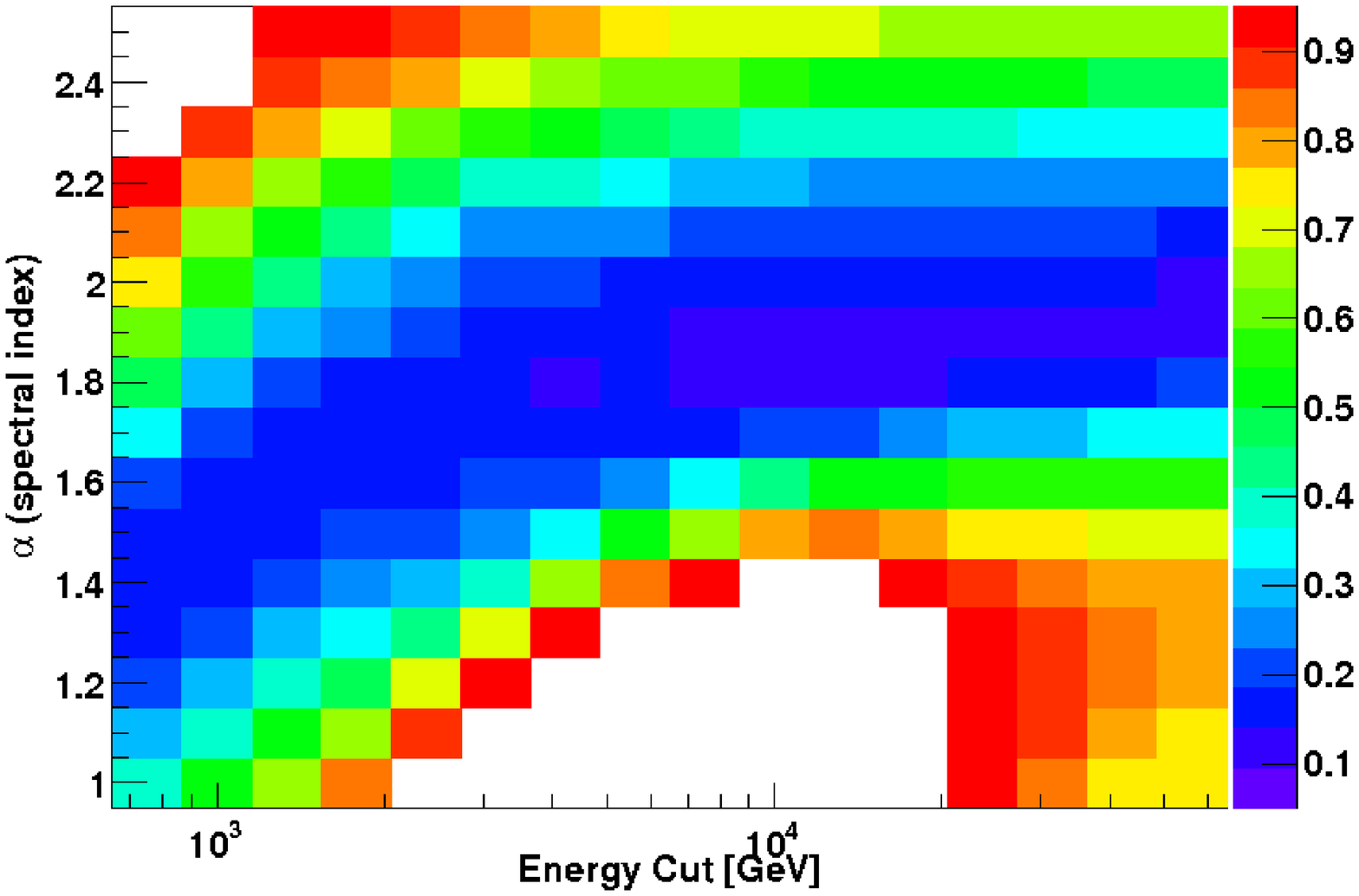}}\quad
    \subfigure{\includegraphics[width=3.4in]{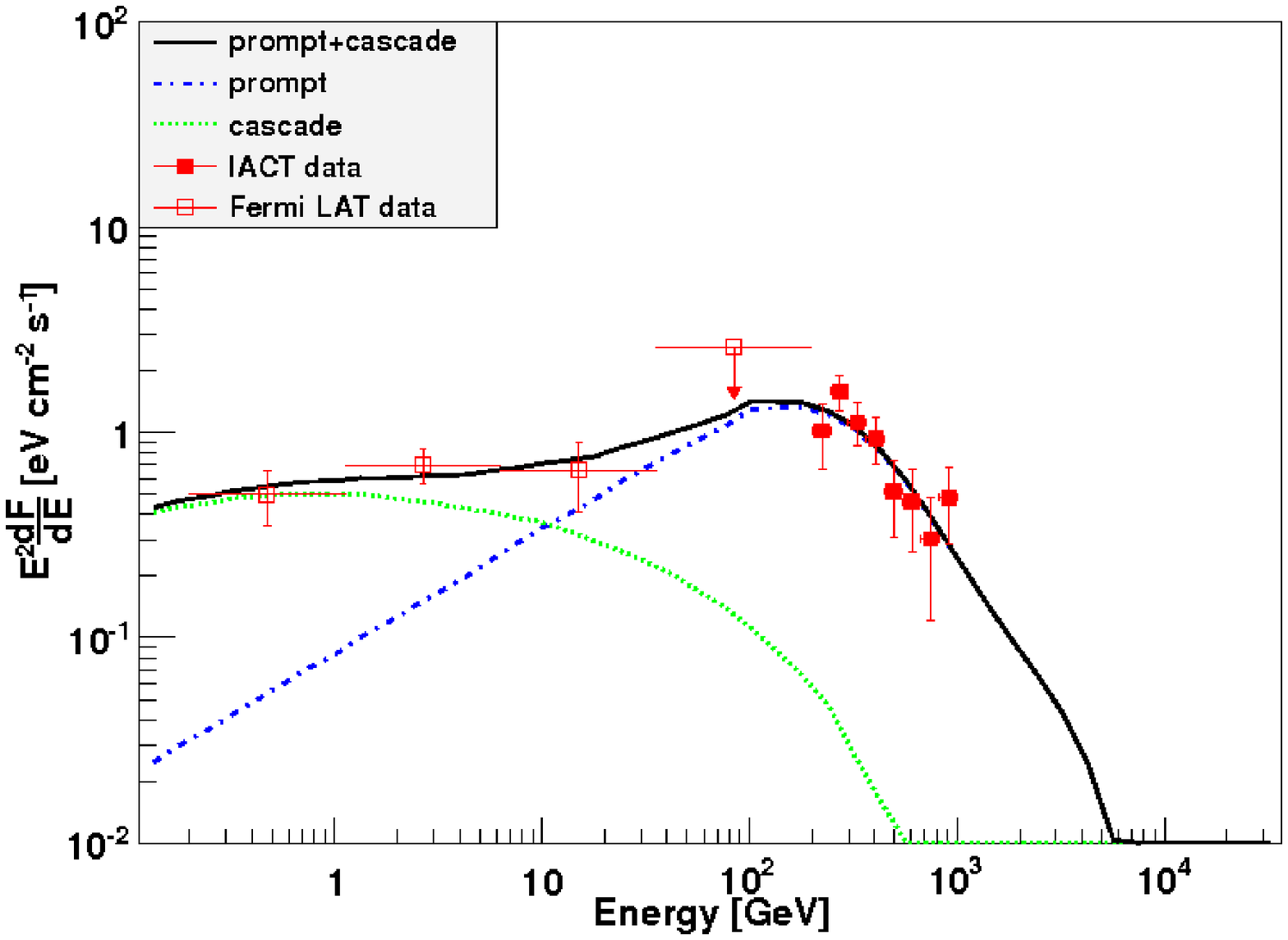} }}
  \caption { \label{H2356_B0} Source H 2356-309 analyzed under the
    assumption of $H0$. (a) (left) Confidence level, or the
    probability of exclusion of the gamma-ray source model (with fixed
    $\alpha$ and $\epsilon_c$), where the remaining three parameters
    of the model ($\gamma$, $\epsilon_B$, and F$_0$, see
    Eq.~\ref{broken_power_law_model}) are chosen so as to minimize $\chi^2$.
    (b) (right) Simulated dFED of the best fit model of $\alpha$ =
    1.8, $\epsilon_c$ = 4.22 TeV, showing both the prompt and
    secondary cascade contributions to the total dFED, along with the
    HE and VHE data.}
\end{figure*}

Observations of H 2356-309 were obtained over the period of June -
September 2004, for a total exposure of 40 hours. The time-averaged
dFED was provided for eight bins over the energy range from 200 GeV -
1.23 TeV (\cite{HESS_H2356}). The flux in the first energy bin is
weakly detected for most simulated models (9 $<$ TS $<$ 25), but the
second and third bins exhibit a strong detection (TS $>$ 25). An upper
limit is derived for the fourth bin in the HE dFED due to a weak
signal present (TS $<$ 9). As illustrated in Figure \ref{H2356_B0},
this source has a very large set of models compatible with the
$H0$. Most of these models, however, suggest that the flux in the two
lowest energy bins of the HE regime is dominated by secondary
radiation.

\begin{figure*}
  \mbox{\subfigure{\includegraphics[width=3.4in]{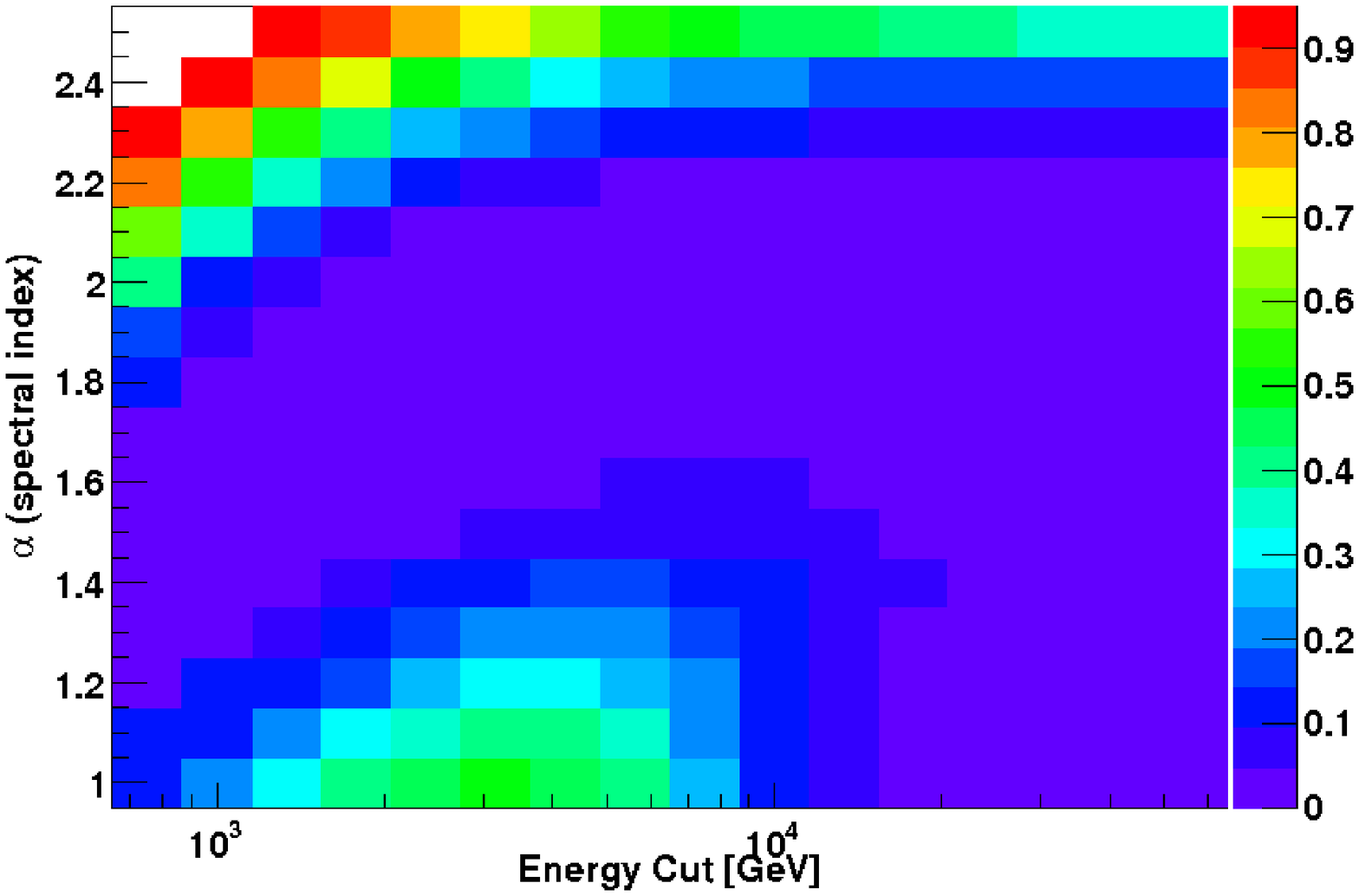}}\quad
    \subfigure{\includegraphics[width=3.4in]{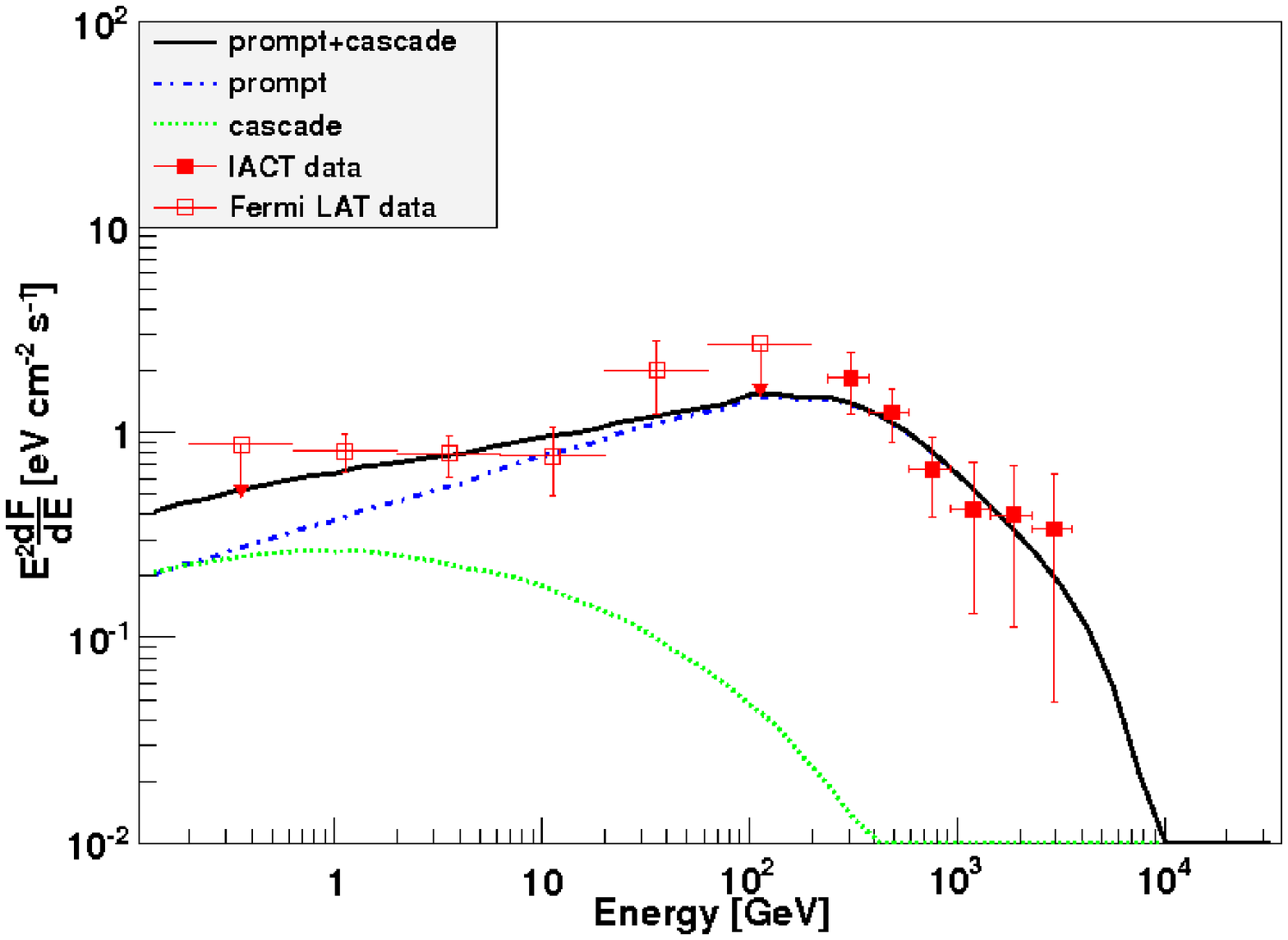} }}
  \caption { \label{rgbj0152_B0} RGB J0152+017 (a) (left) Confidence
    level, or the probability of exclusion of the gamma-ray source
    model (with fixed $\alpha$ and $\epsilon_c$), where the remaining
    three parameters of the model ($\gamma$, $\epsilon_B$, and F$_0$,
    see Eq.~\ref{broken_power_law_model}) are chosen so as to minimize
    $\chi^2$. (b) (right) Simulated dFED of the best fit model of
    $\alpha$ = 1.9, $\epsilon_c$ = 3.16 TeV, showing both the prompt
    and secondary cascade contributions to the total dFED, along with
    the HE and VHE data.}
\end{figure*}

RGB J0152+017 was observed over the period of October 30 - November 14
2007 for a total exposure of 14.7 hours. A time averaged dFED was
reported for six bins over the energy range of 240 GeV - 3.6 TeV
(\cite{HESS_H2356}). The flux in all but the first energy bin of the
\emph{Fermi} data is strongly detected (TS $>$ 25) for the majority of
simulated models. Figure \ref{rgbj0152_B0} a and b illustrate that
this source perhaps provides the weakest constraints on the IGMF with
nearly the entire parameter space of simulated models compatible with
the VHE and HE spectral data. In only some of these models, the HE
part of the spectrum is dominated by the primary emission.

%% file: Discussion.tex

In this paper, we have investigated the HE - VHE energy spectrum of
seven extreme TeV blazars for which radiation in the HE band may be
dominated by the secondary photons produced through cascading. These
sources are characterized by their observed hard spectra in the VHE
band accompanied by redshifts of order $\sim$ 0.1 suggesting a very
large energy output into pair production and subsequent cascading. For
these sources, observations in the HE band can therefore limit the
flux of secondary photons and establish a lower bound on the
IGMF. This strategy has been utilized in several publications
(\cite{Neronov2010, Tavecchio2010, Dermer2010, Essey2010, Dolag2010,
  Taylor2011, HaoHuan2011}) to suggest $B_{IGMF} \geq 10^{-17} -
10^{-18}$ G in the local cascading environment. In contrast to these
studies, we systematically investigated effects of a wide range of
uncertainties using detailed 3D Monte Carlo simulations to conclude
that $H0$ remains compatible with current observations.

Two of these blazars RGB J0710+591 and 1ES 1218+304, had VHE
observations conducted during the \emph{Fermi} mission and the
measured VHE fluxes of these sources were assumed to be representative
of the average VHE activity over the HE measurement period. For four
other sources 1ES 0347-121, 1ES 1101-232, H 2356-309, and RGB
J0152+017, the VHE component was measured prior to the \emph{Fermi}
mission and it was assumed to be representative of the average VHE
activity during the \emph{Fermi} observations. It was found that, for
all of these sources, with \emph{Fermi}-LAT Pass 7 data and a more
general set of intrinsic source models of gamma-ray emission, $H0$
cannot be rejected at $> 95 \%$ confidence level (even with a typical
EBL model, e.g. \cite{Dominguez_ebl_2010} and models referenced
within).

A single source, the extreme blazar 1ES 0229+200, did allow for a
rejection of $H0$ at more than $99\%$ confidence level, under the
standard assumptions as discussed in Section
\ref{subsection_1es0229_analysis}. However, if the EBL model SED is
decreased in the visible and near infrared band to within the
uncertainty limits of measurements based on the galaxy counts, the
$B_{IGMF}$ lower limit cannot be maintained. $H0$ can also be
reconciled with the data if the VHE measurements of its flux conducted
over the period of forty hours are not assumed to be representative of
the past 3.25 years of activity of this source. We must also point out
that a thorough relative energy calibration of the \emph{Fermi}-LAT
and IACTs has not yet been reported and an error in the energy scale
of VHE instruments may have significant impact on the conclusions
regarding $B_{IGMF}$.

\begin{figure}[h]
  \begin{centering} \includegraphics[width=3.4in]{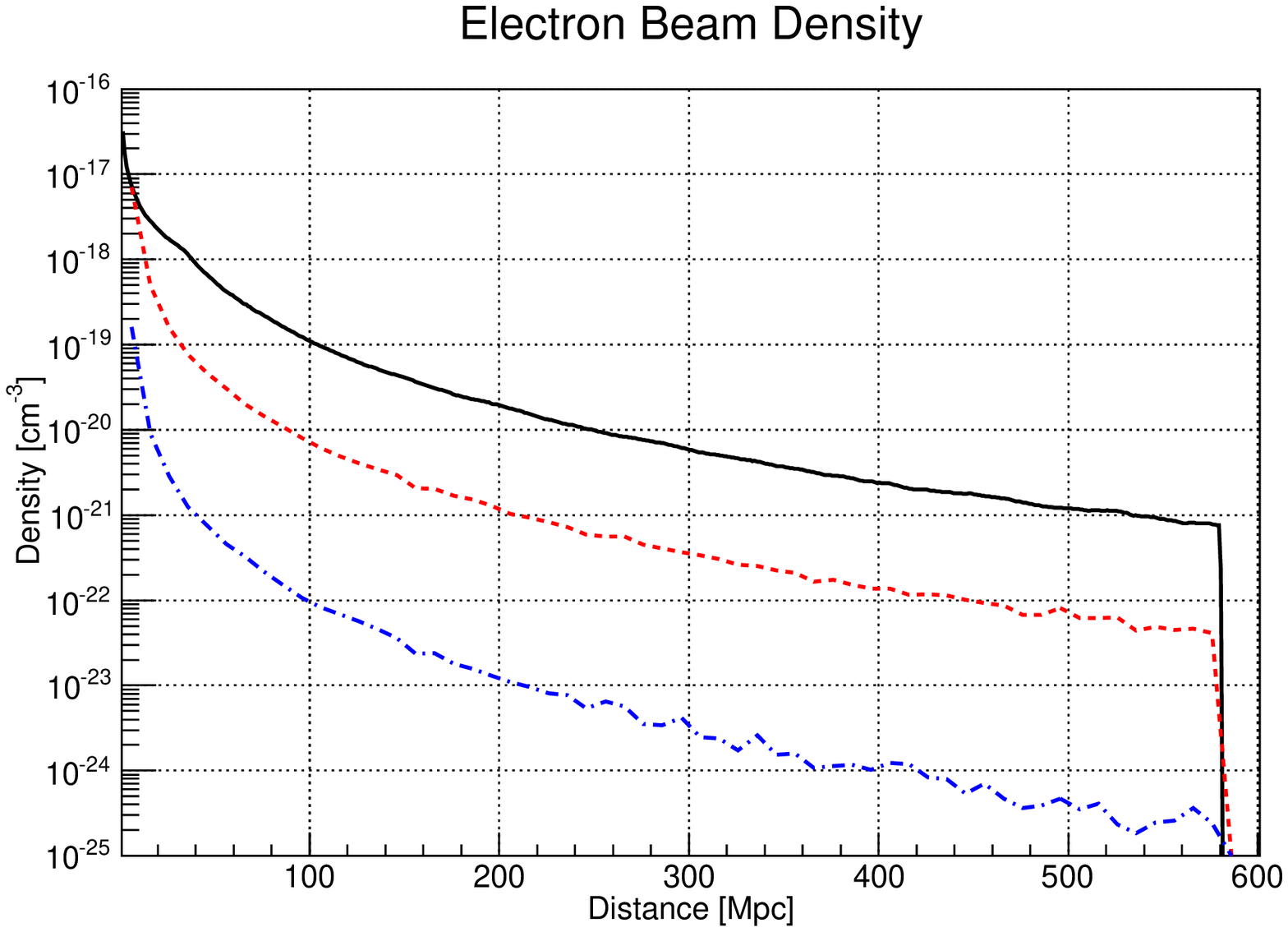} \caption
     { \label{electron_density} Electron Density as a function of
     distance from source 1ES 0229+200, assuming a normal EBL model
     (EBL model 1) and a spectral index in the VHE band ($\alpha$) of
     1.3 and cutoff energy ($\varepsilon_c$) 3.16 TeV. The black
     (solid) line traces the density of electrons with energies $> 10$
     GeV, the red (dashed) line represents the density of electrons
     with energies $> 100$ GeV, and the blue (dot-dashed) line
     corresponds to energies $> 1$ TeV.}  \end{centering}
\end{figure}

\begin{figure}[h]
  \begin{centering} \includegraphics[width=3.4in]{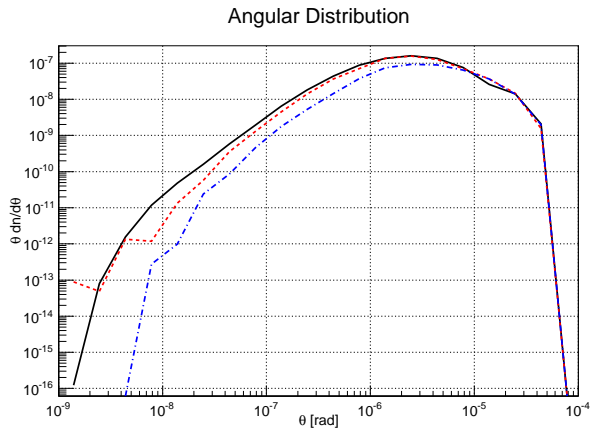} \caption
     { \label{electronAngDist} Electron angular distribution as a
     function of distance from the source 1ES 0229+200, assuming a
     normal EBL model (EBL model 1) and a spectral index in the VHE
     band ($\alpha$) of 1.3 and cutoff energy ($\varepsilon_c$) 3.16
     TeV. The angular distibution of electrons with energies above 10
     GeV at 30 Mpc (black, solid), 100 Mpc (red, dashed), and 300 Mpc
     (blue, dot-dashed). The vicinity of the peak of the distribution
     can be well-represented as a log$_{10}$-normal distribution
     peaking at 10$^{-5.5}$ rad, with $\sigma$ of 0.65.} \end{centering}
\end{figure}

It should also be noted that a claim of the rejection of $H0$ based on
the observational data of a single source should not be considered
robust due to possible variations of the magnetic field in the local
environment of the source. Figure \ref{electron_density} shows the
number density of the secondary electrons from 1ES 0229+200 for a zero
magnetic field as a function of distance from the source for electrons
with energies$> 10$ GeV, $> 100$ GeV, and $> 1$ TeV. It can be seen
that a significant fraction of the secondary gamma-ray flux can be
isotropized if cascading begins in the vicinity of several tens of Mpc
of the source in the magnetic field of filaments which is expected to
be orders of magnitude larger than the magnetic field in the
voids. Recent work by \cite{Sutter_VoidCat} and \cite{Pan_Voids_2012}
on identifying and characterizing the Voids in the Sloan Digital Sky
Survey Data release 7 (\cite{SDSS_Data7}), suggests that the volume
occupied by them accounts for around 40 - 60 \% of the total volume of
the universe. It is therefore critical for studies of the $B_{IGMF}$
to know the exact line of sight distribution of voids particularly in
the few hundred Mpc vicinity of the source.

Another avenue to reduce the secondary gamma-ray flux and reconcile it
with observations and $H0$ has recently been proposed by
\cite{Broderick_instability}. The authors argue that beam-plasma
instabilities could develop through interaction with the electrons and
positrons in the cascade and electrons of the plasma in the voids. If
such a collective interaction between two populations of electrons
indeed exists, the energy dissipation rate of electron positron pairs
into modes of the plasma waves in voids may become significantly
larger than the energy loss through IC scattering.

Parameters of the plasma in the voids are largely unknown. To estimate
an upper limit on the plasma electron number density in the voids, we
assume that the mass of the baryonic matter in the voids does not
exceed the difference between the baryonic mass identified through
analysis of CMB fluctuations and the mass found in the galaxies and
galaxy clusters. The cosmological density of the universe is 5.9
baryons m$^{-3}$ of which 4.6 \% represents baryonic matter. Ninety
percent of the $2.7\times 10^{-7}$ baryons cm$^{-3}$ were identified
in filaments and clusters of galaxies filling about half the volume of
the universe. Therefore, the amount of baryonic matter in the voids
cannot exceed the remaining $\approx$ $0.5 \times$10 \%, making the
density estimate of the electron plasma in the voids $n_e \approx
1.4 \times 10^{-8}$ cm$^{-3}$ or lower. The plasma frequency of these
electrons, $\omega_{p,e} = 2\pi f_{p,e} = \left( 4 \pi n_e e^2/m_e
\right)^{1/2}$, is given by $f_{p,e} = 1\text{ Hz} \left(
  n_e/10^{-8} \right)^{1/2}$. 

To estimate the temperature of the electron plasma, we assume that the
ionization of hydrogen in the voids occurs sometime during the end of
the reionization epoch with redshift z between 6 -- 10. At this point,
the universe became mostly transparent to UV radiation, and UV photons
from ionizing sources in galaxies could propagate to the voids. The
photoionization cross section for hydrogen in the ground state by
photons with energy $\varepsilon > \varepsilon_0 = 13.6$ eV, can be
roughly approximated by $\sigma_i(\varepsilon/\varepsilon_0)^{-3}$
where $\sigma_i$ = $6.3 \times 10^{-18}$ cm$^2$. The ionization is
suppressed at higher photon energies, suggesting that the
characteristic kinetic energy of the plasma electrons acquired by an
electron in the ionization process is of order $\varepsilon_0$. These
electrons undergo cosmological expansion which reduces the kinetic
energy by a factor of (1 + z), and they also transfer energy to CMB
photons through the inverse Compton (IC) interaction. The heating of
electrons by UV radiation accumulated and reprocessed in the EBL is
negligible since the total energy density in the EBL (from the UV to
far-IR) is an order of magnitude less than in the CMB. The average IC
energy loss of non-relativistic electrons is given by

\begin{equation}
\frac{1}{E}\frac{dE}{dt} = \frac{8}{3}\sigma_{\mathrm{T}} c \frac{U_{cmb}}{m_e c^2}
\end{equation}

\noindent where $\sigma_{\mathrm{T}}$ is the Thomson cross section and $U_{cmb}$
is the CMB energy density which at the present epoch is $U_0 = 0.26$
eV/cm$^{3}$. In the context of cosmological expansion, the rate of IC
energy loss of non-relativistic particles is given by

\begin{equation}
\ln \left( \frac{E(z)}{E(0)} \right) = \frac{8}{3} \sigma_{\mathrm{T}} \frac{c}{H_0}\frac{U_0}{m_ec^2} \int_{0}^{z} \frac{(1+z^{\prime})^3 dz^{\prime}}{ \sqrt{\Omega_m(1 + z^{\prime})^3 + \Omega_{\Lambda} } }
\end{equation}

\noindent where $H_0$ is the Hubble constant. For z = 6, the RHS of this
equation is equal to 1.1 and for z = 10, it is equal to 3.4. This,
combined with the reduction of the kinetic energy due to cosmological
expansion, suggests that the average kinetic energy of the plasma
electrons is reduced by a factor of $5 \times 10^{-2}$ (z = 6) or
$3 \times 10^{-3}$ (z = 10), from $\varepsilon_0$, implying that the
present day temperature of the electron plasma in the voids is a few
thousand K.

Given the temperature and density of the electron plasma in the voids,
the screening Debye radius is $\lambda_D = \left( kT/8\pi n_e
e^2 \right)^{1/2}$ = $1.6 \times 10^6 \text{ cm} \sqrt{ \left( T/10^3
\text{ K} \right) \left(10^{-8}\text{cm}^{-3}/n_e \right) }$. We note that
this screening radius is less than the typical distance between
electrons in the beam, as estimated in our simulations shown in
Figure \ref{electron_density}. The bulk of the electrons in the beam
is born near the threshold of pair production when a few hundred GeV
primary photon interacts with the EBL. The typical Lorentz factor of
the outgoing electon is $\sim 2\times 10^{5}$. These electrons
dissipate $\sim 90$ \% of their energy through IC scattering over the
distance of $\sim 30$ Mpc. For distances from the source larger than
this, the typical separation of beam particles is $n_b^{-1/3}$
$\approx$ $10^{6} - 10^{7}$ cm. It appears that electron positron
pairs of the beam are significantly screened by the electrons of the
plasma in the voids, and therefore, collective interactions to
generate a higher rate of energy transfer to plasma oscillations
should be strongly suppressed by screening, which was not accounted
for in \cite{Broderick_instability}. The plasma condition at these
relatively small distances from the source ($\sim$ few Mpc where the
density of the beam may be higher) is likely to be different from
those assumed here for the voids, and the Debye radius estimated may
not be applicable in this region.

One may argue that this screening effect can be alleviated if the
density of electrons in the voids is considerably lower than the given
upper bound estimate. Based on the results of simulations of density
fluctuations quoted in \cite{Miniati_Elyiv_Voids}, the present-day
density of the electrons in objects with spatial scales of several
tens of Mpc might be as low as $\approx$ $4 \times 10^{-10}$
cm$^{-3}$. This would make the Debye radius, $\lambda_D \approx 10^7$ cm,
which suggests that screening effects remain critically important for
the rate of the energy losses into the excitation of plasma waves. To
estimate another critical paramter in the problem, namely, the effect
of the angular distribution of electrons in the beam, we consider a
hydrodynamical approximation of the beam-plasma instability for the
two populations of electrons with zero temperatures and one beam at
relative velocity c. The dispersion relation for these waves is given
by

\begin{equation}
  1 = \frac{\omega^2_{p,e}}{\omega^2} + \frac{\omega^2_{p,b}}{ (
    \textbf{k} \cdot \textbf{c} - \omega )^2 } = \frac{1}{(x+iy)^2} +
  \frac{\kappa^2}{(z-x-iy)^2},
\end{equation}

\noindent where $\omega_{p,b}$ is the plasma frequency of the
electrons and positrons in the beam, and where we introduced the
dimensionless variables $x = Re(\omega)/\omega_{p,e}$, $y=
Im(\omega)/\omega_{p,e}$, $z = \textbf{k} \cdot
\textbf{c}/\omega_{p,e}$ and $\kappa = \omega_{p,b}/\omega_{p,e} =
\sqrt{n_b/n_e} \ll 1$. The real and imaginary part of this equation
can be separated, and the unstable (y $\ne$ 0) solution should satisfy

\begin{equation}
z^2 = \frac{\left( 1 + \varepsilon^2 \right)}{\left(1 -
 \varepsilon^2 \right)} \frac{\left( 1 - \kappa \varepsilon \right)
 \left( \varepsilon - \kappa \right) }{\varepsilon},
\end{equation}

\begin{equation}
\frac{y^2}{x^2} = \varepsilon\frac{\kappa - \varepsilon^3}{1 -
  \kappa \varepsilon},
\end{equation} 

\noindent where $\varepsilon = \sqrt{z/x - 1}$. The solution exists
for $\kappa < \varepsilon < \kappa^{1/3}$ and for $0 < y/x <
3^{1/2}(\kappa/4)^{2/3}$. Since $\kappa \ll 1$, this would require
that $\left\lvert z \right\rvert/ \left\lvert x \right\rvert \sim 1$
and that $\left\lvert z \right\rvert \leq 1$. The last condition can
only be satisfied for plasma waves with wave vector precisely
perpendicular to the beam velocity. The cosine of this angle must
satisfy the following condition $\cos\theta$ $\leq (\lambda_D / c
\omega_{p,e})(k\lambda_D)^{-1} = (k_BT / 2m_ec^2)^{1/2}
(k\lambda_D)^{-1}$ = $2.9 \times 10^{-4} (T/10^3 \text{K})^{1/2} (k
\lambda_D)^{-1}$, in which $k \lambda_D \gg 1$ to enable collective
interaction of plasma particles without significant screening. For the
lowest estimate for $n_e$ in the voids, we assume that
$k \lambda_D \sim$ few, and therefore $\cos\theta$ may need to be less
than $10^{-4}$, but not significatnly less than this. 

According to our simulations, the angular distribution of electrons in
the beam is shown in Fig.~\ref{electronAngDist}. The bulk of the
electrons is created with characteristic angles of order $1/\gamma$
$\approx$ $5\times 10^{-6}$. Although these electrons lose their
energy through IC scattering, they retain the angular distribution of
the parent particle, due to a small energy transfer to the CMB
photons. Therefore, the angular distribution is relatively independent
from the distance to the source, as illustrated in
Fig.~\ref{electronAngDist}. Under the condition of the lowest electron
density in the voids, maintaining perpendicularity of the wave vector
of the plasma waves with respect to the beam, appears to be possible,
and the approximation of the beam as completely collimated seems
applicable here. All particles of the beam in this reactive regime of
interaction will be involved into the energy transfer to the plasma
waves making energy losses potentially higher. If $k \lambda_D$
becomes larger than about a few tens, then plasma waves will interact
collectively only with a fraction of beam electrons in the kinetic
regime of interaction, which has been pointed out
in \cite{Miniati_Elyiv_Voids}. The energy losses in this regime are
expected to be lower. Based on the parameters of the electron beam
which we derived based on simulations, and reasonable assumptions
about properties of electron plasma in the voids, we conclude that all
effects of Debye screening and kinetic vs. reactive descriptions of
two beam instabilities are important. Perhaps the most detailed to
date study of the relaxation of beam plasma instabilities in cosmic
voids has been reported in \cite{Miniati_Elyiv_Voids} with the
conclusion that the relativistic pair beams of blazars remain stable
on timescales much longer than the characteristic IC cooling time of
electrons, and collective plasma-beam interaction effects in the voids
are negligible. We find that the screening effects not accounted for
in this work will only further validate their conclusions. However,
the available parameter space in the regime of very low plasma density
in the voids may enable the reactive interaction regime and therefore
increase the energy loss rate.

An alternative approach to reconcile a lack of the secondary HE
radiation in the observational data of extreme blazars is to suggest
that a significant part of the VHE emission originates from
proton-initiated cascades, rather than from direct photons, if CR
protons were also accelerated by the same source
(\cite{Essey_protons}). The universe is nearly transparent to
$10^{16}$ - $10^{19}$ eV protons on spatial scales of hundreds of Mpc
and therefore, electromagnetic cascades can be initiated by these
particles in a random location along the line of sight. Crossing small
regions of intense magnetic fields such as those present in clusters
and filaments represent difficulties for this mechanism since they
rapidly destroy the correlations between the cosmic ray directions and
the source. Moreover, it is likely a formidable task to devise a
mechanism by which cosmic rays have been accelerated to ultra high
energies in the source by intense magnetic fields satisfying the
Hillas condition and then highly collimated along a particular
direction, eventually entering regions of potentially extremely small
magnetic fields of voids without being significantly disrupted by the
intermediate magnetic fields. If this mechanism takes place in nature,
it has two distinct observational characteristics, namely that the VHE
radiation produced should show little evidence for variability and
correlations with other wavelength bands such as X-ray, etc. and most
importantly, should be accompanied by higher energy gamma-rays, with
energies above 10s of TeV in the VHE spectra of extreme blazars with z
$\geq$ 0.1.  No such observational evidence has been collected so far
to necessitate considerations of this scenario, which also lacks an
explanation of the origin of the collimation. This mechanism has been
recently studied (\cite{Murase_2012_UHECR}) in the context of 1ES
0229+200 with the conclusion that the detection of larger than 25 TeV
photons would provide an indication of acceleration of ultra high
energy cosmic rays in this source.

In summary, we have considered the observational data of 7 extreme TeV
blazars and performed detailed simulations of cascading in the
intergalactic magnetic field of the cosmic voids. We find for all
sources, except for 1ES 0229+200, no evidence to claim exclusion of
$H0$ ($B_{IGMF} = 0$ hypothesis), due to existing multiple systematic
uncertainties in source models. Furthermore, for the case of 1ES
0229+200, we find no definitive evidence that its data is in
contradiction with the zero IGMF hypothesis, when astrophysical (EBL,
local magnetic field environment) and systematic uncertainties (VHE
duty cycle) are accounted for.

\bigskip

We thank the anonymous referee for their critical remarks and
important suggestions, which improved the paper significantly. The
simulations were performed on the UCLA hoffman2 cluster as well as the
Joint Fermilab - KICP Supercomputing Cluster.  V.V.V. gratefully
acknowledges grants from UCLA and S.P.W. gratefully acknowledges
grants from Fermilab, Kavli Institute for Cosmological Physics, and
the University of Chicago. This research is supported by the
U.S. National Science Foundation under grants no. (PHY-0969948,
PHY-0422093, and PHY-0969529).

%% file: IGMF_Paper.bbl
\begin{thebibliography}{66}
\expandafter\ifx\csname natexlab\endcsname\relax\def\natexlab#1{#1}\fi

\bibitem[{{Abazajian} {et~al.}(2009){Abazajian}, {Adelman-McCarthy},
  {Ag{\"u}eros}, {Allam}, {Allende Prieto}, {An}, {Anderson}, {Anderson},
  {Annis}, {Bahcall}, \& et~al.}]{SDSS_Data7}
{Abazajian}, K.~N., {et~al.} 2009, \apjs, 182, 543

\bibitem[{{Acciari} {et~al.}(2009){Acciari}, {Aliu}, {Arlen}, {Beilicke},
  {Benbow}, {Bradbury}, {Buckley}, {Bugaev}, {Butt}, {Byrum}, {Celik},
  {Cesarini}, {Ciupik}, {Chow}, {Cogan}, {Colin}, {Cui}, {Daniel}, {Ergin},
  {Falcone}, {Fegan}, {Finley}, {Fortin}, {Fortson}, {Furniss}, {Gillanders},
  {Grube}, {Guenette}, {Gyuk}, {Hanna}, {Hays}, {Holder}, {Horan}, {Hui},
  {Humensky}, {Imran}, {Kaaret}, {Karlsson}, {Kertzman}, {Kieda}, {Kildea},
  {Konopelko}, {Krawczynski}, {Krennrich}, {Lang}, {LeBohec}, {Maier},
  {McCann}, {McCutcheon}, {Moriarty}, {Mukherjee}, {Nagai}, {Niemiec}, {Ong},
  {Pandel}, {Perkins}, {Pohl}, {Quinn}, {Ragan}, {Reyes}, {Reynolds}, {Rose},
  {Schroedter}, {Sembroski}, {Smith}, {Steele}, {Swordy}, {Toner}, {Valcarcel},
  {Vassiliev}, {Wagner}, {Wakely}, {Ward}, {Weekes}, {Weinstein}, {White},
  {Williams}, {Wissel}, {Wood}, \& {Zitzer}}]{VERITAS_1ES1218_2009}
{Acciari}, V.~A., {et~al.} 2009, \apj, 695, 1370

\bibitem[{{Acciari} {et~al.}(2010{\natexlab{a}}){Acciari}, {Aliu}, {Beilicke},
  {Benbow}, {Boltuch}, {B{\"o}ttcher}, {Bradbury}, {Bugaev}, {Byrum},
  {Cesarini}, {Ciupik}, {Cogan}, {Cui}, {Dickherber}, {Duke}, {Falcone},
  {Finley}, {Finnegan}, {Fortson}, {Furniss}, {Galante}, {Gall}, {Gibbs},
  {Guenette}, {Gillanders}, {Godambe}, {Grube}, {Hanna}, {Hui}, {Humensky},
  {Imran}, {Kaaret}, {Karlsson}, {Kertzman}, {Kieda}, {Krawczynski},
  {Krennrich}, {Lang}, {LeBohec}, {Maier}, {McArthur}, {McCann}, {Moriarty},
  {Nagai}, {Ong}, {Otte}, {Pandel}, {Perkins}, {Pichel}, {Pohl}, {Quinn},
  {Ragan}, {Reyes}, {Reynolds}, {Roache}, {Rose}, {Schroedter}, {Sembroski},
  {Smith}, {Steele}, {Swordy}, {Theiling}, {Thibadeau}, {Vassiliev}, {Vincent},
  {Wakely}, {Weekes}, {Weinstein}, {Weisgarber}, {Williams}, \& {VERITAS
  Collaboration}}]{VERITAS_1ES1218_2010}
---. 2010{\natexlab{a}}, \apjl, 709, L163

\bibitem[{{Acciari} {et~al.}(2010{\natexlab{b}}){Acciari}, {Aliu}, {Arlen},
  {Aune}, {Bautista}, {Beilicke}, {Benbow}, {B{\"o}ttcher}, {Boltuch},
  {Bradbury}, \& et~al.}]{VERITAS_RGBJ0710}
---. 2010{\natexlab{b}}, \apjl, 715, L49

\bibitem[{{Aharonian} {et~al.}(2006{\natexlab{a}}){Aharonian}, {Akhperjanian},
  {Bazer-Bachi}, {Beilicke}, {Benbow}, {Berge}, {Bernl{\"o}hr}, {Boisson},
  {Bolz}, {Borrel}, {Braun}, {Breitling}, {Brown}, {Chadwick}, {Chounet},
  {Cornils}, {Costamante}, {Degrange}, {Dickinson}, {Djannati-Ata{\"i}},
  {Drury}, {Dubus}, {Emmanoulopoulos}, {Espigat}, {Feinstein}, {Fontaine},
  {Fuchs}, {Funk}, {Gallant}, {Giebels}, {Gillessen}, {Glicenstein}, {Goret},
  {Hadjichristidis}, {Hauser}, {Hauser}, {Heinzelmann}, {Henri}, {Hermann},
  {Hinton}, {Hofmann}, {Holleran}, {Horns}, {Jacholkowska}, {de Jager},
  {Kh{\'e}lifi}, {Klages}, {Komin}, {Konopelko}, {Latham}, {Le Gallou},
  {Lemi{\`e}re}, {Lemoine-Goumard}, {Leroy}, {Lohse}, {Martin},
  {Martineau-Huynh}, {Marcowith}, {Masterson}, {McComb}, {de Naurois}, {Nolan},
  {Noutsos}, {Orford}, {Osborne}, {Ouchrif}, {Panter}, {Pelletier}, {Pita},
  {P{\"u}hlhofer}, {Punch}, {Raubenheimer}, {Raue}, {Raux}, {Rayner}, {Reimer},
  {Reimer}, {Ripken}, {Rob}, {Rolland}, {Rowell}, {Sahakian}, {Saug{\'e}},
  {Schlenker}, {Schlickeiser}, {Schuster}, {Schwanke}, {Siewert}, {Sol},
  {Spangler}, {Steenkamp}, {Stegmann}, {Tavernet}, {Terrier}, {Th{\'e}oret},
  {Tluczykont}, {van Eldik}, {Vasileiadis}, {Venter}, {Vincent}, {V{\"o}lk}, \&
  {Wagner}}]{Aharonian_Nature_2006}
{Aharonian}, F., {et~al.} 2006{\natexlab{a}}, \nat, 440, 1018

\bibitem[{{Aharonian} {et~al.}(2006{\natexlab{b}}){Aharonian}, {Akhperjanian},
  {Bazer-Bachi}, {Beilicke}, {Benbow}, {Berge}, {Bernl{\"o}hr}, {Boisson},
  {Bolz}, {Borrel}, {Braun}, {Breitling}, {Brown}, {B{\"u}hler},
  {B{\"u}sching}, {Carrigan}, {Chadwick}, {Chounet}, {Cornils}, {Costamante},
  {Degrange}, {Dickinson}, {Djannati-Ata{\"i}}, {O'C.~Drury}, {Dubus},
  {Egberts}, {Emmanoulopoulos}, {Espigat}, {Feinstein}, {Ferrero}, {Fontaine},
  {Funk}, {Funk}, {Gallant}, {Giebels}, {Glicenstein}, {Goret},
  {Hadjichristidis}, {Hauser}, {Hauser}, {Heinzelmann}, {Henri}, {Hermann},
  {Hinton}, {Hofmann}, {Holleran}, {Horns}, {Jacholkowska}, {de Jager},
  {Kh{\'e}lifi}, {Komin}, {Konopelko}, {Latham}, {Le Gallou}, {Lemi{\`e}re},
  {Lemoine-Goumard}, {Lohse}, {Martin}, {Martineau-Huynh}, {Marcowith},
  {Masterson}, {McComb}, {de Naurois}, {Nedbal}, {Nolan}, {Noutsos}, {Orford},
  {Osborne}, {Ouchrif}, {Panter}, {Pelletier}, {Pita}, {P{\"u}hlhofer},
  {Punch}, {Raubenheimer}, {Raue}, {Rayner}, {Reimer}, {Reimer}, {Ripken},
  {Rob}, {Rolland}, {Rowell}, {Sahakian}, {Saug{\'e}}, {Schlenker},
  {Schlickeiser}, {Schwanke}, {Sol}, {Spangler}, {Spanier}, {Steenkamp},
  {Stegmann}, {Superina}, {Tavernet}, {Terrier}, {Th{\'e}oret}, {Tluczykont},
  {van Eldik}, {Vasileiadis}, {Venter}, {Vincent}, {V{\"o}lk}, {Wagner}, \&
  {Ward}}]{HESS_H2356}
---. 2006{\natexlab{b}}, \aap, 455, 461

\bibitem[{{Aharonian} {et~al.}(2006{\natexlab{c}}){Aharonian}, {Akhperjanian},
  {Bazer-Bachi}, {Beilicke}, {Benbow}, {Berge}, {Bernl{\"o}hr}, {Boisson},
  {Bolz}, {Borrel}, {Braun}, {Breitling}, {Brown}, {B{\"u}hler},
  {B{\"u}sching}, {Carrigan}, {Chadwick}, {Chounet}, {Cornils}, {Costamante},
  {Degrange}, {Dickinson}, {Djannati-Ata{\"i}}, {O'C.~Drury}, {Dubus},
  {Egberts}, {Emmanoulopoulos}, {Espigat}, {Feinstein}, {Ferrero}, {Fiasson},
  {Fontaine}, {Funk}, {Funk}, {Gallant}, {Giebels}, {Glicenstein}, {Goret},
  {Hadjichristidis}, {Hauser}, {Hauser}, {Heinzelmann}, {Henri}, {Hermann},
  {Hinton}, {Hofmann}, {Holleran}, {Horns}, {Jacholkowska}, {de Jager},
  {Kh{\'e}lifi}, {Komin}, {Konopelko}, {Kosack}, {Latham}, {Le Gallou},
  {Lemi{\`e}re}, {Lemoine-Goumard}, {Lohse}, {Martin}, {Martineau-Huynh},
  {Marcowith}, {Masterson}, {McComb}, {de Naurois}, {Nedbal}, {Nolan},
  {Noutsos}, {Orford}, {Osborne}, {Ouchrif}, {Panter}, {Pelletier}, {Pita},
  {P{\"u}hlhofer}, {Punch}, {Raubenheimer}, {Raue}, {Rayner}, {Reimer},
  {Reimer}, {Ripken}, {Rob}, {Rolland}, {Rowell}, {Sahakian}, {Saug{\'e}},
  {Schlenker}, {Schlickeiser}, {Schwanke}, {Sol}, {Spangler}, {Spanier},
  {Steenkamp}, {Stegmann}, {Superina}, {Tavernet}, {Terrier}, {Th{\'e}oret},
  {Tluczykont}, {van Eldik}, {Vasileiadis}, {Venter}, {Vincent}, {V{\"o}lk},
  {Wagner}, \& {Ward}}]{Hess_crab_06}
---. 2006{\natexlab{c}}, \aap, 457, 899

\bibitem[{{Aharonian} {et~al.}(2007{\natexlab{a}}){Aharonian}, {Akhperjanian},
  {Bazer-Bachi}, {Beilicke}, {Benbow}, {Berge}, {Bernl{\"o}hr}, {Boisson},
  {Bolz}, {Borrel}, {Braun}, {Brion}, {Brown}, {B{\"u}hler}, {B{\"u}sching},
  {Boutelier}, {Carrigan}, {Chadwick}, {Chounet}, {Coignet}, {Cornils},
  {Costamante}, {Degrange}, {Dickinson}, {Djannati-Ata{\"i}}, {O'C.~Drury},
  {Dubus}, {Egberts}, {Emmanoulopoulos}, {Espigat}, {Farnier}, {Feinstein},
  {Ferrero}, {Fiasson}, {Fontaine}, {Funk}, {Funk}, {F{\"u}{\ss}ling},
  {Gallant}, {Giebels}, {Glicenstein}, {Gl{\"u}ck}, {Goret}, {Hadjichristidis},
  {Hauser}, {Hauser}, {Heinzelmann}, {Henri}, {Hermann}, {Hinton}, {Hoffmann},
  {Hofmann}, {Holleran}, {Hoppe}, {Horns}, {Jacholkowska}, {de Jager},
  {Kendziorra}, {Kerschhaggl}, {Kh{\'e}lifi}, {Komin}, {Kosack}, {Lamanna},
  {Latham}, {Le Gallou}, {Lemi{\`e}re}, {Lemoine-Goumard}, {Lohse}, {Martin},
  {Martineau-Huynh}, {Marcowith}, {Masterson}, {Maurin}, {McComb}, {Moulin},
  {de Naurois}, {Nedbal}, {Nolan}, {Noutsos}, {Olive}, {Orford}, {Osborne},
  {Panter}, {Pelletier}, {Petrucci}, {Pita}, {P{\"u}hlhofer}, {Punch},
  {Ranchon}, {Raubenheimer}, {Raue}, {Rayner}, {Ripken}, {Rob}, {Rolland},
  {Rosier-Lees}, {Rowell}, {Sahakian}, {Santangelo}, {Saug{\'e}}, {Schlenker},
  {Schlickeiser}, {Schr{\"o}der}, {Schwanke}, {Schwarzburg}, {Schwemmer},
  {Shalchi}, {Sol}, {Spangler}, {Spanier}, {Steenkamp}, {Stegmann}, {Superina},
  {Tam}, {Tavernet}, {Terrier}, {Tluczykont}, {van Eldik}, {Vasileiadis},
  {Venter}, {Vialle}, {Vincent}, {V{\"o}lk}, {Wagner}, \&
  {Ward}}]{HESS_1ES1101_232}
---. 2007{\natexlab{a}}, \aap, 470, 475

\bibitem[{{Aharonian} {et~al.}(2007{\natexlab{b}}){Aharonian}, {Akhperjanian},
  {Barres de Almeida}, {Bazer-Bachi}, {Behera}, {Beilicke}, {Benbow},
  {Bernl{\"o}hr}, {Boisson}, {Bolz}, {Borrel}, {Braun}, {Brion}, {Brown},
  {B{\"u}hler}, {Bulik}, {B{\"u}sching}, {Boutelier}, {Carrigan}, {Chadwick},
  {Chounet}, {Clapson}, {Coignet}, {Cornils}, {Costamante}, {Dalton},
  {Degrange}, {Dickinson}, {Djannati-Ata{\"i}}, {Domainko}, {O'C.~Drury},
  {Dubois}, {Dubus}, {Dyks}, {Egberts}, {Emmanoulopoulos}, {Espigat},
  {Farnier}, {Feinstein}, {Fiasson}, {F{\"o}rster}, {Fontaine}, {Funk},
  {F{\"u}{\ss}ling}, {Gallant}, {Giebels}, {Glicenstein}, {Gl{\"u}ck}, {Goret},
  {Hadjichristidis}, {Hauser}, {Hauser}, {Heinzelmann}, {Henri}, {Hermann},
  {Hinton}, {Hoffmann}, {Hofmann}, {Holleran}, {Hoppe}, {Horns},
  {Jacholkowska}, {de Jager}, {Jung}, {Katarzy{\'n}ski}, {Kendziorra},
  {Kerschhaggl}, {Kh{\'e}lifi}, {Keogh}, {Komin}, {Kosack}, {Lamanna},
  {Latham}, {Lemi{\`e}re}, {Lemoine-Goumard}, {Lenain}, {Lohse}, {Martin},
  {Martineau-Huynh}, {Marcowith}, {Masterson}, {Maurin}, {Maurin}, {McComb},
  {Moderski}, {Moulin}, {de Naurois}, {Nedbal}, {Nolan}, {Ohm}, {Olive}, {de
  O{\~n}a Wilhelmi}, {Orford}, {Osborne}, {Ostrowski}, {Panter}, {Pedaletti},
  {Pelletier}, {Petrucci}, {Pita}, {P{\"u}hlhofer}, {Punch}, {Ranchon},
  {Raubenheimer}, {Raue}, {Rayner}, {Renaud}, {Ripken}, {Rob}, {Rolland},
  {Rosier-Lees}, {Rowell}, {Rudak}, {Ruppel}, {Sahakian}, {Santangelo},
  {Schlickeiser}, {Sch{\"o}ck}, {Schr{\"o}der}, {Schwanke}, {Schwarzburg},
  {Schwemmer}, {Shalchi}, {Sol}, {Spangler}, {Stawarz}, {Steenkamp},
  {Stegmann}, {Superina}, {Tam}, {Tavernet}, {Terrier}, {van Eldik},
  {Vasileiadis}, {Venter}, {Vialle}, {Vincent}, {Vivier}, {V{\"o}lk}, {Volpe},
  {Wagner}, {Ward}, {Zdziarski}, \& {Zech}}]{HESS_1ES0347}
---. 2007{\natexlab{b}}, \aap, 473, L25

\bibitem[{{Aharonian} {et~al.}(2007{\natexlab{c}}){Aharonian}, {Akhperjanian},
  {Barres de Almeida}, {Bazer-Bachi}, {Behera}, {Beilicke}, {Benbow},
  {Bernl{\"o}hr}, {Boisson}, {Bolz}, {Borrel}, {Braun}, {Brion}, {Brown},
  {B{\"u}hler}, {Bulik}, {B{\"u}sching}, {Boutelier}, {Carrigan}, {Chadwick},
  {Chounet}, {Clapson}, {Coignet}, {Cornils}, {Costamante}, {Dalton},
  {Degrange}, {Dickinson}, {Djannati-Ata{\"i}}, {Domainko}, {O'C.~Drury},
  {Dubois}, {Dubus}, {Dyks}, {Egberts}, {Emmanoulopoulos}, {Espigat},
  {Farnier}, {Feinstein}, {Fiasson}, {F{\"o}rster}, {Fontaine}, {Funk},
  {F{\"u}{\ss}ling}, {Gallant}, {Giebels}, {Glicenstein}, {Gl{\"u}ck}, {Goret},
  {Hadjichristidis}, {Hauser}, {Hauser}, {Heinzelmann}, {Henri}, {Hermann},
  {Hinton}, {Hoffmann}, {Hofmann}, {Holleran}, {Hoppe}, {Horns},
  {Jacholkowska}, {de Jager}, {Jung}, {Katarzy{\'n}ski}, {Kendziorra},
  {Kerschhaggl}, {Kh{\'e}lifi}, {Keogh}, {Komin}, {Kosack}, {Lamanna},
  {Latham}, {Lemi{\`e}re}, {Lemoine-Goumard}, {Lenain}, {Lohse}, {Martin},
  {Martineau-Huynh}, {Marcowith}, {Masterson}, {Maurin}, {Maurin}, {McComb},
  {Moderski}, {Moulin}, {de Naurois}, {Nedbal}, {Nolan}, {Ohm}, {Olive}, {de
  O{\~n}a Wilhelmi}, {Orford}, {Osborne}, {Ostrowski}, {Panter}, {Pedaletti},
  {Pelletier}, {Petrucci}, {Pita}, {P{\"u}hlhofer}, {Punch}, {Ranchon},
  {Raubenheimer}, {Raue}, {Rayner}, {Renaud}, {Ripken}, {Rob}, {Rolland},
  {Rosier-Lees}, {Rowell}, {Rudak}, {Ruppel}, {Sahakian}, {Santangelo},
  {Schlickeiser}, {Sch{\"o}ck}, {Schr{\"o}der}, {Schwanke}, {Schwarzburg},
  {Schwemmer}, {Shalchi}, {Sol}, {Spangler}, {Stawarz}, {Steenkamp},
  {Stegmann}, {Superina}, {Tam}, {Tavernet}, {Terrier}, {van Eldik},
  {Vasileiadis}, {Venter}, {Vialle}, {Vincent}, {Vivier}, {V{\"o}lk}, {Volpe},
  {Wagner}, {Ward}, {Zdziarski}, \& {Zech}}]{HESS_1ES0229}
---. 2007{\natexlab{c}}, \aap, 475, L9

\bibitem[{{Aharonian} {et~al.}(2008){Aharonian}, {Akhperjanian}, {Barres de
  Almeida}, {Bazer-Bachi}, {Behera}, {Beilicke}, {Benbow}, {Bernl{\"o}hr},
  {Boisson}, {Borrel}, {Braun}, {Brion}, {Brucker}, {B{\"u}hler}, {Bulik},
  {B{\"u}sching}, {Boutelier}, {Carrigan}, {Chadwick}, {Chaves}, {Chounet},
  {Clapson}, {Coignet}, {Cornils}, {Costamante}, {Dalton}, {Degrange},
  {Dickinson}, {Djannati-Ata{\"i}}, {Domainko}, {O'C.~Drury}, {Dubois},
  {Dubus}, {Dyks}, {Egberts}, {Emmanoulopoulos}, {Espigat}, {Farnier},
  {Feinstein}, {Fiasson}, {F{\"o}rster}, {Fontaine}, {F{\"u}{\ss}ling},
  {Gabici}, {Gallant}, {Giebels}, {Glicenstein}, {Gl{\"u}ck}, {Goret},
  {Hadjichristidis}, {Hauser}, {Hauser}, {Heinzelmann}, {Henri}, {Hermann},
  {Hinton}, {Hoffmann}, {Hofmann}, {Holleran}, {Hoppe}, {Horns},
  {Jacholkowska}, {de Jager}, {Jung}, {Katarzy{\'n}ski}, {Kaufmann},
  {Kendziorra}, {Kerschhaggl}, {Khangulyan}, {Kh{\'e}lifi}, {Keogh}, {Komin},
  {Kosack}, {Lamanna}, {Latham}, {Lenain}, {Lohse}, {Martin},
  {Martineau-Huynh}, {Marcowith}, {Masterson}, {Maurin}, {McComb}, {Moderski},
  {Moulin}, {Naumann-Godo}, {de Naurois}, {Nedbal}, {Nekrassov}, {Nolan},
  {Ohm}, {Olive}, {de O{\~n}a Wilhelmi}, {Orford}, {Osborne}, {Ostrowski},
  {Panter}, {Pedaletti}, {Pelletier}, {Petrucci}, {Pita}, {P{\"u}hlhofer},
  {Punch}, {Quirrenbach}, {Raubenheimer}, {Raue}, {Rayner}, {Renaud}, {Rieger},
  {Ripken}, {Rob}, {Rosier-Lees}, {Rowell}, {Rudak}, {Ruppel}, {Sahakian},
  {Santangelo}, {Schlickeiser}, {Sch{\"o}ck}, {Schr{\"o}der}, {Schwanke},
  {Schwarzburg}, {Schwemmer}, {Shalchi}, {Sol}, {Spangler}, {Stawarz},
  {Steenkamp}, {Stegmann}, {Superina}, {Tam}, {Tavernet}, {Terrier}, {van
  Eldik}, {Vasileiadis}, {Venter}, {Vialle}, {Vincent}, {Vivier}, {V{\"o}lk},
  {Volpe}, {Wagner}, {Ward}, {Zdziarski}, \& {Zech}}]{HESS_RGBJ0152}
---. 2008, \aap, 481, L103

\bibitem[{{Aharonian} {et~al.}(1994){Aharonian}, {Coppi}, \&
  {Voelk}}]{AharonianCoppi94}
{Aharonian}, F.~A., {Coppi}, P.~S., \& {Voelk}, H.~J. 1994, \apjl, 423, L5

\bibitem[{{Ahlers}(2011)}]{Ahlers2011}
{Ahlers}, M. 2011, \prd, 84, 063006

\bibitem[{{Albert} {et~al.}(2006){Albert}, {Aliu}, {Anderhub}, {Antoranz},
  {Armada}, {Asensio}, {Baixeras}, {Barrio}, {Bartelt}, {Bartko}, {Bastieri},
  {Bavikadi}, {Bednarek}, {Berger}, {Bigongiari}, {Biland}, {Bisesi}, {Bock},
  {Bretz}, {Britvitch}, {Camara}, {Chilingarian}, {Ciprini}, {Coarasa},
  {Commichau}, {Contreras}, {Cortina}, {Curtef}, {Danielyan}, {Dazzi}, {De
  Angelis}, {de los Reyes}, {De Lotto}, {Domingo-Santamar{\'{\i}}a}, {Dorner},
  {Doro}, {Errando}, {Fagiolini}, {Ferenc}, {Fern{\'a}ndez}, {Firpo}, {Flix},
  {Fonseca}, {Font}, {Galante}, {Garczarczyk}, {Gaug}, {Giller}, {Goebel},
  {Hakobyan}, {Hayashida}, {Hengstebeck}, {H{\"o}hne}, {Hose}, {Jacon},
  {Kalekin}, {Kranich}, {Laille}, {Lenisa}, {Liebing}, {Lindfors}, {Longo},
  {L{\'o}pez}, {L{\'o}pez}, {Lorenz}, {Lucarelli}, {Majumdar}, {Maneva},
  {Mannheim}, {Mariotti}, {Mart{\'{\i}}nez}, {Mase}, {Mazin}, {Meucci},
  {Meyer}, {Miranda}, {Mirzoyan}, {Mizobuchi}, {Moralejo}, {Nilsson},
  {O{\~n}a-Wilhelmi}, {Ordu{\~n}a}, {Otte}, {Oya}, {Paneque}, {Paoletti},
  {Pasanen}, {Pascoli}, {Pauss}, {Pavel}, {Pegna}, {Persic}, {Peruzzo},
  {Piccioli}, {Poller}, {Prandini}, {Rhode}, {Rico}, {Riegel}, {Rissi},
  {Robert}, {R{\"u}gamer}, {Saggion}, {S{\'a}nchez}, {Sartori}, {Scalzotto},
  {Schmitt}, {Schweizer}, {Shayduk}, {Shinozaki}, {Shore}, {Sidro},
  {Sillanp{\"a}{\"a}}, {Sobczy{\'n}ska}, {Stamerra}, {Stark}, {Takalo},
  {Temnikov}, {Tescaro}, {Teshima}, {Tonello}, {Torres}, {Torres}, {Turini},
  {Vankov}, {Vardanyan}, {Vitale}, {Wagner}, {Wibig}, {Wittek}, \&
  {Zapatero}}]{MAGIC_1ES1218_2006}
{Albert}, J., {et~al.} 2006, \apjl, 642, L119

\bibitem[{{Berta} {et~al.}(2010){Berta}, {Magnelli}, {Lutz}, {Altieri},
  {Aussel}, {Andreani}, {Bauer}, {Bongiovanni}, {Cava}, {Cepa}, {Cimatti},
  {Daddi}, {Dominguez}, {Elbaz}, {Feuchtgruber}, {F{\"o}rster Schreiber},
  {Genzel}, {Gruppioni}, {Katterloher}, {Magdis}, {Maiolino}, {Nordon},
  {P{\'e}rez Garc{\'{\i}}a}, {Poglitsch}, {Popesso}, {Pozzi}, {Riguccini},
  {Rodighiero}, {Saintonge}, {Santini}, {Sanchez-Portal}, {Shao}, {Sturm},
  {Tacconi}, {Valtchanov}, {Wetzstein}, \& {Wieprecht}}]{Berta_Herschel_2010}
{Berta}, S., {et~al.} 2010, \aap, 518, L30

\bibitem[{{B{\'e}thermin} {et~al.}(2010){B{\'e}thermin}, {Dole}, {Beelen}, \&
  {Aussel}}]{Bethermin_Spitzer_2010}
{B{\'e}thermin}, M., {Dole}, H., {Beelen}, A., \& {Aussel}, H. 2010, \aap, 512,
  A78

\bibitem[{{Biermann}(1950)}]{Biermann1950}
{Biermann}, L. 1950, Zeitschrift Naturforschung Teil A, 5, 65

\bibitem[{{Blasi} {et~al.}(1999){Blasi}, {Burles}, \& {Olinto}}]{Blasi1999}
{Blasi}, P., {Burles}, S., \& {Olinto}, A.~V. 1999, \apjl, 514, L79

\bibitem[{{Broderick} {et~al.}(2012){Broderick}, {Chang}, \&
  {Pfrommer}}]{Broderick_instability}
{Broderick}, A.~E., {Chang}, P., \& {Pfrommer}, C. 2012, \apj, 752, 22

\bibitem[{{Carilli} \& {Taylor}(2002)}]{Carilli2002}
{Carilli}, C.~L., \& {Taylor}, G.~B. 2002, \araa, 40, 319

\bibitem[{{Dermer} {et~al.}(2011){Dermer}, {Cavadini}, {Razzaque}, {Finke},
  {Chiang}, \& {Lott}}]{Dermer2010}
{Dermer}, C.~D., {Cavadini}, M., {Razzaque}, S., {Finke}, J.~D., {Chiang}, J.,
  \& {Lott}, B. 2011, \apjl, 733, L21

\bibitem[{{Dolag} {et~al.}(2009){Dolag}, {Kachelrie{\ss}}, {Ostapchenko}, \&
  {Tom{\`a}s}}]{Dolag09}
{Dolag}, K., {Kachelrie{\ss}}, M., {Ostapchenko}, S., \& {Tom{\`a}s}, R. 2009,
  \apj, 703, 1078

\bibitem[{{Dolag} {et~al.}(2011){Dolag}, {Kachelriess}, {Ostapchenko}, \&
  {Tom{\`a}s}}]{Dolag2010}
{Dolag}, K., {Kachelriess}, M., {Ostapchenko}, S., \& {Tom{\`a}s}, R. 2011,
  \apjl, 727, L4+

\bibitem[{{Dole} {et~al.}(2006){Dole}, {Lagache}, {Puget}, {Caputi},
  {Fern{\'a}ndez-Conde}, {Le Floc'h}, {Papovich}, {P{\'e}rez-Gonz{\'a}lez},
  {Rieke}, \& {Blaylock}}]{Dole_Spitzer_2006}
{Dole}, H., {et~al.} 2006, \aap, 451, 417

\bibitem[{{Dom{\'{\i}}nguez} {et~al.}(2011){Dom{\'{\i}}nguez}, {Primack},
  {Rosario}, {Prada}, {Gilmore}, {Faber}, {Koo}, {Somerville},
  {P{\'e}rez-Torres}, {P{\'e}rez-Gonz{\'a}lez}, {Huang}, {Davis},
  {Guhathakurta}, {Barmby}, {Conselice}, {Lozano}, {Newman}, \&
  {Cooper}}]{Dominguez_ebl_2010}
{Dom{\'{\i}}nguez}, A., {et~al.} 2011, \mnras, 410, 2556

\bibitem[{{Elbaz} {et~al.}(2002){Elbaz}, {Cesarsky}, {Chanial}, {Aussel},
  {Franceschini}, {Fadda}, \& {Chary}}]{Elbaz_ISOCAM_2002}
{Elbaz}, D., {Cesarsky}, C.~J., {Chanial}, P., {Aussel}, H., {Franceschini},
  A., {Fadda}, D., \& {Chary}, R.~R. 2002, \aap, 384, 848

\bibitem[{{Elyiv} {et~al.}(2009){Elyiv}, {Neronov}, \& {Semikoz}}]{ENS09}
{Elyiv}, A., {Neronov}, A., \& {Semikoz}, D.~V. 2009, \prd, 80, 023010

\bibitem[{{Essey} {et~al.}(2011{\natexlab{a}}){Essey}, {Ando}, \&
  {Kusenko}}]{Essey2010}
{Essey}, W., {Ando}, S., \& {Kusenko}, A. 2011{\natexlab{a}}, Astroparticle
  Physics, 35, 135

\bibitem[{{Essey} {et~al.}(2011{\natexlab{b}}){Essey}, {Kalashev}, {Kusenko},
  \& {Beacom}}]{Essey_protons}
{Essey}, W., {Kalashev}, O., {Kusenko}, A., \& {Beacom}, J.~F.
  2011{\natexlab{b}}, \apj, 731, 51

\bibitem[{{Eungwanichayapant} \& {Aharonian}(2009)}]{Eungwan09}
{Eungwanichayapant}, A., \& {Aharonian}, F. 2009, International Journal of
  Modern Physics D, 18, 911

\bibitem[{{Fazio} {et~al.}(2004){Fazio}, {Hora}, {Allen}, {Ashby}, {Barmby},
  {Deutsch}, {Huang}, {Kleiner}, {Marengo}, {Megeath}, {Melnick}, {Pahre},
  {Patten}, {Polizotti}, {Smith}, {Taylor}, {Wang}, {Willner}, {Hoffmann},
  {Pipher}, {Forrest}, {McMurty}, {McCreight}, {McKelvey}, {McMurray}, {Koch},
  {Moseley}, {Arendt}, {Mentzell}, {Marx}, {Losch}, {Mayman}, {Eichhorn},
  {Krebs}, {Jhabvala}, {Gezari}, {Fixsen}, {Flores}, {Shakoorzadeh}, {Jungo},
  {Hakun}, {Workman}, {Karpati}, {Kichak}, {Whitley}, {Mann}, {Tollestrup},
  {Eisenhardt}, {Stern}, {Gorjian}, {Bhattacharya}, {Carey}, {Nelson},
  {Glaccum}, {Lacy}, {Lowrance}, {Laine}, {Reach}, {Stauffer}, {Surace},
  {Wilson}, {Wright}, {Hoffman}, {Domingo}, \& {Cohen}}]{Fazio_IRAC_2004}
{Fazio}, G.~G., {et~al.} 2004, \apjs, 154, 10

\bibitem[{{Franceschini} {et~al.}(2008){Franceschini}, {Rodighiero}, \&
  {Vaccari}}]{Franceschini2008}
{Franceschini}, A., {Rodighiero}, G., \& {Vaccari}, M. 2008, \aap, 487, 837

\bibitem[{Gould \& Schr\'eder(1967)}]{Gould1967}
Gould, R.~J., \& Schr\'eder, G.~P. 1967, Phys. Rev., 155, 1404

\bibitem[{{Grasso} \& {Rubinstein}(2001)}]{GrassoRubinstein01}
{Grasso}, D., \& {Rubinstein}, H.~R. 2001, \physrep, 348, 163

\bibitem[{{Huan} {et~al.}(2011){Huan}, {Weisgarber}, {Arlen}, \&
  {Wakely}}]{HaoHuan2011}
{Huan}, H., {Weisgarber}, T., {Arlen}, T., \& {Wakely}, S.~P. 2011, \apjl, 735,
  L28+

\bibitem[{{Ichiki} {et~al.}(2008){Ichiki}, {Inoue}, \& {Takahashi}}]{Ichiki08}
{Ichiki}, K., {Inoue}, S., \& {Takahashi}, K. 2008, \apj, 682, 127

\bibitem[{Jelley(1966)}]{Jelley1966}
Jelley, J.~V. 1966, Phys. Rev. Lett., 16, 479

\bibitem[{{Kneiske} \& {Dole}(2010)}]{Kneiske2010}
{Kneiske}, T.~M., \& {Dole}, H. 2010, \aap, 515, A19

\bibitem[{{Kronberg}(1994)}]{Kronberg94}
{Kronberg}, P.~P. 1994, \nat, 370, 179

\bibitem[{{Kronberg} {et~al.}(2001){Kronberg}, {Dufton}, {Li}, \&
  {Colgate}}]{Kronberg2001}
{Kronberg}, P.~P., {Dufton}, Q.~W., {Li}, H., \& {Colgate}, S.~A. 2001, \apj,
  560, 178

\bibitem[{{Kronberg} \& {Perry}(1982)}]{Kronberg1982}
{Kronberg}, P.~P., \& {Perry}, J.~J. 1982, \apj, 263, 518

\bibitem[{{Kulsrud} \& {Zweibel}(2008)}]{Kulsrud_Zweibel_2008}
{Kulsrud}, R.~M., \& {Zweibel}, E.~G. 2008, Reports on Progress in Physics, 71,
  046901

\bibitem[{{Madau} \& {Pozzetti}(2000)}]{MadauPozzetti_2000}
{Madau}, P., \& {Pozzetti}, L. 2000, \mnras, 312, L9

\bibitem[{{Matsuura} {et~al.}(2011){Matsuura}, {Shirahata}, {Kawada},
  {Takeuchi}, {Burgarella}, {Clements}, {Jeong}, {Hanami}, {Khan}, {Matsuhara},
  {Nakagawa}, {Oyabu}, {Pearson}, {Pollo}, {Serjeant}, {Takagi}, \&
  {White}}]{Matsuura_Akari_2011}
{Matsuura}, S., {et~al.} 2011, \apj, 737, 2

\bibitem[{{Mazin} \& {Raue}(2007)}]{MazinRaue07}
{Mazin}, D., \& {Raue}, M. 2007, \aap, 471, 439

\bibitem[{{Miniati} \& {Elyiv}(2012)}]{Miniati_Elyiv_Voids}
{Miniati}, F., \& {Elyiv}, A. 2012, ArXiv e-prints

\bibitem[{{Murase} {et~al.}(2012){Murase}, {Dermer}, {Takami}, \&
  {Migliori}}]{Murase_2012_UHECR}
{Murase}, K., {Dermer}, C.~D., {Takami}, H., \& {Migliori}, G. 2012, \apj, 749,
  63

\bibitem[{{Murase} {et~al.}(2008){Murase}, {Takahashi}, {Inoue}, {Ichiki}, \&
  {Nagataki}}]{Murase08}
{Murase}, K., {Takahashi}, K., {Inoue}, S., {Ichiki}, K., \& {Nagataki}, S.
  2008, \apjl, 686, L67

\bibitem[{{Neronov} {et~al.}(2010){Neronov}, {Semikoz}, {Kachelriess},
  {Ostapchenko}, \& {Elyiv}}]{Neronov_DegreeScaleJets_2011}
{Neronov}, A., {Semikoz}, D., {Kachelriess}, M., {Ostapchenko}, S., \& {Elyiv},
  A. 2010, \apjl, 719, L130

\bibitem[{{Neronov} \& {Semikoz}(2006)}]{NeronovSemikoz06}
{Neronov}, A., \& {Semikoz}, D.~V. 2006, ArXiv Astrophysics e-prints

\bibitem[{{Neronov} \& {Semikoz}(2009)}]{NeronovSemikoz_Sensitivity09}
---. 2009, \prd, 80, 123012

\bibitem[{{Neronov} \& {Vovk}(2010)}]{Neronov2010}
{Neronov}, A., \& {Vovk}, I. 2010, Science, 328, 73

\bibitem[{{Pan} {et~al.}(2012){Pan}, {Vogeley}, {Hoyle}, {Choi}, \&
  {Park}}]{Pan_Voids_2012}
{Pan}, D.~C., {Vogeley}, M.~S., {Hoyle}, F., {Choi}, Y.-Y., \& {Park}, C. 2012,
  \mnras, 421, 926

\bibitem[{{Papovich} {et~al.}(2004){Papovich}, {Dole}, {Egami}, {Le Floc'h},
  {P{\'e}rez-Gonz{\'a}lez}, {Alonso-Herrero}, {Bai}, {Beichman}, {Blaylock},
  {Engelbracht}, {Gordon}, {Hines}, {Misselt}, {Morrison}, {Mould},
  {Muzerolle}, {Neugebauer}, {Richards}, {Rieke}, {Rieke}, {Rigby}, {Su}, \&
  {Young}}]{Papovich_Spitzer_2004}
{Papovich}, C., {et~al.} 2004, \apjs, 154, 70

\bibitem[{{Perkins} \& {VERITAS Collaboration}(2010)}]{PerkinsPoster2010}
{Perkins}, J.~S., \& {VERITAS Collaboration}. 2010, in Bulletin of the American
  Astronomical Society, Vol.~42, AAS/High Energy Astrophysics Division \#11,
  708

\bibitem[{{Plaga}(1995)}]{Plaga95}
{Plaga}, R. 1995, \nat, 374, 430

\bibitem[{{Ryu} {et~al.}(2008){Ryu}, {Kang}, {Cho}, \& {Das}}]{Ryu_2008}
{Ryu}, D., {Kang}, H., {Cho}, J., \& {Das}, S. 2008, Science, 320, 909

\bibitem[{{Sutter} {et~al.}(2012){Sutter}, {Lavaux}, {Wandelt}, \&
  {Weinberg}}]{Sutter_VoidCat}
{Sutter}, P.~M., {Lavaux}, G., {Wandelt}, B.~D., \& {Weinberg}, D.~H. 2012,
  ArXiv e-prints

\bibitem[{{Tavecchio} {et~al.}(2011){Tavecchio}, {Ghisellini}, {Bonnoli}, \&
  {Foschini}}]{Tavecchio_extreme_2011}
{Tavecchio}, F., {Ghisellini}, G., {Bonnoli}, G., \& {Foschini}, L. 2011,
  \mnras, 414, 3566

\bibitem[{{Tavecchio} {et~al.}(2010){Tavecchio}, {Ghisellini}, {Foschini},
  {Bonnoli}, {Ghirlanda}, \& {Coppi}}]{Tavecchio2010}
{Tavecchio}, F., {Ghisellini}, G., {Foschini}, L., {Bonnoli}, G., {Ghirlanda},
  G., \& {Coppi}, P. 2010, \mnras, 406, L70

\bibitem[{{Taylor} {et~al.}(2011){Taylor}, {Vovk}, \& {Neronov}}]{Taylor2011}
{Taylor}, A.~M., {Vovk}, I., \& {Neronov}, A. 2011, \aap, 529, A144

\bibitem[{{Urry} \& {Padovani}(1995)}]{Urry_Padovani_1995}
{Urry}, C.~M., \& {Padovani}, P. 1995, \pasp, 107, 803

\bibitem[{{Vassiliev}(2000)}]{Vassiliev2000}
{Vassiliev}, V.~V. 2000, Astroparticle Physics, 12, 217

\bibitem[{{Vovk} {et~al.}(2012){Vovk}, {Taylor}, {Semikoz}, \&
  {Neronov}}]{Vovk_EBL_IGMF_2012}
{Vovk}, I., {Taylor}, A.~M., {Semikoz}, D., \& {Neronov}, A. 2012, \apjl, 747,
  L14

\bibitem[{{Widrow}(2002)}]{Widrow02}
{Widrow}, L.~M. 2002, Reviews of Modern Physics, 74, 775

\bibitem[{{Zweibel}(2006)}]{Zweibel2006}
{Zweibel}, E.~G. 2006, Astronomische Nachrichten, 327, 505

\end{thebibliography}
